\documentclass[12pt]{article}

\usepackage{amstext,amsmath,amssymb,amsfonts,bbm}
\usepackage[latin1]{inputenc}
\usepackage{epsfig}
\usepackage{dsfont}
\usepackage{hyperref}
\usepackage{amsthm}
\usepackage{subfigure}
\usepackage{color}
\usepackage{multirow}
\usepackage{psfrag}
\usepackage{graphicx}
\usepackage{extarrows}

\theoremstyle{plain}

\theoremstyle{definition}

\usepackage{amstext,amsmath,amssymb,amsfonts,bbm,slashed}
\usepackage[latin1]{inputenc}
\usepackage{epsfig}
\usepackage{hyperref}
\usepackage{amsthm}
\usepackage{subfigure}
\usepackage{color}
\usepackage{multirow}
\usepackage{calc}

\usepackage{stmaryrd}

\usepackage{cancel}

\usepackage{authblk}

\newcommand{\bea}{\begin{eqnarray}}
\newcommand{\eea}{\end{eqnarray}}
\newcommand{\beq}{\begin{equation}}
\newcommand{\eeq}{\end{equation}}

\begin{document}

\newcommand{\A}{A}
\newcommand{\M}{M}

\title{Polchinski's exact renormalisation group for tensorial theories: Gau\ss ian universality and power counting}
\author[1]{Thomas Krajewski\thanks{thomas.krajewski@cpt.univ-mrs.fr}}
\author[1,2,3]{Reiko Toriumi\thanks{torirei@gmail.com}}
\affil[1]{\small \sl Centre de Physique Th\'eorique, CNRS UMR 7332, Aix-Marseille Universit\'e,  Marseille, France}
\affil[2]{\small \sl Department of Physics and Astronomy, San Francisco State University, San Francisco, CA 94132 USA }
\affil[3]{\small \sl Department of Physics, University of California, Berkeley,  CA 94720 USA}

\maketitle

\abstract{In this paper, we use the exact renormalisation in the context of tensor models and tensorial group field theories. As a byproduct, we rederive Gau\ss ian universality for random tensors and provide a general power counting for Abelian tensorial field theories with a closure constraint, leading us to a only five renormalizable theories.}

%%%%%%%%%%%%%%%%%%%%%%%%%%%%%
\section{Introduction}
%%%%%%%%%%%%%%%%%%%%%%%%%%%%%

Constructing a quantum theory of gravity remains one of the most tantalising open problems in theoretical physics. There are currently multiple approaches: string theory, loop quantum gravity, non commutative quantum field theory, asymptotic safety, dynamical triangulations, and Regge calculus, to name a few.
Here we deal with 
tensor models and group field theories 
that have initially been introduced in the context of quantum gravity, merging ideas from loop quantum gravity,  dynamical triangulations and Regge calculus (see for instance \cite{tensortrack} and \cite{GFT} for a recent review).

Tensor models are higher dimensional generalisations of matrix models whose basic variable is a rank $D$ tensor $T_{i_{1}\dots i_{D}}$ instead of a matrix $M_{i_{1}i_{2}}$. Matrices are useful in two dimensional quantum gravity because their perturbative expansions are sums over ribbon graphs that reproduce triangulated surfaces. Perturbative expansions of higher rank tensors provide sums over $D$-dimensional triangulations which may be considered as a first step towards a path integral over geometries. In 
group field theories, 
the discrete indices are replaced by group elements (typically the rotation or Lorentz group) whose perturbative expansion reproduces a sum over spin foam amplitudes.

While tensor models and group field theories have been known since more than two decades, decisive breakthroughs only occurred in the recent years, thanks to the advent of colored models \cite{colored}. Among these important results, let us mention the Gau\ss ian universality for random tensors \cite{universality} and the construction of renormalisable tensorial group field theories, first in the Abelian case \cite{renormabelian}, then in the non Abelian case \cite{renormnonabelian}.

In this letter, we are interested in the applications of the exact renormalisation group equation (ERGE), in the form derived by Polchinski \cite{Polchinski}, to tensor models and tensorial field theories. This is a continuation of our previous work \cite{Corfu}, where we have formulated Polchinski's ERGE in the tensorial context. In particular, we revisit two of the aforementioned results: Gau\ss ian universality and the Abelian power counting for theories with a closure constraint. While these results are known, we want to stress that the technique we use here is radically different: we do not compute any Feynman graph but simply use the scaling properties of the ERGE. 

 Note that the ERGE (in the form of Wetterich's equation) has also been studied (\cite{astrid1} \cite{astrid2}) and used to investigate the fixed point structure of tensor models \cite{tensorERGE} and  tensorial field theories \cite{4epsilon} and 
\cite{Lahoche}. A discrete formulation of the flow equation, based on multiscale analysis, has also been propsed in \cite{Discreterenormalisation}.

It is organised as follows. In section 2 we present tensor models and their effective actions. Then, we derive the ERGE in section 3 and prove Gau\ss ian universality in section 4. Section 5 deals with tensorial field theories and their power counting.  An appendix collects some explicit low order computations in the invariant tensor models, for $D=3$ and $D=4$.

%%%%%%%%%%%%%%%%%%%%%%%%%%%%%
\section{Random tensor models and their effective actions}
%%%%%%%%%%%%%%%%%%%%%%%%%%%%%

Random tensors are generalisations of random matrices whose basic variables are a rank $D$ complex tensor $\Phi_{i_{1},\dots,i_{D}}$ and its complex conjugate $\overline{\Phi}_{i_{1},\dots,i_{D}}$, with indices $i_{k}\in\left\{1,\dots,N\right\}$, the integer $N$ being the size of the tensor. The tensors are not supposed to obey any invariance properties under permutations of its indices, as is the case in colored models. 

In general, the main objects of interest are the expectation values of certain observables $O$,
\begin{equation}
\langle O\rangle =\frac{1}{Z}\int d\overline{\Phi}d\Phi\, 
O(\overline{\Phi},\Phi )\,\exp\left\{-\overline{\Phi}\cdot C^{-1}\cdot\Phi+S(\overline{\Phi},\Phi)\right\},
\end{equation}
where $Z=\int d\overline{\Phi}d\Phi \exp\left\{-\overline{\Phi}\cdot C^{-1}\cdot\Phi+S(\overline{\Phi},\Phi)\right\}$ is the partition function. $C$ is the covariance and $S$ the interacting part of the action. Because the tensor is complex, it is convenient to consider its complex conjugate $\overline{\Phi}_{\overline{i}_{1},\dots,\overline{i}_{D}}$
as an independent variable. As a convention, we always denote by a bar the indices of $\overline{i}$, and is another index independent of $i$.

The precise form of the argument of the exponential is dictated by the fact that there are particularly simple tensor models which are the Gau\ss ian ones. The latter are defined by retaining only quadratic terms in the argument of the exponential. Introducing a multi-index $I=(i_{1},\dots,i_{D})$, the latter are defined by the quadratic term
\begin{equation}
\overline{\Phi}\cdot C^{-1}\cdot\Phi=\sum_{I,\overline{I}}
\overline{\Phi}_{\overline{I}}C^{-1}_{\overline{I},I}\Phi_{I}
=\sum_{i_{1},\dots,i_{D},\overline{i}_{1},\dots,\overline{i}_{D}}
\overline{\Phi}_{\overline{i}_{1},\dots,\overline{i}_{D}}C^{-1}_{\overline{i}_{1},\dots,\overline{i}_{D},i_{1},\dots,i_{D}}\Phi_{i_{1},\dots,i_{D}}.
\end{equation}
In the Gau\ss ian case, the expectation value of products of tensors is evaluated using Wick's theorem
\begin{equation}
{\cal N}^{-1}\int  d\overline{\Phi}d\Phi \exp\left\{-\overline{\Phi}\cdot C^{-1}\cdot\Phi\right\}
\Phi_{I_{1}}\dots\Phi_{I_{n}} \overline{\Phi}_{J_{1}}\dots\overline{\Phi}_{J_{n}} 
=\sum_{\sigma\in\mathfrak{S}_{n}}
C_{I_{1},J_{\sigma(1)}}
\dots C_{I_{n},J_{\sigma(n)}},
\label{Wick}
\end{equation}
and vanishes if there is not an equal number of $\Phi$ and $\overline{\Phi}$.
$\sigma\in\mathfrak{S}_{n}$ is a permutation of $n$ elements, {\it{i.e.,}} a bijection from $\left\{1,2.\dots,n\right\}$ onto itself. ${\cal N}=\int  d\overline{\Phi}d\Phi \exp\left\{-\overline{\Phi}\cdot C^{-1}\cdot\Phi\right\}$ is a normalisation factor and the positive hermitian matrices $C_{I,J}$ and $C^{-1}_{I,J}$ are inverse of the other. Note that the covariance $C$ is enough to define a Gau\ss ian integration and the latter makes sense even if $C$ is not invertible, tensors in the kernel of $C_{I,J}$ are simply set to 0.  This can be easily done
by adding a $\epsilon>0$ to $C$ to get $C_{I,J}+\epsilon\delta_{I,J}$ instead of $C_{I,J}$. In that case, $\epsilon$ can be chosen in such a way that $C_{I,J}+\epsilon\delta_{I,J}$ is invertible. Then, the limit $\epsilon\rightarrow 0$ provides the Gau\ss ian integration formula \eqref{Wick}, {whether} $C$ is invertible or not.

We are interested in perturbations of a Gau\ss ian measure using a suitable class of functions $S(\overline{\Phi},\Phi)$ that can be constructed using bubbles. Recall that for a rank $D$ tensor, a bubble ${\cal B}$ is $D$-colored bipartite graph. This means that there is an equal number of white and black vertices and that any edge connects a black vertex and a white vertex, Moreover, we assume that exactly $D$ edges are incident to each vertex and that every edge can be assigned a color in $\left\{1,\dots,D\right\}$ in such a way that all the edges incident to a given vertex carry different colors. Note that we do not impose any connectivity requirements on the bubbles ${\cal B}$. Then, the action is expanded over bubbles as
\begin{equation}
S(\Phi,\overline{\Phi})=\sum_{\cal B}
\text{Tr}_{\cal B,\lambda_{\cal B}}(\Phi,\overline{\Phi})\quad
\text{with}\quad \text{Tr}_{\cal B,\lambda_{\cal B}}(\Phi,\overline{\Phi})=\frac{1}{\sigma_{\cal B}}
\sum_{\left\{i_{e},\overline{i}_{e}\right\}}
\lambda_{{\cal B}}(\left\{i_{e},\overline{i}_{e}\right\})
\prod_{v}\Phi_{I_{\cal B}(v)}
\prod_{\overline{v}}\overline{\Phi}_{\overline{I}_{\cal B}(\overline{v})},
\label{tensor_bubble}
\end{equation}
where the sum runs over the pair of indices $\left\{i_{e},\overline{i}_{e}\right\}$  ranging from 1 to $N$ for  each edge of ${\cal B}$.  $I_{\cal B}(v)$ (resp. $\overline{I}_{\cal B}(\overline{v})$) is the $D$-tuple of indices $\left\{i_{e}\right\}$ (resp. $\left\{\overline{i}_{e}\right\}$) pertaining to the lines of color $1,\dots,D$ attached to the white vertex $v$ (resp. black vertex $\overline{v}$). The symmetry factor $\sigma_{\cal B}$ is equal to the order of the symmetry group of ${\cal B}$ (transformations of edges and vertices that preserve incidence relations and colorings) and is included for further convenience. Note that we generically consider bubbles differing solely by their color assignment as different. If not, we can always collect contributions from such bubbles into a single term. Finally, $\lambda_{\cal B}(\left\{i_{e},\overline{i}_{e}\right\})$ plays a role analogue to a coupling constant. The notation suggests that this generalises the expansion of a matrix interaction over traces of products of $MM^{\dagger}$.

If we require the action to be invariant under the $\text{U}(N)^{D}$ transformation
\begin{equation}
\Phi_{i_{1},\dots,i_{D}}\rightarrow\sum_{j_{1},\dots,j_{D}}U^{1}_{i_{1}j_{1}}\cdots U^{D}_{i_{D}j_{D}}\Phi_{j_{1},\dots,j_{D}},
\qquad
\overline{\Phi}_{i_{1},\dots,i_{D}}\rightarrow\sum_{j_{1},\dots,j_{D}}\overline{U}^{1}_{i_{1}j_{1}}\cdots \overline{U}^{D}_{i_{D}j_{D}}\overline{\Phi}_{j_{1},\dots,j_{D}},
\end{equation}
then the couplings must be expressed in terms of the unique rank 2 invariant tensor, 
$\lambda_{\cal B}(\left\{i_{e}\right\})=\lambda_{B}\prod_{e}\delta_{i_{e},\overline{i}_{e}}$, with $\lambda_{\cal B}$ a scalar. This is the conventional setting for random tensors which we refer to as the invariant random tensors. Our generalisation has been introduced to encompass the case of a covariance that depends on indices (with a view to write a renormalisation group equation) and to extend easily to tensorial field theories (see section \ref{GFT:sec}).

\begin{figure}
\centering
\begin{subfigure}[Dipole graph (Gau\ss ian measure)]{
\parbox{3.5cm}{\includegraphics[width=3cm]{dipole}}
\hspace{1cm}$
\sum_{i_{a}}\overline{T}_{i_{1},i_{2},i_{3}}
T _{i_{1}i_{2}i_{3}}
$}\\
\vspace{0.5cm}
\end{subfigure}\hspace{2cm}
\begin{subfigure}[Degree 6 interaction]{
\parbox{3cm}{\includegraphics[width=3cm]{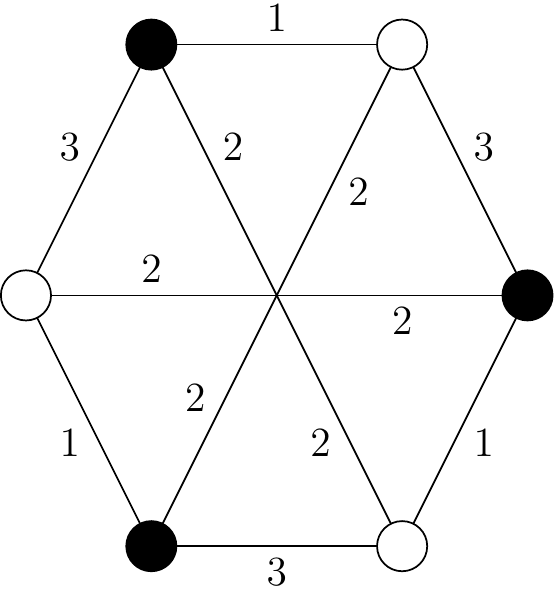}}
\hspace{1cm}
$
\sum_{i_{a},\,j_{b},\,k_{c}}
\overline{T}_{i_{1}i_{2}i_{3}}\overline{T}_{j_{1}j_{2}j_{3}}\overline{T}_{k_{1}k_{2}k_{3}}
T_{i_{1}k_{2}j_{3}}T_{j_{1}i_{2}k_{3}}T_{k_{1}j_{2}i_{3}}
$}
\end{subfigure}
\caption{Melonic bubbles and their invariants}
\end{figure}

Let us sketch how non Gau\ss ian integral can be expanded as a perturbation of the Gau\ss ian one over $(D\!+\!1)$-colored bipartite graphs ${\cal G}$ as
\begin{multline}
\int d\overline{\Phi}d\Phi\,
\Phi_{I_{1}}\dots\Phi_{I_{n}}\overline{\Phi}_{\overline{I}_{1}}\dots\overline{\Phi}_{\overline{I}_{n}}\exp\left\{-\overline{\Phi}\cdot C^{-1}\cdot\Phi+S(\overline{\Phi},\Phi)\right\}
=
{ \sum_{{\cal G}}\frac{1}{\sigma_{\cal G}} A_{\cal G}(\{\lambda_{\cal B}\}, \{{I}_{k},\overline{I}_{k}\}).}
\end{multline}

\begin{figure}[h]
 \centering
     \begin{minipage}[t]{.8\textwidth}
      \centering
\def\svgwidth{0.5\columnwidth}
\tiny{
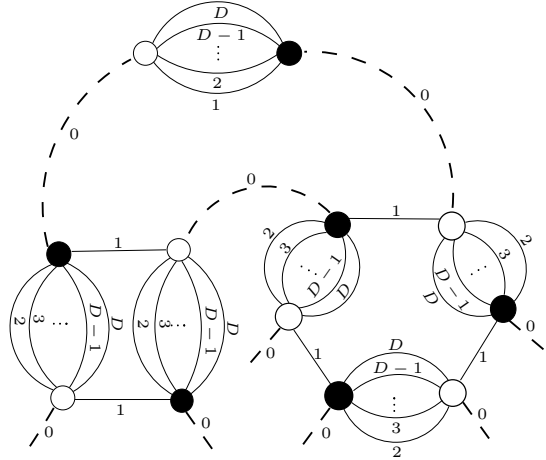
}
\caption{\small  A graph with 6 external color $0$ edges and $3$ bubbles.}
\label{fig:graph} 
\end{minipage}
\end{figure}

This is obtained by expanding $\exp S(\overline{\Phi},\Phi)$ and using Wick's theorem \eqref{Wick}. Each occurrence of a covariance between two tensors $\overline{\Phi}$ and $\Phi$ leads to a new color $0$ line which is only a half-edge if one of the tensors pertains to the 
{n-point function}
observables $\Phi_{I_{1}}\dots\Phi_{I_{n}}\overline{\Phi}_{\overline{I}_{1}}\dots\overline{\Phi}_{\overline{I}_{n}}$. 
The color $0$ edges connect with covariances $C$ the various bubbles involving colors $\left\{1,\dots,D\right\}$ 
 so that we obtain a graph with $D+1$ colors and, possibly, external color 0 edges (see Fig. \ref{fig:graph} for an example).

Finally, we sum over all indices of the bubbles ${\cal B}$ and covariances following 
{ the graph ${\cal G}$}, with the indices of the half-edges held fixed. It is nothing but the conventional Feynman graph expansion adapted to random tensors. This is the initial formulation of colored models \cite{colored} while the formulation we use has been introduced in  \cite{uncoloring}.

This expansion can be given the following simplicial interpretation, useful in the context of discretised approaches to quantum gravity and random geometry. Each 
bubble ${\cal B}$
with $D$ colors 
{represents} a triangulation of dimension $D-1$ (space) whose simplices of dimension $D-1$ are the vertices of 
${\cal B}$
attached along their $D-2$ faces following the edges of 
${\cal B}$. 
Similarly, {a graph  {${\cal G}$} }
with $D+1$ colors represents a triangulation of dimension 
{$D+1$}
 (space-time). Finally, the contraction of two vertices $v,\overline{v}$ in 
{ ${\cal G}$}
 joined by a line of color 0 is defined by removing  the color 0 line, $v$, and $\overline{v}$ and joining the remaing half edges of the same colors in $1,\dots,D$. Contracting all the pairs of vertices joined by a color $0$ line and removing the color $0$ half edge, we obtain 
a bubble ${\cal B}$
 with $D$ colors. At the level of triangulations, this bubble corresponds to the boundary, ${\cal B}=\partial{\cal G}$.

%%%%%%%%%%%%%%%%%%%%%%%%%%%%%
\section{Exact renormalisation group for tensors}
%%%%%%%%%%%%%%%%%%%%%%%%%%%%%

In complete analogy with the case of quantum field theory, let us define the effective action for a tensor model as
\begin{equation}
S_{t,t_{0}}(\Phi,\overline{\Phi})
=\log\int \frac{d\overline{\Psi}d\Psi}{{\cal N}_{t,t_{0}}}\,\exp\left\{ -
\overline{\Psi}\cdot C_{t,t_{0}}^{-1}\cdot\Psi+
S_{0}(\Phi+\Psi,\overline{\Phi}+\overline{\Psi})\right\},
\label{effective_tensor}
\end{equation}
with ${\cal N}_{t,t_{0}}=\int d\overline{\Psi}d\Psi\,\exp\ 
\overline{\Psi}\cdot C_{t,t_{0}}^{-1}\cdot\Psi$ a suitable normalisation factor. The paramater $t$ allows us to follow the flow of effective actions, starting from a bare action at $t_{0}$. In conventional QFT, $t=\log(\Lambda/\mu)$, where $\Lambda$ is a varying cut-off and $\mu$ is a given reference scale.  We assume that the covariance can be written as $C_{t,t_{0}}=\int_{t_{0}}^{t}dsK_{s}$ so that the semi-group property holds, and
\begin{equation}
S_{t_{2},t_{0}}(\Phi,\overline{\Phi})
=\log\int \frac{d\overline{\Psi}d\Psi}{{\cal N}_{t_{2},t_{1}}}\,\exp\left\{ -
\overline{\Psi}\cdot C_{t_{2},t_{1}}^{-1}\cdot\Psi+
S_{t_{1},t_{0}}(\Phi+\Psi,\overline{\Phi}+\overline{\Psi})\right\},
\end{equation}
with $C_{t_{2},t_{1}}=C_{t_{2},t_{0}}-C_{t_{1},t_{0}}$
In particular, $S_{t_{0}}=S_{0}$ can be identified with the bare action in quantum field theory and $S_{t}$ is an effective action obtained by a partial integration over some modes of the tensors. The semi-group property means that integrating from $t_{0}$ to $t_{2}$ with covariance $C_{t_{2},t_{0} }$ and the action $S_{t_{0}}$ is equivalent to integrating from $t_{1}$ to $t_{2}$ with covariance $C_{t_{2},t_{1}}$ and the effective action 
$S_{t_{1},t_{0}}$.

The effective action obeys the following differential equation, which is an infinitesimal version of the semi-group equation with $t_{1}=t$ and $t_{2}=t+\delta t$, 
\begin{equation}
\frac{\partial S}{\partial t}=\sum_{I,\overline{I}}\, K_{I,\overline{I}}\Bigg(
\frac{\partial S}{\partial {\Phi}_{I}}\frac{\partial S}{\partial { \overline{\Phi}}_{\overline{I}}}+\frac{\partial^{2}S}{\partial {\Phi}_{I}\partial {\overline{\Phi}}_{\overline{I}}}
\Bigg),
\label{Polchinski_tensor}
\end{equation}
where we have dropped the explicit dependence of $S$ on $t$ and $t_{0}$ and of $K$ on $t$.
\begin{figure}
\centering
\[
\frac{\partial}{\partial t}\,\,\parbox{1.5cm}{\includegraphics[width=1.5cm]{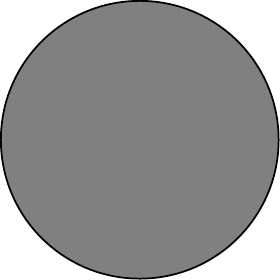}}
\quad=\quad
\parbox{2cm}{\includegraphics[width=2cm]{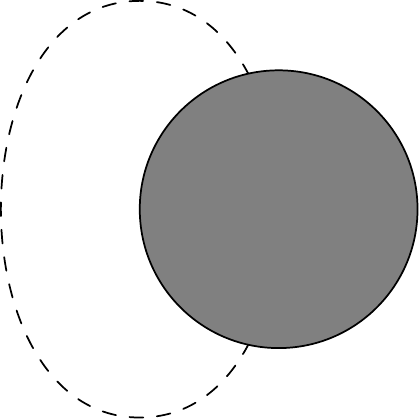}}
\quad+\quad\parbox{3cm}{\includegraphics[width=3cm]{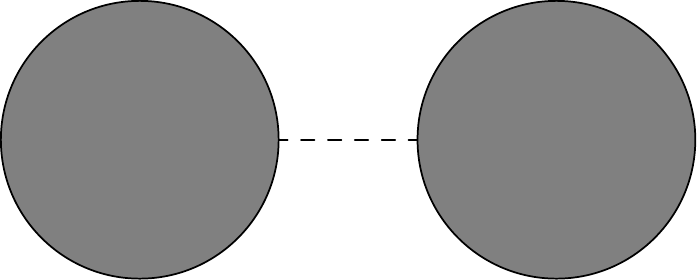}}\]
\caption{A graphical interpretation of the Polchinski's equation}
\end{figure}

This is a straightforward adaptation of the Polchinski's equation in quantum field theory (see \cite{Polchinski}) where we have used the fact that $\frac{\partial C_{t,t_{0}}}{\partial t}=K_{t}$. Let us also note that there is no factor $\frac{1}{2}$ because we are working with a complex tensor.

At a formal perturbative level, this differential equation together with the boundary condition $S_{t_{0}}=S_{0}$ is equivalent to the integral representation 
\eqref{effective_tensor}. In order to use it in an efficient way, it is necessary to translate it into a system of differential equation for the bubble couplings $\lambda_{B}(\{i_{e},\overline{i}_{e}\})$.

To this aim, let us define for 
$k\in\{0,\dots,D\}$ 
a 
{$k$-cut $c$} in a bubble ${\cal B}$ as a subset of edges 
$\left\{e_{1},\dots,e_{k}\right\}$
of ${\cal B}$ with different colors. The cut bubble ${\cal B}_{c}$ is the bubble obtained from ${\cal B}$ by cutting the $k$ edges
$\left\{e_{1},\dots,e_{{ k}}\right\}$ 
 into half-edges, attaching to them a new black $\overline{v}$ and white vertex $v$ and joining $v$ and $\overline{v}$ by 
$D-k$ edges
{carrying the colors not in $\left\{e_{1},\dots,e_{k}\right\}$. }

This ensures that ${\cal B}_{c}$ is a bubble with $D$ colors. In particular, if $c$ is a 0-cut, ${\cal B}_{c}$ is just the disjoint union of ${\cal B}$ with a dipole. A 1-cut on an edge $e$ is just the insertion on $e$ of a pair of vertices joined by $D-1$ edges carrying the colors different from that of $e$.
\begin{figure}
\centering
\begin{subfigure}[A { 3-cut}]{
\parbox{3cm}{\includegraphics[width=3cm]{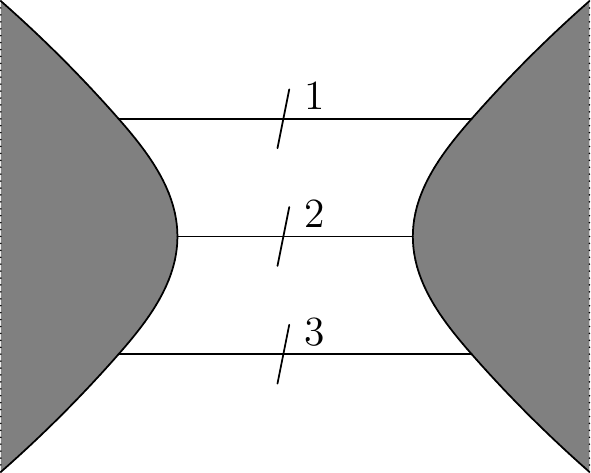}}
$\quad\rightarrow\quad$\parbox{5cm}{\includegraphics[width=5cm]{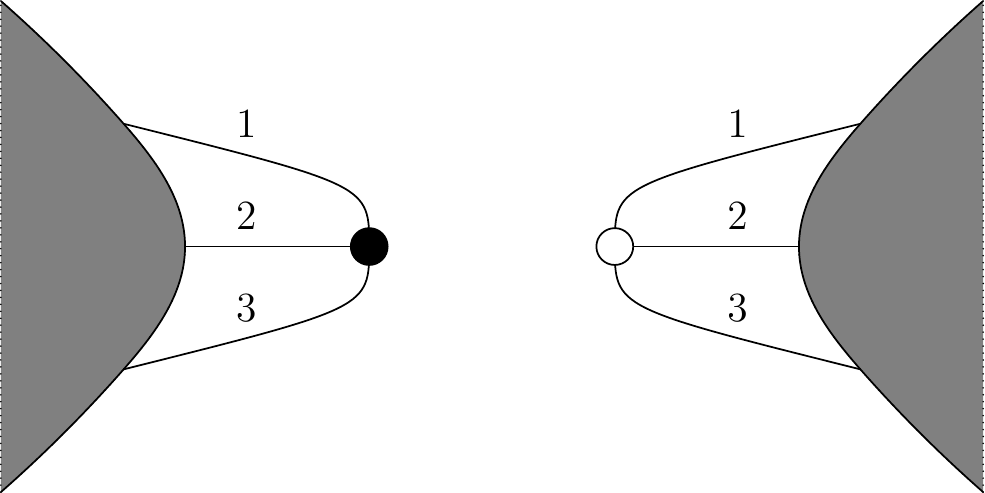}}}
\end{subfigure}
\begin{subfigure}[A 2-cut]{
\parbox{3cm}{\includegraphics[width=3cm]{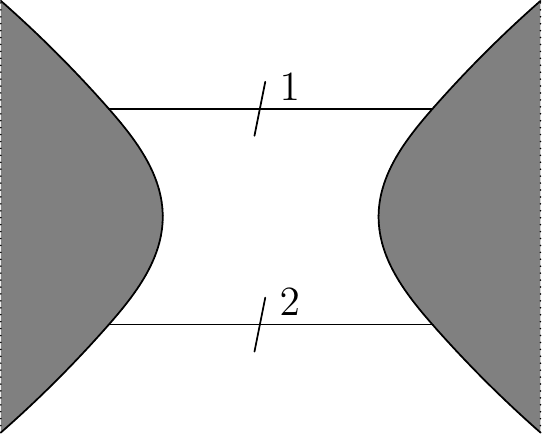}}
$\quad\rightarrow\quad$\parbox{5cm}{\includegraphics[width=5cm]{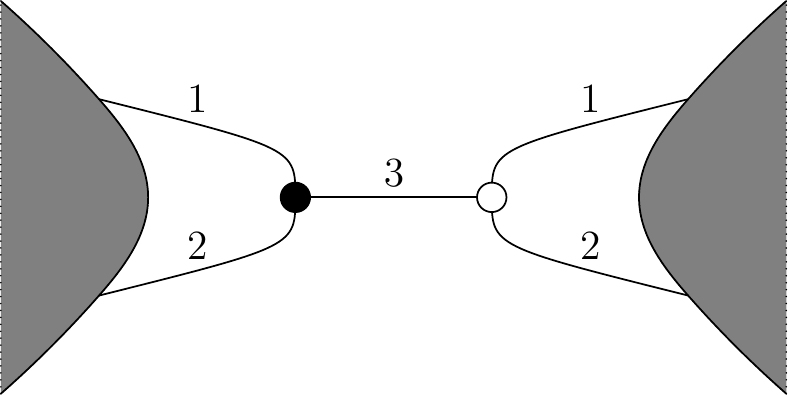}}}
\end{subfigure}
\caption{Examples of cut operations}
\end{figure}
Substituting the bubble expansion \eqref{tensor_bubble} into the flow equation \eqref{Polchinski_tensor}, we arrive at the system of differential equations governing the bubble couplings
{
\begin{align}
&\frac{\partial \lambda_{\cal B}(\left\{i_{e},\overline{i}_{e}\right\})}{\partial t}
=
\sum_{k=0}^{D}\quad
\sum_{k\text{-cut } c}\quad
\sum_{\left\{j_{l},\overline{j}_{l}\right\}}
\quad
K (\left\{j_{l},\overline{j}_{l}\right\})
\quad
\lambda_{B_{c}}
\big( \big\{\left\{i_{e},\overline{i}_{e}\right\}_{e\in{\cal B}},
\left\{j_{l},\overline{j}_{l}\right\}
\big\}\big)\label{bubble_ERGE}
\\
&+\hskip-0.5cm\sum_{D\text{-cut }c\atop
\kappa({\cal B}_{c})>\kappa({\cal B})}
\sum_{\substack {{\cal B}', {\cal B}'' \\
{\cal B}_{c}={\cal B}'\cup {\cal B}'', \\ v\in{\cal B}', \overline{v}\in{\cal B}''}}
\sum_{\left\{j_{l},\overline{j}_{l}\right\}}
K(\left\{j_{l},\overline{j}_{l}\right\})
\lambda_{\cal B'}\big(\big\{
\left\{i_{e}
\right\}_{e\in {\cal B'}},
\left\{\overline{i}_{e}
\right\}_{e\in {\cal B'}-c},
\left\{\overline{j}_{l}\right\}
\big\}\big)
\lambda_{\cal B''}
\big(\big\{
\left\{\overline{i}_{e}
\right\}_{e\in {\cal B}''},
\left\{i_{e}
\right\}_{e\in {\cal B}''-c},
\left\{j_{l}\right\}
\big\}\big).
\nonumber
\end{align}
}
In this equation, the sum runs over cuts $c$ and $D$ pairs of  tensor indices $\left\{j_{l},\overline{j}_{l}\right\}_{1\leq l\leq D}$. The tensor indices attached to the bubble ${\cal B}$ are naturally attached to corresponding  half-edges in ${\cal B}_{c}$ and are completed by the new indices $\left\{j_{l},\overline{j}_{l}\right\}_{1\leq l\leq D}$. 
 The second term involves a summation over $D$-cut that increases the number of connected components $\kappa$ and over ways of writing ${\cal B}_{c}$ as a disjoint union of a bubble ${\cal B}'$ containing $v$ and a bubble ${\cal B}''$ containing $\overline{v}$.  {In the equation,the bubbles ${\cal B'}-c$ and ${\cal B''}-c$ are the bubbles ${\cal B}'$ and ${\cal B}'$ with the edge $c$ removed}.

These equations have been derived in a very general framework, not assuming unitary invariance. This may be useful to implement a renormalisation group idea in the context of tensor models, if 
they are 
assumed to distinguish between "high" and "low" modes.

%%%%%%%%%%%%%%%%%%%%%%%%%%%%%
\section{Gau\ss ian limit at large $N$ limit revisited}
\label{sec:largeN}
%%%%%%%%%%%%%%%%%%%%%%%%%%%%%

As an application of the previous formalism, let us show how it can be used to give another proof of Gurau's Gau\ss ian universality result, see \cite{universality}. It states that at leading order in $N$ when $N\rightarrow\infty$, the expectation value of a bubble in a random tensor model for $D>2$ is Gau\ss ian
\begin{align}
\langle \text{Tr}_{\cal B}(\overline{\Psi},\Psi)\rangle&=
\frac{1}{Z}\int d\overline{\Phi}d\Phi \,
\text{Tr}_{\cal B}(\overline{\Phi},\Phi)\,
\exp\left\{-N^{D-1}\Psi\cdot\Psi+S_{0}(\overline{\Psi},\Psi)\right\}\nonumber\\
&=
c_{{\cal B}}N^{\varpi_{\cal B}}G^{\frac{v_{\cal B}}{2}}+...\label{scaling}
\end{align} 
where $...$ represents subdominant terms when $N\rightarrow\infty$ and $v_{\cal B}$ the number of vertices of ${\cal B}$. Here, we work with unitarily invariant random tensors so that the action only involves invariants

\begin{equation}
S_{0}(\overline{\Psi}, \Psi)=\sum_{{\cal B}}
N^{D-\kappa_{\cal B}}\frac{\lambda^{0}_{\cal B}}{\sigma_{\cal B}}
\text{Tr}_{\cal B}(\overline{\Psi},\Psi),
\label{trace_invariants}
\end{equation}
where {$\lambda^{0}_{\cal B}$ is a set of parameters necessary to define $S_0$, and}
$\kappa_{\cal B}$ is the number of connected components of ${\cal B}$ and the invariants are defined by
\begin{equation}
\text{Tr}_{\cal B}(\overline{\Psi},\Psi)=
\sum_{\left\{i_{e},\overline{i}_{e}\right\}}
\prod_{e}\delta_{i_{e},\overline{i}_{e}}
\prod_{v}\Psi_{I_{\cal B}(v)}
\prod_{\overline{v}}\overline{\Psi}_{\overline{I}_{\cal B}(\overline{v})}.
\end{equation}

Because of unitary invariance, the 
couplings do not depend on edge indices $\{i_{e},\overline{i}_{e}\}$. The large $N$ behavior  singles out the so-called melonic bubbles which may be defined as follows: a bubble is melonic if for every white vertex $v$, there is a black vertex $\overline{v}$ such that removing $v$ and $\overline{v}$ increases the number of connected components by $D-1$. For melonic bubbles, 
$c_{\cal B}=1$ and  $\varpi_{\cal B}=\kappa_{\cal B}$ while for non melonic bubbles we have $\varpi_{\cal B}<\kappa_{\cal B}$. Finally, $G$ only depends on the melonic couplings in $S_{0}$ and can be obtained by solving the equation
\begin{equation}
1-G=-\sum_{{\cal B}}{\frac{v_{\cal B}\lambda^{0}_{\cal B}}{2\sigma_{{\cal B}}}}\,G^{\frac{v_{\cal B}}{2}},
\end{equation}
where the sum runs over all melonic bubbles in $S_{0}$. This result can be derived from the generating function defined by
\begin{equation}
Z(J,\overline{J})=
\frac{1}{Z}\int d\overline{\Psi}d\Psi\, 
\exp\left\{-N^{D-1}\Psi\cdot\Psi+S_{0}(\overline{\Psi},\Psi)+\overline{J}\cdot\Psi+\overline{\Psi}\cdot J\right\},
\end{equation}
where $J$ is also a rank $D$ complex tensor. It is normalised in such a way that $Z(0,0)=1$.

Using the effective action and its flows, we give an alternative proof of these results. We show that a leading order in $N$ this generating function is Gau\ss ian
\begin{equation}
Z(J,\overline{J})=\exp\left\{\frac{\overline{J}\cdot G\cdot J}{N^{D-1}}+...\right\},
\end{equation}
where the dots represent subdominant terms when $N\rightarrow\infty$ and $G$ only depends on the melonic couplings $\lambda_{\cal B}^{0}$. The cornerstone of our proof is the system of differential equations for the bubble couplings, \eqref{flow_invariants}.

To begin with, let us first show that this generating function is related to the effective action by a simple change of variables. We define an effective action with a diagonal covariance $C=\frac{t}{N^{D-1}}\delta_{I,\overline{I}}$ 
and set $t_{0}=0$ for simplification.  Thus, the effective action, normalised by the free energy $F=\log Z$ reads
\begin{equation}
\exp\left\{
S_{t}(\Phi,\overline{\Phi})-S_{t}(0,0)
\right\}
=\frac{1}{Z}\int d\overline{\Psi}d\Psi\,
\exp\left\{-\frac{N^{D-1}}{t}\overline{\Psi}\cdot\Psi+
S_{0}(\Phi+\Psi,\overline{\Phi}+\overline{\Psi})
\right\}.
\label{eq:actiont}
\end{equation}
In this {section, the parameter $t$ controls the flow of effective actions and does not directly refer to a cut-off, since there is no such notion of high or low momenta in this example ({see Section \ref{GFT:sec})} for a field theoretical example, where the flow is controlled by a true cut-off). 

Then, let us shift the integration variables 
$\Psi\rightarrow\Psi-\Phi$ and $\overline\Psi\rightarrow\overline\Psi-\overline\Phi$ so that we get
\begin{multline}
\exp\left\{
S_{t}(\Phi,\overline{\Phi})-S_{t}(0,0)
+\frac{N^{D-1}}{t}\Phi\cdot\Phi\right\}
=\\
\frac{1}{Z}\int d\overline{\Psi}d\Psi\,\exp\left\{-\frac{N^{D-1}}{t}
\overline{\Psi}\cdot\Psi+
S_{0}(\Psi,\overline{\Psi})
+\frac{N^{D-1}}{t}\overline{\Psi}\cdot\Phi
+\frac{N^{D-1}}{t}\overline{\Phi}\cdot\Psi
\right\}.
\end{multline}
In this equation, the parameter $t$ is just a bookkeeping device, the term of order $t^{n}$ contains the contribution of all connected graphs ${\cal G}$ with $n$ edges.

Introducing the sources $J=\frac{\Phi}{N^{D-1}}$ and $\overline J=\frac{\overline\Phi}{N^{D-1}}$, the effective action at $t=1$ is related to the generating function by
\begin{equation}
Z(J,\overline J)
=
\exp\left\{
S_{t=1}\left(\frac{J}{N^{D-1}},\frac{\overline{J}}{N^{D-1}}\right) - S_{t=1}(0,0)
+\frac{1}{N^{D-1}}J\cdot \overline J\right\}.\label{generating_effective}
\end{equation}
In order to understand the large $N$ behavior of the effective action, it is convenient to expand it over trace invariants as in \eqref{trace_invariants} 
{
\begin{equation}
S_{t}(\Phi, \overline{\Phi})=\sum_{{\cal B}}
N^{D-\kappa_{\cal B}}\frac{{\lambda}_{\cal B}(t)}{\sigma_{\cal B}}
\text{Tr}_{\cal B}(\Phi, \overline{\Phi}).
\end{equation}
}

The couplings $\lambda_{\cal B}(t)$ obey a system of differential equations that can be deduced from \eqref{Polchinski_tensor}. Because of unitary invariance, it greatly simplifies and the sum over indices reduces to a factor of 
$N^{D-k}$
 for a 
$k$-cut.
Moreover, our normalisation in powers of $N$ of the covariance and of the couplings allows us to have only constant or negative powers of $N$. Thus, the evolution of a couplings reads
\begin{align}
\frac{\partial\lambda_{\cal B}}{\partial t}=\sum_{k=0}^{D}\quad
\sum_{k\text{-cut } c}\quad
N^{\kappa_{{\cal B},c}-k} 
\lambda_{B_{c}}+
\sum_{D\text{-cut }c\atop
{\kappa({\cal B}_{c})>\kappa({\cal B})}
}
\sum_{{\cal B}',{\cal B}''\atop
{\cal B}_{c}={\cal B}'\cup {\cal B}'',v\in{\cal B}', \overline{v}\in{\cal B}''}
\lambda_{\cal B'}
\lambda_{\cal B''},\label{flow_invariants}
\end{align}
where $\kappa_{{\cal B},c}$ is defined as the number of connected components of ${\cal B}$ containing edges of the cut, except for a 
{$D$-cut} that increases the number of connected components of ${\cal B}$, in which case $\kappa_{{\cal B},c}=0$. We always have $k\geq \kappa_{{\cal B},c}$ and $k\geq \kappa_{{\cal B},c}$ if
only all edges of the cut belong to different connected components of ${\cal B}$. In particular, this holds for a 0-cut and a 1-cut.

There are two important consequences of this result. First, because the exponent of $N$ is negative or 0, the couplings $\lambda_{\cal B}$ only contain constants or negative powers of $N$, so that they remain bounded as $N\rightarrow\infty$. Therefore, as $N\rightarrow\infty$, only the quadratic term survives in \eqref{generating_effective}
\begin{equation}
\exp\left\{
S_{t=1}\left(\frac{\overline J}{N^{D-1}},\frac{J}{N^{D-1}}\right) - S_{t=1}(0,0)
+
\frac{1}{N^{D-1}}\overline J\cdot J\right\}
=\exp\left\{\frac{G}{N^{D-1}}J\cdot\overline{J}
+...\right\},
\end{equation}
where as usual the dots stand for subleading powers of $N$. Comparing both sides shows that the effective covariance reads
\begin{equation}
G=1+ \lim_{N\rightarrow\infty} \lambda_{\includegraphics[width=0.5cm]{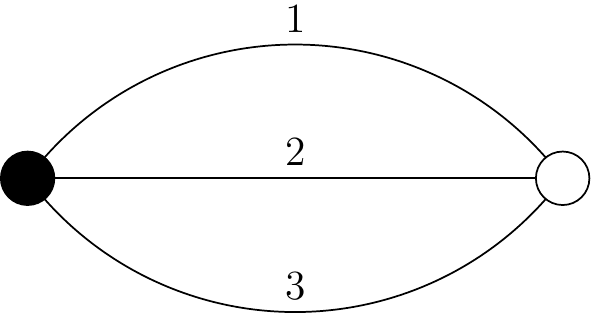}}(t=1)\,.
\end{equation}
Indeed, after the substitutions $J=\frac{\Phi}{N^{D-1}}$ and $\overline J=\frac{\overline\Phi}{N^{D-1}}$ in \eqref{generating_effective}, a bubble coupling $\lambda_{\cal B}$ scales like $N^{D-\kappa_{\cal B}-v_{\cal B}(D-1)}$. As $N\rightarrow\infty$, the leading term corresponds to the minimal value of $\kappa_{\cal B}=1$ and $v_{\cal B}=1$. Only the dipole \includegraphics[width=0.7cm]{dipole} obeys this requirement, so that the leading term is Gau\ss ian.

Second, consider 
the large $N$ limit in \eqref{flow_invariants}
\begin{align}
\frac{\partial\widetilde{\lambda}_{\cal B}}{\partial t}=\sum_{k=0}^{D}\quad
\sum_{k\text{-cut } c\atop
\text{in $k$ distinct connected components of ${\cal B}$}}\quad
\widetilde{\lambda}_{B_{c}}+
\sum_{D\text{-cut }c\atop
{\kappa({\cal B}_{c})>\kappa({\cal B})}}
\sum_{{\cal B}',{\cal B}''\atop
{\cal B}_{c}={\cal B}'\cup {\cal B}'',v\in{\cal B}', \overline{v}\in{\cal B}''}
\sum_{\left\{j_{l},\overline{j}_{l}\right\}}
\widetilde{\lambda}_{\cal B'}
\widetilde{\lambda}_{\cal B''},
\label{Polchinski_unitary}
\end{align} 
{where} $\widetilde{\lambda}_{\cal B}=\lim_{N\rightarrow\infty}\lambda_{\cal B}$.
We refer to 
{Appendices}
\ref{example3} and \ref{example4} for a few low order examples in $D=3$ and $D=4$.

Since we are interested in the dipole coupling $\widetilde{\lambda}_{\includegraphics[width=0.5cm]{dipole.pdf}}$, we may start solving this system. Even if this involves an infinite number of equations, at a given order in the initial conditions $\lambda_{\cal B}(0)=\lambda_{\cal B}^{0}$,  there are only finitely many terms involved if the action $S_{0}$ we start with contains a finite number of bubble couplings. 
  
 It is instrumental to notice that starting with the dipole coupling $\widetilde{\lambda}_{\includegraphics[width=0.5cm]{dipole.pdf}}$, only melonic couplings are involved at leading order in $1/N$. This is easily established by induction: If ${\cal B}$ is melonic, so are ${\cal B}_{c}$, ${\cal B}'$ and ${\cal B}''$. Therefore, at leading order in $1/N$, $\widetilde{\lambda}_{\includegraphics[width=0.5cm]{dipole.pdf}}(t)$ only depends on the initial couplings $\lambda_{\cal B}(0)$ for ${\cal B}$ melonic. Then, the expectation value of an arbitrary bubble can be derived from the generating function using the identity
 \begin{equation}
 \langle \text{Tr}_{\cal B}(\overline{\Psi},\Psi)\rangle=
 \text{Tr}_{\cal B}\bigg(\frac{\partial }{\partial J},\frac{\partial }{\partial \overline J}\bigg)
Z(J,\overline J)\bigg|_{J=\overline J=0}.
 \label{differential}
 \end{equation} 
 Because at leading order 
$Z(J,\overline J)$ 
is Gau\ss ian, this simply amounts to sum over all possible Wick contraction on the invariant $\text{Tr}_{\cal B}(\overline{\Psi},\Psi)$ with a covariance $G$. It leads to
\begin{equation}
 \langle \text{Tr}_{\cal B}(\overline{\Psi},\Psi)\rangle
=
{c_{{\cal B}} }
N^{\zeta_{\cal B}}
\bigg(\frac{G}{N^{D-1}}\bigg)^{\frac{v_{\cal B}}{2}},
\label{expectation}
\end{equation} 
where 
{$c_{\cal B}$}
 is the number of Wick contraction giving a leading order contribution of $N^{\zeta_{\cal B}}$. At fixed number of vertices $v_{\cal B}$, $\zeta_{\cal B}$ is maximal for a melonic bubble. This is established by contracting pairs of conjugate vertices in a melonic bubble. This unique Wick contraction leads to  $\zeta_{\cal B}=(D-1)\frac{v_{\cal B}}{2}+\kappa_{\cal B}$,
which provides the announced scaling law for a melonic bubble given in \eqref{scaling}.

Finally, let us compute the effective covariance $G$. It is known that this can be achieved using the Schwinger-Dyson equations \cite{bonzom}. Alternatively, one may observe that, using the flow equation \eqref{Polchinski_unitary} that  $\frac{\partial \lambda_{\emptyset}}{\partial t}= \lambda_{\includegraphics[width=0.5cm]{dipole.pdf}}$. The empty bubble (with $\kappa_{\emptyset}=0$) is related to the free energy 
\begin{align}
N^{D}\lambda_{\emptyset}&=
S_{t}(0,0)\\
&=\log \int d\overline{\Psi}d\Psi\,\exp\left\{-\frac{N^{D-1}}{t}
\overline{\Psi}\cdot\Psi+
S_{0}(\Psi,\overline{\Psi})
\right\}
-
\log \int d\overline{\Psi}d\Psi\,\exp\left\{-\frac{N^{D-1}}{t}
\overline{\Psi}\cdot\Psi\right\},
\end{align}
where the last term arises from the normalisation. By rescaling the $\Psi$ and $\overline{\Psi}$, one immediately shows that 
$S_{t}(0,0)$
is invariant under the rescaling of  $t\rightarrow s^{-2}t$ and of the couplings in $S_{0}$, $\lambda_{\cal B}^{0}\rightarrow s^{v_{\cal B}}\lambda^{0}_{\cal B}$. This leads to the differential equation 
\begin{equation}
-2t\frac{\partial 
S_{t}(0,0)}{\partial t}
+
\sum_{\cal B}v_{\cal B}\lambda_{\cal B}^{0}
\frac{\partial}{\partial \lambda_{\cal B}^{0}}
S_{t}(0,0)
=0
\end{equation}
which is equivalent to \begin{equation}
{-2}
tN^{D}\lambda_{\includegraphics[width=0.5cm]{dipole.pdf}}(t)
+\sum_{\cal B}v_{\cal B}
\frac{\lambda_{\cal B}^{0}N^{D-\kappa_{\cal B}}}{\sigma_{\cal B}}
\langle\text{Tr}_{\cal B}(\Psi,\overline{\Psi})\rangle
=0.
\end{equation}
Then, using \eqref{expectation}, we obtain at $t=1$ and for $N\rightarrow\infty$

\begin{equation}
G-1
-
\sum_{\cal B}v_{\cal B}
\frac{\lambda_{\cal B}^{0}}{ 2 \sigma_{\cal B}}
G^{v_{\cal B}}=0
\end{equation}
where we have discarded non melonic couplings in the large $N$ limit since the are non leading. This equation can be used to determine $G$ as a formal power series in the bubble couplings $\lambda_{\cal B}^{0}$.

The arguments we have presented hold for $D>2$ only. In the matrix model case, the situation is more involved. Indeed, even if \eqref{Polchinski_unitary} still holds and the generating function 
$Z(J,\overline{J})$ 
is still Gau\ss ian, the expectation value of a bubble is no longer given by Wick contrition only. Indeed, in this case the differential operator in \eqref{differential} can generate higher powers of $N$ in such a way that non quadratic terms admit a finite limit.

%%%%%%%%%%%%%%%%%%%%%%%%%%%%%
\section{Tensorial group field theory}
%%%%%%%%%%%%%%%%%%%%%%%%%%%%%
\label{GFT:sec}
Tensorial group field theories are specific quantum field theories defined on $D$ copies of a group, usually $\text{SU}(2)$, $\text{U(1)}^{d}$, ${\Bbb R}^{d}$ or $\text{SL}_{2}({\Bbb C})$. In quantum gravity applications, $D$ denotes the space-time dimension and $d$ is the dimension of the group. Tensorial group field theories are constructed  from tensor models by replacing tensors by tensorial fields
\begin{equation}
\Phi_{i_{1},\dots,i_{D}}\rightarrow
\Phi(g_{1},\dots,g_{D})\qquad
\overline\Phi_{i_{1},\dots,i_{D}}\rightarrow
\overline\Phi(\overline g_{1},\dots,\overline g_{D}).
\end{equation}
Note that $\overline{g}$ is not the complex conjugate of $g$ but an independent group element that is an argument of the complex conjugate field $\overline\Phi$.

The effective action is expanded over bubbles as in \eqref{tensor_bubble}, except that indices are replaced by group elements and sums by integration over the group using the Haar measure, 
\begin{equation}
S(\Phi,\overline{\Phi})=\sum_{\cal B}
\lambda_{\cal B}(\Phi,\overline{\Phi})=\frac{1}{\sigma_{\cal B}}
\int \prod_{e}dg_{e}d\overline{g}_{e}
\lambda_{{\cal B}}(\left\{g_{e}\overline{g}_{e}^{-1}\right\})
\prod_{v}\Phi(G_{\cal B}(v))
\prod_{\overline{v}}\overline{\Phi}(\overline{G}_{\cal B}(\overline{v})),
\end{equation}
where capital letters $G_{\cal B}(v)$ and $\overline{G}_{\cal B}(\overline{v})$ indicate the $D$-tuplets of group elements (ordered by their colors) incident to $v$ and $\overline  v$ in ${\cal B}$. We also assume that the couplings and the covariance only depend on the products $g_{e}\overline{g}_{e}^{-1}$. In some cases it is also assumed that the action and the propagator are invariant under the gauge transformations $g_{e}\rightarrow  h_{v}g_{e}$ and $\overline g_{e}\rightarrow  \overline h_{\overline v}\overline g_{e}$, where $v$ (resp. $\overline{v}$) is the white vertex (resp. black vertex) attached to $e$. This translates into the following conditions obeyed by the covariance and by the couplings
\begin{align}
C(\left\{g_{e}\overline{g}_{e}^{-1}\right\})=
C(\left\{hg_{e}\overline{g}_{e}^{-1}\overline{h}^{-1}\right\})\qquad
\lambda_{\cal B}(\left\{g_{e}\overline{g}_{e}^{-1}\right\})=
\lambda_{\cal B}(\left\{h_{v_{\cal B}(e)}g_{e}\overline{g}_{e}^{-1}\overline{h}^{-1}_{\overline{v}_{\cal B}(e)}\right\}),
\end{align}
where $v_{\cal B}(e)$ (resp. $\overline v_{\cal B}(e)$) is the white (resp. black) vertex of ${\cal B}$ attached to $e$. Finally, it is convenient to write the covariance as the integral of a product of $D$ averaged heat kernels $H_{\alpha}$ on the group. We also use a more conventional notation for quantum fields, writing $t=\log (\Lambda/ \mu)$
and interpreting $\Lambda$ as a momentum space cut-off. The covariance reads
\begin{equation}
C_{\Lambda,\Lambda_{0}} \! \left(\left\{g_{e}\overline{g}_{e}^{-1}\right\}\right)=\int_{\frac{1}{\Lambda_{0}^{2}}}^{\frac{1}{\Lambda^{2}}}d\alpha \, K_{\alpha} \!\! \left(\left\{g_{e}\overline{g}_{e}^{-1}\right\}\right)
\quad
\text{with}
\quad
K_{\alpha} \!\! \left(\left\{g_{e}\overline{g}_{e}^{-1}\right\}\right)
=
\int dhd\overline{h} \!\!
\prod_{1\leq i\leq D} \!\!\! H_{\alpha}(hg_{i}\overline{g}_{i}^{-1}\overline{h}^{-1}).
\end{equation}
In the last equation, $H_{\alpha}(g)$ refers to the heat kernel on the group at time $\alpha$. It is the solution of the heat equation $\frac{\partial H_{\alpha}}{\partial \alpha}=\Delta H_{\alpha}$ such that $\lim_{\alpha\rightarrow0}H_{\alpha}(g)=\delta(g)$, with $\delta$ the Dirac distribution concentrated
at
the identity. Then, the ERGE for the effective action can be written as

\begin{equation}
\Lambda\frac{\partial S}{\partial \Lambda}=-\frac{2}{\Lambda^{2}}
\int\prod_{1\leq i\leq D}dg_{i}d\overline{g}_{i}
K_{\frac{1}{\Lambda^{2}}}\left(\left\{g_{i}\overline{g}_{i}^{-1}\right\}\right)\bigg(
\frac{\delta S}{\delta \Phi (G)}\frac{\delta S}{\delta \overline{\Phi}(\overline G)}+\frac{\delta^{2}S}{\delta{\Phi}(G)\delta\overline{\Phi}(\overline{G})}
\Bigg).
\label{Polchinski_GFT}
\end{equation}
In the $\text{SU}(2)$ case which is involved in many spin-foam models, it is possible to expand all the bubble couplings on spin networks, see \cite{Corfu},  rather than just bubbles.

In this article, we are mostly interested in deriving scaling properties of the bubble couplings from the ERGE. For simplicity, we restrict our attention to Abelian models. Elements of  $\text{U(1)}$ are written as $g=\text{e}^{2\text{i}\pi\frac{\theta}{L}}$ with $\theta\in[0,L]$, with $L$ being a length. The propagator is given by 
\begin{equation}
C_{\Lambda, \Lambda_{0}}(\{\theta_{i}-\overline{\theta}_{i}\})
=
\int_{\frac{1}{\Lambda_{0}^{2}}}^{\frac{1}{\Lambda^{2}}}d\alpha\quad
\sum_{\left\{p_{i}\right\}\in{L\Bbb Z}^{dD}}
\exp-\left\{\alpha\sum_{i}p_{i}^{2}+\text{i}\sum p_{i}(\theta_{i}-\overline\theta_{i})\right\}
\delta_{\sum p_{i},0},
\end{equation}
where we have enforced the condition $\sum_{i} p_{i}=0$ so that the theory is invariant under the gauge transformation of the field, $\Phi(g_{1},...,g_{D})\rightarrow
\Phi(hg_{1},...,hg_{D})$ and $\overline{\Phi}(\overline{g}_{1},...,\overline{g}_{D})\rightarrow
\overline{\Phi}(\overline{h}\overline{g}_{1},...,\overline{h}\overline{g}_{D})$.

At first sight, the bubble couplings depend on $D$ group elements for each bubble vertex, so that there is a total of $v_{\cal B}D$ group elements. It is also possible to consider that there are two group elements $\theta_{e}$ and $\overline{\theta}_{e}$ per edge. Then, the bubble coupling only depend on the differences,  $\lambda_{\cal B}(\{\theta_{e}-\overline{\theta}_{e}\})$. By Fourier transform, there is a single momentum per edge (because only the differences $\theta_{e}-\overline{\theta}_{e}$), which we conventionally orient from the black to white vertices. Furthermore, gauge invariance is equivalent to the constraint ${
 \sum_{\text{\tiny $e$ incident to $v$}}}\,p_{e}=0$ that has to be implemented at every vertex $v$.

The exact renormalisation group equation written in terms of bubble couplings is 
{
\begin{align}
&\Lambda\frac{\partial \lambda_{\cal B}(\left\{p_{e}\right\})}{\partial \Lambda}
=
-\frac{2}{\Lambda^{2}}\sum_{k=0}^{D} \,
\sum_{k\text{-cut } c} \,
\sum_{\left\{p_{l}\right\}_{l\notin c}}
\delta_{\sum_{l\notin c}p_{l}+\sum_{e\in c}p_{e},0}\,
\text{e}^{-\frac{\sum_{l\notin c}p^{2}_{l}+\sum_{e\in c}p_{e}^{2}}{\Lambda^{2}}}\,
\lambda_{B_{c}}
(\{\left\{p_{e}\right\}_{e\in{\cal B}},
\left\{p_{l}\right\}_{l\notin c}
\})\nonumber
\\
&-\frac{2}{\Lambda^{2}}\sum_{D\text{-cut }c\atop
\kappa({\cal B}_{c})>\kappa({\cal B})}
\sum_{{\cal B}',{\cal B}''\atop
{\cal B}_{c}={\cal B}'\cup {\cal B}'',v\in{\cal B}', \overline{v}\in{\cal B}''}
\quad
\delta_{\sum_{e\in c}p_{e},0}\,
\text{e}^{-\frac{\sum_{e\in c}p_{e}^{2}}{\Lambda^{2}}}\,
\lambda_{\cal B'}(\left\{
p_{e}
\right\}_{e\in {\cal B'}})
\lambda_{\cal B''}
(\left\{{p}_{e}
\right\}_{e\in {\cal B}''}), \label{abelian_ERGE}
\end{align}
}
As in \eqref{bubble_ERGE}, we perform $k$-cuts with $k\in\left\{0,...,D\right\}$. The first term generates the analogue of the loops in QFT. It involves a summation over the momenta associated with the colors not in the cut. The second one corresponds to the QFT trees and involves  $D$-cuts that separate the bubbles into ${\cal B}'$ and ${\cal B}''$. Note  that this is nothing but the tensorial exact renormalisation group equation \eqref{bubble_ERGE}, with the constraint that the 2 edge indices coincide, $i_{e}=\overline{i}_{e}=p_{e}$ and take their values in $L{\Bbb Z}^{d}$.

In the last part of this section, we  aim at addressing the issue of the renormalisability of the model, from the point of view of the exact renormalisation group equation \eqref{abelian_ERGE}. This implies {the need of}
the definition of a proper scaling dimension of the bubble coupling, which we first present from a heuristic point of view. To ease the discussion, we assume $L$ to be large enough ($1/L\ll \Lambda$) so that momenta can be treated as continuous variables, then we can trade sums for integrals
\begin{equation}
\sum_{p}\,\rightarrow\,L^{d}\int dp\qquad
\text{and}\qquad
\delta_{\sum{p_{i}},0}\,\rightarrow\,\frac{1}{L^{d}}\delta(\sum{p_{i}}).
\end{equation}

Otherwise, a more precise but cumbersome analysis can be performed using the Poisson resummation. Finally, we also allow derivatives of the bubble couplings with respect to momenta and rotational invariance in momentum space. This last hypothesis is fulfilled if we assume that the initial couplings are invariant under rotation since the propagator is.

In the remainder of this section, we introduce the canonical dimension for bubble couplings. We motivate our choice by a heuristic analysis. This is later justified by the fact that dimensionless couplings obey an equation with only negative powers of the cut-off, see eq. \eqref{dimensionless_ERGE}.

The normalisation of the effective action at $\Phi=\overline{\Phi}=0$ involves a subtraction of the divergent quantity $\log\det C_{\Lambda,\Lambda_{0}}$ which behaves for large $\Lambda$ as $(L\Lambda)^{d(D-1)}$ up to an additive constant, as is easily seen using a free field theory. Therefore, it is natural to assume that the action behaves as 
$S(0,0)\sim (L\Lambda)^{d(D-1)}$ 
for large $\Lambda$, up to a multiplicative constant. As far as the scaling of $\Lambda$ is concerned, we attribute to the action the scaling dimension $[S]=d(D-1)$, even if it is dimensionless ({\it{i.e.,}} invariant if we rescale both $\Lambda\rightarrow s\Lambda$ and $L\rightarrow L/s$). 

Moreover, our procedure involves integrations over shells of momenta $p\sim\Lambda$, so that $[p]=1$. Then, the dimension of the field can be obtained by examining the quadratic terms 
\begin{equation}
[S] =d(D-1)+ 2 +2[\Phi],
\end{equation} 
where we recall that there are $dD-d=d(D-1)$ momenta ($D$ momenta in ${\mathbb Z}^{d}/L$ and one constraint). 
The first term on the RHS is the dimension of the integration over momenta and the second term is the dimension of the Laplacian.
Because we also have {$[S]=d(D-1)$, }
the scaling dimension in $\Lambda$ of the field  is $[\Phi]=-1$. This does not mean that $[\Phi]$ has the same 
units as
$1/\Lambda$, but rather that we should scale $\Phi$ as $\Lambda^{-1}$ if we want to have a well defined limit for large $\Lambda$ at fixed $L$.

The dimension of a bubble coupling obeys a similar equation
\begin{equation}
[S] =\delta_{\cal B}+d\big(e_{\cal B}-v_{\cal B}+\kappa_{\cal B}\big)+v_{\cal B}[\Phi],
\end{equation} 
where $\delta_{\cal B}=[\lambda_{{\cal B}}]$ is the dimension of the bubble coupling $\lambda_{\cal B}$. We denote by $e_{\cal B}$ the number of edges, by $v_{\cal B}$ the number of vertices and by $\kappa_{\cal B}$ is the number of connected components of ${\cal B}$. 
$d(e_{\cal B}-v_{\cal B}+\kappa_{\cal B})+v_{\cal B}[\Phi]$ 
representing
the number of independent integrations over momenta in ${\Bbb R}^{d}$ since there is one such momentum for each edge, one constraint at each vertex and a remaining constraint for each connected component. Taking into account the relation $2e_{\cal B}=Dv_{\cal B}$, it leads to the following scaling dimension for a bubble coupling
\begin{equation}
\delta_{{\cal B}}= d(D-1)-d\kappa_{\cal B}-\big(d(D-2)-2\big) \frac{v_{\cal B}}{2}.
\end{equation}

Since a bubble coupling depends on several momenta, it is helpful to perform a Taylor expansion to order $n$. This leads to a sum of homogenous polynomials of total degree less than $n$,

\begin{equation}
\lambda_{\cal B} (\left\{p_{e}\right\})
=\sum_{n}
\lambda_{{\cal B},n}\times\text{homogeneous polynomial of degree $n$ in momenta}.
\end{equation}
Accordingly, the dimension of coefficients of the Taylor expansion are $\delta_{{\cal B},n}=\delta_{\cal B}-n$. In particular, we recover the expected dimension for a mass term $\delta_{\includegraphics[width=0.5cm]{dipole.pdf},0}=2$ and for the kinetic term 
$\delta_{\includegraphics[width=0.5cm]{dipole.pdf},2}=0$. Moreover, for a connected bubble, it reduces to the scaling dimension deduced from an analysis of the degree of divergence in \cite{Carrozza:2013mna}. In a slightly different context, a similar analysis has also been performed in 
\cite{tensorERGE}.
 
To understand the scaling properties of the interactions, let us introduce the analogue of dimensionless bubble couplings $u_{\cal B}$ defined by  
\begin{equation}
\lambda_{\cal B}(\left\{p_{e}\right\},\Lambda)=
\Lambda^{\delta_{\cal B}}u_{\cal B}(\left\{q_{e}\right\},\Lambda) \qquad\text{with $q_{e}=\frac{p_{e}}{\Lambda}$}.
\end{equation} 
Note that the couplings depend explicitly on the scale $\Lambda$ and on momenta. In term of these new couplings, the exact renormalisation group equation \eqref{abelian_ERGE} takes the form

\begin{align}
&\Lambda\frac{\partial u_{\cal B}(\left\{q_{e}\right\})}{\partial \Lambda}
=-\delta_{\cal B}\,u_{\cal B}(\{q_{e}\})+\sum_{e}q_{e}\frac{\partial u_{\cal B}(\left\{q_{e}\right\})}{\partial q_{e}}\nonumber\\
&
-\!\frac{2}{\Lambda^{2}}\sum_{k=0}^{D}\,
\sum_{k\text{-cut } c} 
\! \frac{L^{d(D-k-1)}}
{
{\Lambda^{d(k-
{
\kappa_{{\cal B},c})}
}
}
}\int \!\! \prod_{l\notin c} dq_{l}\,
\delta\Big( \!\sum_{l\notin c}q_{l}+\! \sum_{e\in c}q_{e}\!\Big)\,
\text{e}^{- \! \big( \! \sum_{l\notin c}q^{2}_{l}+\sum_{e\in c}q_{e}^{2}\big)}\,
u_{B_{c}} \!
( \! \{\left\{q_{e}\right\}_{e\in{\cal B}},
\left\{q_{l}\right\}_{l\notin c} \!
\} \!)\nonumber
\\
&-\frac{2}{\Lambda^{2}}\sum_{D\text{-cut }c\atop
\kappa({\cal B}_{c})>\kappa({\cal B})}
\sum_{{\cal B}',{\cal B}''\atop
{\cal B}_{c}={\cal B}'\cup {\cal B}'',v\in{\cal B}', \overline{v}\in{\cal B}''}
\quad
L^{-d}\, \delta\big(\sum_{e\in c}q_{e}\big)\,
\text{e}^{-\sum_{e\in c}q_{e}^{2}}\,
u_{\cal B'}(\left\{
q_{e}
\right\}_{e\in {\cal B'}})
u_{\cal B''}
(\left\{{q}_{e}
\right\}_{e\in {\cal B}''}),
\label{dimensionless_ERGE}
\end{align}
with 
$\kappa_{{\cal B},c}$ the number of connected components of ${\cal B}$ containing edges of the cut, except if we cut $D$ colors and disconnect the graph, in which case $\kappa_{{\cal B},c}=0$.
 In order to derive \eqref{dimensionless_ERGE} we insert  $\lambda_{\cal B}(\left\{p_{e}\right\},\Lambda)=\Lambda^{\delta_{\cal B}}u_{\cal B}(\left\{q_{e}\right\},\Lambda)$ in \eqref{abelian_ERGE} with $q_{e}=\frac{p_{e}}{\Lambda}$.

The first two terms on the RHS in {\eqref{dimensionless_ERGE}} follows from
\begin{equation}
\Lambda
\frac{\partial\lambda_{\cal B}(\left\{p_{e}\right\},\Lambda)}{\partial \Lambda}=
\delta_{\cal B}\Lambda^{\delta_{\cal B}}u_{\cal B}(\left\{q_{e}\right\},\Lambda)
-\Lambda^{\delta_{\cal B}}\sum_{e}q_{e}\frac{\partial }{\partial q_{e}}u_{\cal B}(\left\{q_{e}\right\},\Lambda)+
\Lambda^{\delta_{\cal B}}\,\Lambda\frac{\partial u_{\cal B}(\left\{q_{e}\right\},\Lambda)}{\partial \Lambda}.
\end{equation}
To understand the origin of the powers of $\Lambda$ in the third term in  {\eqref{dimensionless_ERGE}}, observe that a $k$-cut on 
$\kappa_{{\cal B},c}$ 
different connected components gathers these connected components into a single one. Thus, 
$\kappa_{{\cal B}_{c}}=\kappa_{\cal B}-\kappa_{{\cal B},c}+1$. 
Since ${\cal B}_{c}$ has two more vertices, we have $\delta_{{\cal B}_{c}}=\delta_{\cal B}+d\kappa_{{\cal B},c}-(D-1)d+2$. In addition, there are $D-k$ integrations over $p_{l}$ for $l\notin c$ and one constraint, so that, after the change of variables $p_{e}\rightarrow q_{e}$ we get an extra power $\Lambda^{d(D-k)-d}$.  Taking all terms into account, the exponent of $\Lambda$ is $\delta_{{\cal B}_{c}}+d(D-k)-d-2-\delta_{\cal B}=d(\kappa_{{\cal B},c}-k)$, which is the announced result.

The last term involves no power of $\Lambda$ since $\kappa_{\cal B}=\kappa_{{\cal B}'}+\kappa_{{\cal B}''}-1$, so that $\delta_{\cal B}=\delta_{{\cal B}'}+\delta_{{\cal B}''}-2$.
{

Note that $\Lambda$ only appears with a non positive exponent identical to that of $N$ in the tensor model \eqref{flow_invariants}. It is also useful to notice that the operator $\sum_{e}q_{e}\frac{\partial}{\partial q_{e}}$ is diagonalised by homogeneous polynomials in $\left\{q_{e}\right\}$ with eigenvalues being the total degree of the polynomials since it is the generator of momentum rescaling.

\section{Renormalisable Abelian tensorial group field theories with a closure constraint}

In analogy with ordinary quantum field theory (see \cite{Polchinski}), this equation can serve as a basis for identifying the renormalisable interactions in tensorial group field theories. Indeed, the group field theory flow equation  \eqref{dimensionless_ERGE} differs from the quantum field theory flow equation only by some non positive powers of the cut-off. Indeed, the ERGE for dimensionless variables \eqref{dimensionless_ERGE} can be written as
\begin{equation}
\Lambda\frac{\partial u_{\cal B}}{\partial\Lambda}
=\sum_{n\geq 0}\frac{\beta^{(n)}_{\cal B}(u_{\cal B})}{\Lambda^{n}}
 \end{equation}
where the functionals $\beta^{(n)}_{\cal B}$ are independent of $\Lambda$. In quantum field theory, one usually has only $\beta^{0}_{\cal B}$. In our case, there are, for a given ${\cal B}$, finitely many negative powers of $\Lambda$. These are not specific to tensorial field theories but can be traced back to the second scale $L$. However, it is not possible to take the infinite volume limit $\Lambda\rightarrow 0$ in a naive way because of the closure constraint.

Therefore, it is reasonable to expect these terms in the flow equation to have no effect  for large $\Lambda$ and 
 do not alter the power counting. This follows from a standard argument in quantum field theory. Suppose that we momentarily disregard the non linear terms in the flow equation \eqref{dimensionless_ERGE}, 
 the latter reduces to $\Lambda\frac{\partial u_{\cal B}}{\partial \Lambda}
=-\delta_{\cal B}u_{\cal B}$. With a boundary condition at $\Lambda_{0}$, it leads to the solution $u_{\cal B}(\Lambda)=(\Lambda/\Lambda_{0})^{-\delta_{\cal B}}u_{\cal B}(\Lambda_{0})$ if the boundary conditions are imposed at the high energy scale $\Lambda_{0}$. Therefore, variables with positive dimensions  diverge as $\Lambda_{0}$ is taken to infinity. Then, renormalisation amounts to impose boundary condition at a low energy scale $\Lambda_{r}$ for those variables. Finally, it may be shown (see \cite{Polchinski}) that this picture remains valid in perturbation theory when the non linear corrections are added.  Indeed, one can inductively prove that $u_{\cal B}$ obeys some inequalities whose leading term is the previous scaling law corrected by some polynomial in $\log(\Lambda/\Lambda_{0})$.

Admitting that this picture remains valid in our context, only the interactions with a positive scaling dimension in $\Lambda$ require renormalisation. Consequently, the only renormalisable (for simplicity we do not distinguish between the superrenromalisable ($\delta>0$) and the just renromalisable ($\delta=0$) cases) interactions are the homogeneous monomials of degree $n$ in the momenta of a bubble ${\cal B}$ whose dimension 
obeys
\begin{equation}
\delta_{{\cal B},n}=
(D-1)d-\kappa_{\cal B}\,d-\big[(D-2)d-2\big]\frac{v_{\cal B}}{2}-n\geq 0.
\label{dimension}
\end{equation}
In particular, among all the renormalisable  interactions, we always have the mass term $\delta_{\includegraphics[width=0.5cm]{dipole.pdf},0}=2$ and for the kinetic term 
$\delta_{\includegraphics[width=0.5cm]{dipole.pdf},2}=0$.

Let us end by listing all renormalisable interactions in Abelian models with a closure constraint. The condition $\delta_{\cal B}\geq 0$ implies $v_{\cal B}\leq 2+\frac{4}{d(D\!-\!2)-2}$. Non trivial interactions involve bubbles with at most four vertices, so that we are left with only two possibilities: $d(D-2)=4$ and $d(D-2)=3$, which leads to the following 5 renormalizable theories. This list, with connected interaction only, was first obtained in the study of a non abelian model \cite{renormnonabelian} using the multiscale analysis of Feynman graphs. Superrenormalisable abelian models (with $D=4$ and $d=1$ so that $d(D-2)=2$, thus leading to arbitrary polynomial interactions) have also been studied in \cite{renormabeliancarro}.

\begin{itemize}
\item $D=3$ and $d=4$ so that $\delta=8-4\kappa-v$. The renormalisable interactions is quartic and melonic. 

\begin{center}
\parbox{1cm}{\includegraphics[width=1cm]{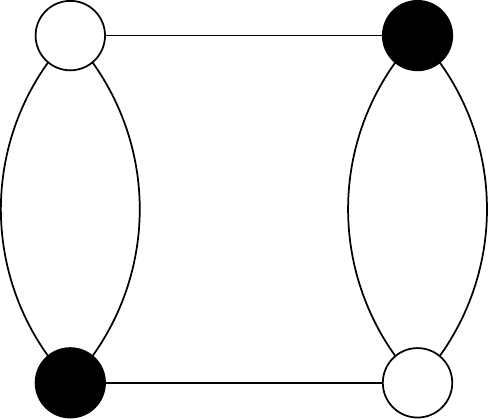}}
\,($\delta=0$)
\end{center}
The fixed point structure of a non abelian version of this model has been studied in \cite{4epsilon}
\item $D=4$ and $d=2$ so that $\delta=6-2\kappa-v$. The renormalisable interactions are quartic 
\begin{center}
\parbox{1cm}{\includegraphics[width=1cm]{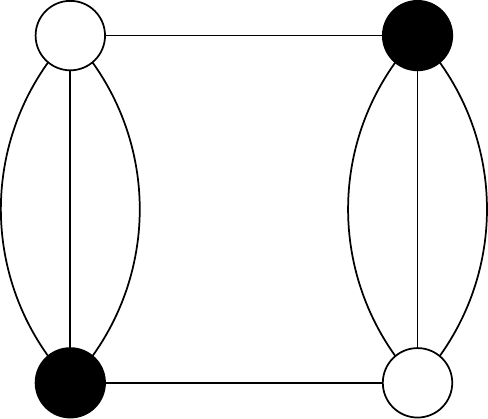}}
\,($\delta=0$),\quad
\parbox{1cm}{\includegraphics[width=1cm]{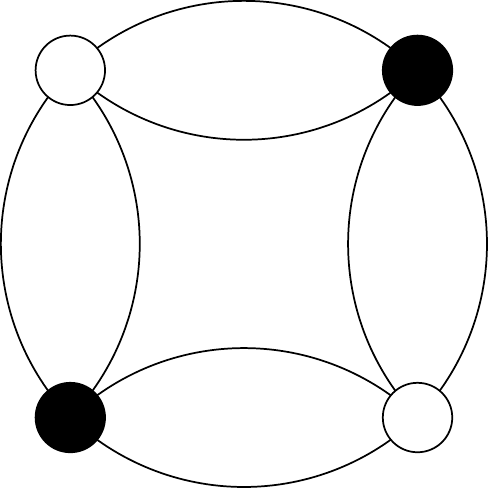}}
\,($\delta=0$)
\end{center}
Note that the second interaction is not melonic (called necklace in \cite{necklace}).

\item $D=6$ and $d=1$ so that $\delta=5-\kappa-v$. The renormalisable interactions are quartic

\begin{center}
\parbox{1.2cm}{\includegraphics[width=1.2cm]{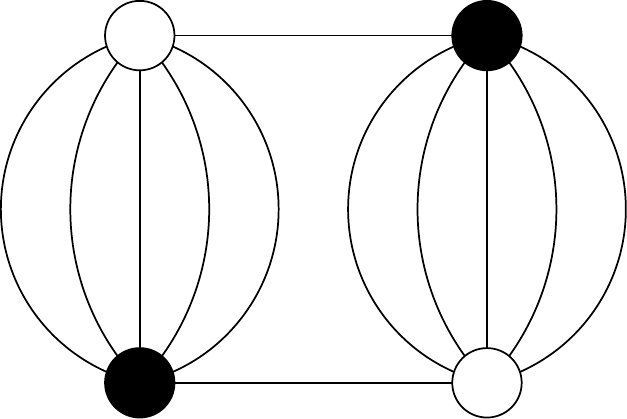}}
\,($\delta=0$),\quad
\parbox{1cm}{\includegraphics[width=1cm]{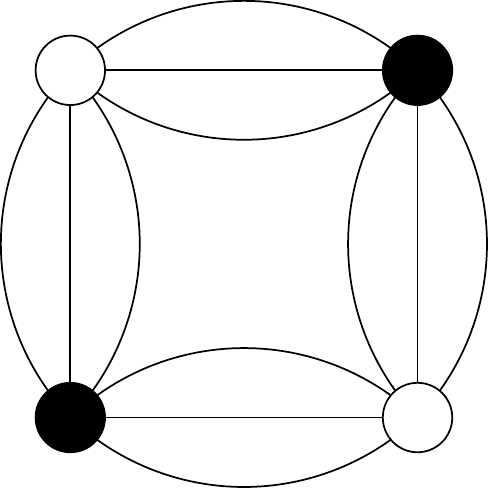}}
\,($\delta=0$),
\quad
\parbox{1cm}{\includegraphics[width=1cm]{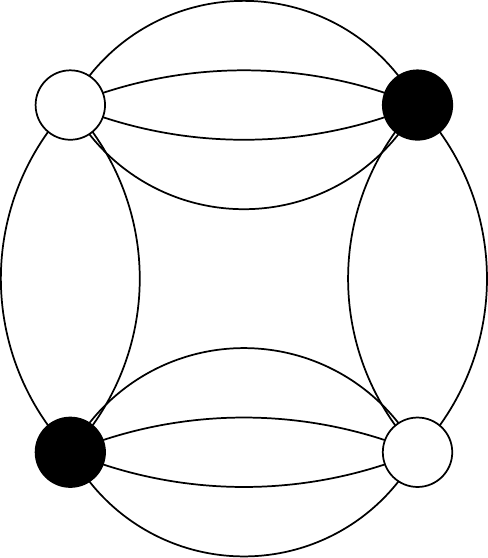}}
\,($\delta=0$).
\end{center}

The last two interactions are not melonic. This model was shown to be renormalizable in \cite{Fabien} and its fixed point structure was further investigated in \cite{Lahoche}.

\item $D=3$ and $d=3$ so that $\delta=6-3\kappa-v/2$. The renormalisable interactions are quartic and sextic

\begin{center}
\parbox{1cm}{\includegraphics[width=1cm]{quarticmelon3}}
\,($\delta=1$),\quad
\parbox{1cm}{\includegraphics[width=1cm]{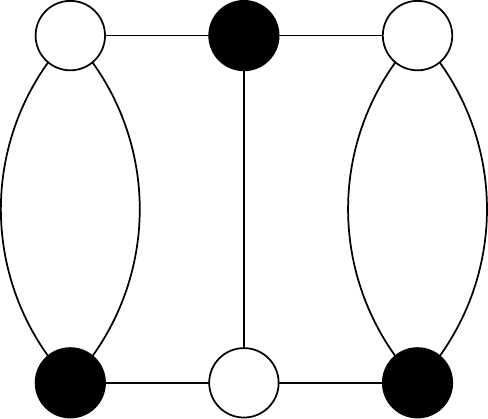}}
\,($\delta=0$),
\quad
\parbox{1cm}{\includegraphics[width=1cm]{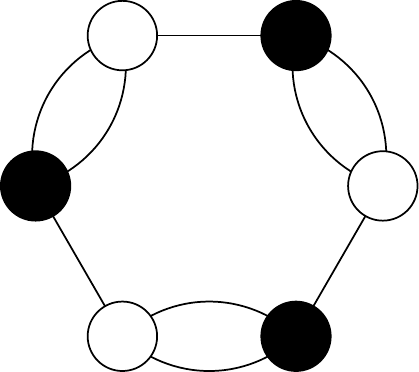}}
\,($\delta=0$),\quad
\parbox{1cm}{\includegraphics[width=1cm]{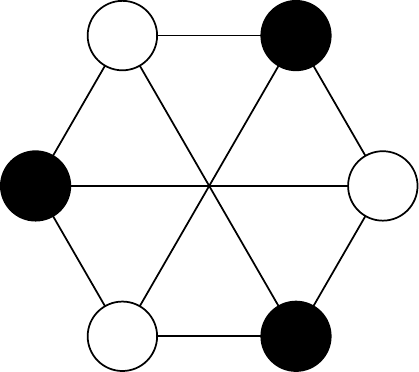}}
\,($\delta=0$).

\end{center}
The first term is in fact superrenormalisable ($\delta>0$) and the last one is not melonic. Its non Abelian counterpart is the tensorial $\text{SU(2)}$ group field theory in $D=3$ which corresponds to three dimensional Euclidian quantum gravity without the cosmological constant, first shown to be renormalisable in \cite{renormnonabelian}. 

\item $D=5$ and $d=1$ so that $\delta=4-\kappa-v/2$. The renormalisable interactions are quartic and sextic. We first have the melonic ones

\begin{center}
\parbox{1.5cm}{\includegraphics[width=1.5cm]{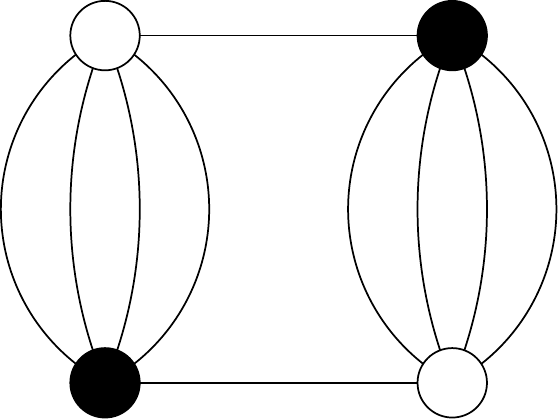}}
\,($\delta=1$),\quad
\parbox{1.5cm}{\includegraphics[width=1.5cm]{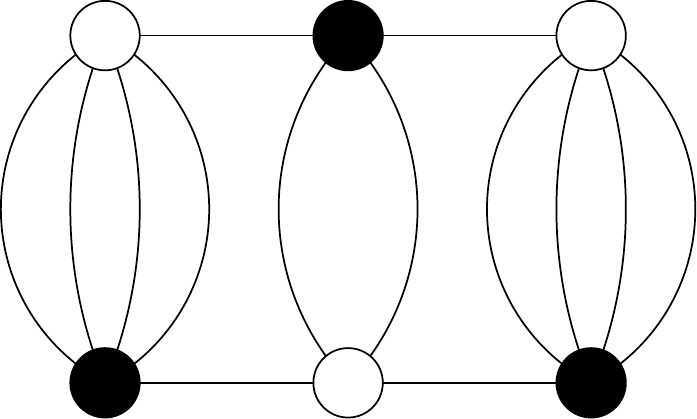}}
\,($\delta=0$),
\quad
\parbox{1cm}{\includegraphics[width=1cm]{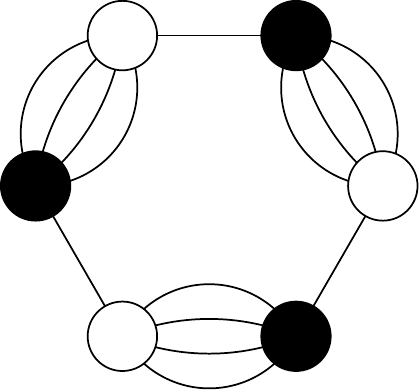}}
\,($\delta=0$),\quad
\parbox{1.5cm}{\includegraphics[width=1.5cm]{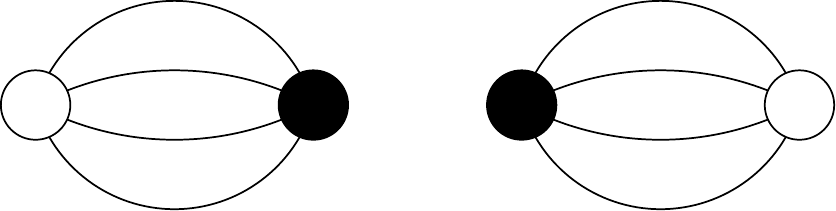}}
\,($\delta=0$).

\end{center}
The last interaction is not connected ($\kappa=2$). This model, with the non connected interaction, was shown to be renormalizable in \cite{Fabien}. Besides, there are also non melonic renormalisable interactions.

\end{itemize}

Note that all these bubble couplings with $v\geq 4$ have dimension 0 or 1, so that there is no possibility of adding a derivative coupling. Indeed, the latter corresponds to a term with $n>0$ in \eqref{dimension}, which would lead to a negative dimension, the case $n=1$ being forbidden by rotational symmetry. Of course, we could always add quadratic terms (mass and kinetic terms) to the list of interaction. Then, the kinetic term can play the role of a derivative coupling.  

It is also worthwhile to notice that some of the renormalisable interactions are not necessary melonic. As for unitarily invariant random tensors, non melonic interaction appear with a negative power of $\Lambda$ in the flow of renormalisable melonic interactions.  Therefore, they do not require renormalisation if the bare theory only contains melonic interactions. However, if the bare theory contains non melonic renormalisable interactions, the latter requires renormalisation. A rigorous analysis of the role of the non melonic couplings, based on bounds on the aforementioned bounds on the couplings, will be presented in a separate publication.

\section{Conclusion and outlook}

In this paper, we have introduced an exact 
{renormalisation}
group equation for tensor models and tensorial field theories, in the form proposed by Polchinski in the context of quantum field theories. We have expanded the Wilsonian effective action on bubble couplings $\lambda_{\cal B}$ and formulated a system of differential equations for these couplings, both for invariant tensors \eqref{flow_invariants} and Abelian tensorial group field theories with {a} closure constraint \eqref{abelian_ERGE}. These differential 
{equations} have provided us with two applications.

\begin{itemize}
\item We have rederived, using the flow equation, Gurau's universality theorem for invariant random tensors. This study could be pursued by further investigating the phase structure of random tensor models.

\item In the case of Abelian tensorial group field theories, we have introduced a suitable notion of the dimension for bubble couplings, which led us to a list of only five renormalisable theories. Of course, this is only a first step and non Abelian models with a simplicity constraint should be considered.
\end{itemize}

To conclude, let us emphasise that the originality of our method lies in the fact that we have not evaluated a single $(D+1)$-dimensional Feynman graphs. Our equations only involve $D$-dimensional graphs, which {collect} the contribution of all space-time triangulations with a fixed boundary.   Therefore, it is technically simpler than evaluating the Feynman graphs and may be applied to a large class of models, including models with a simplicity constraint. 
 
Furtermore, let us note that although our study is exact (no truncations of the equations have been used), it remains perturbative. Non perturbative studies, based on truncations to a couple of bubble couplings, have been recently presented in the literature \cite{tensorERGE}, \cite{4epsilon} and \cite{Lahoche}.    These works used Wetterich's formulation of the ERGE, more suitable than Polchinski's for truncations. For tensor models, Wetterich's equation can also be  formulated using cuts and lead to a system of differential equations for the couplings.

\appendix
%%%%%%%%%%%%%%

\baselineskip 16pt \oddsidemargin 0pt \evensidemargin 0pt \topmargin
0pt \headheight 0pt \headsep 0pt \footskip 32pt \textheight
40\baselineskip \advance \textheight by \topskip \textwidth 470pt
\makeatletter

%%%%%%%%%%%%%%

\section*{Acknowledgments}

T.K. thanks ANR JCJC CombPhysMat2Tens grant for partial support and R. Gurau for interesting and helpful discussions. T.K. and R.T. thank ESI Vienna for hospitality during the program "Combinatorics, Geometry and Physics".

\section{ERGE for low order bubble couplings}

In this appendix, we give explicit evolution equations for low order bubble couplings in invariant tensor models. We order terms in decreasing powers of $N$.

\subsection{Couplings in rank $D=3$ tensors models}

\label{example3}

\bea
{\partial \over \partial t}  \lambda_{\includegraphics[width=0.7cm]{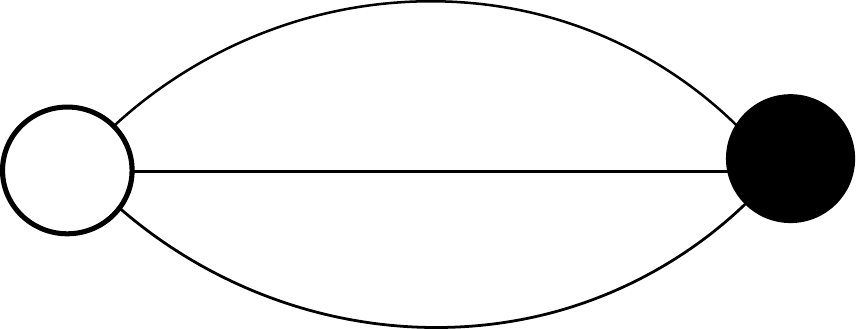}}
&=&
\big[
\lambda_{{\includegraphics[width=0.7cm]{eq2point1}} \; {\includegraphics[width=0.7cm]{eq2point1}}}
\big]\big \vert_{0 \; {\rm{cut}}}
+
\big[
3 \;
\lambda_{\includegraphics[width=0.7cm]{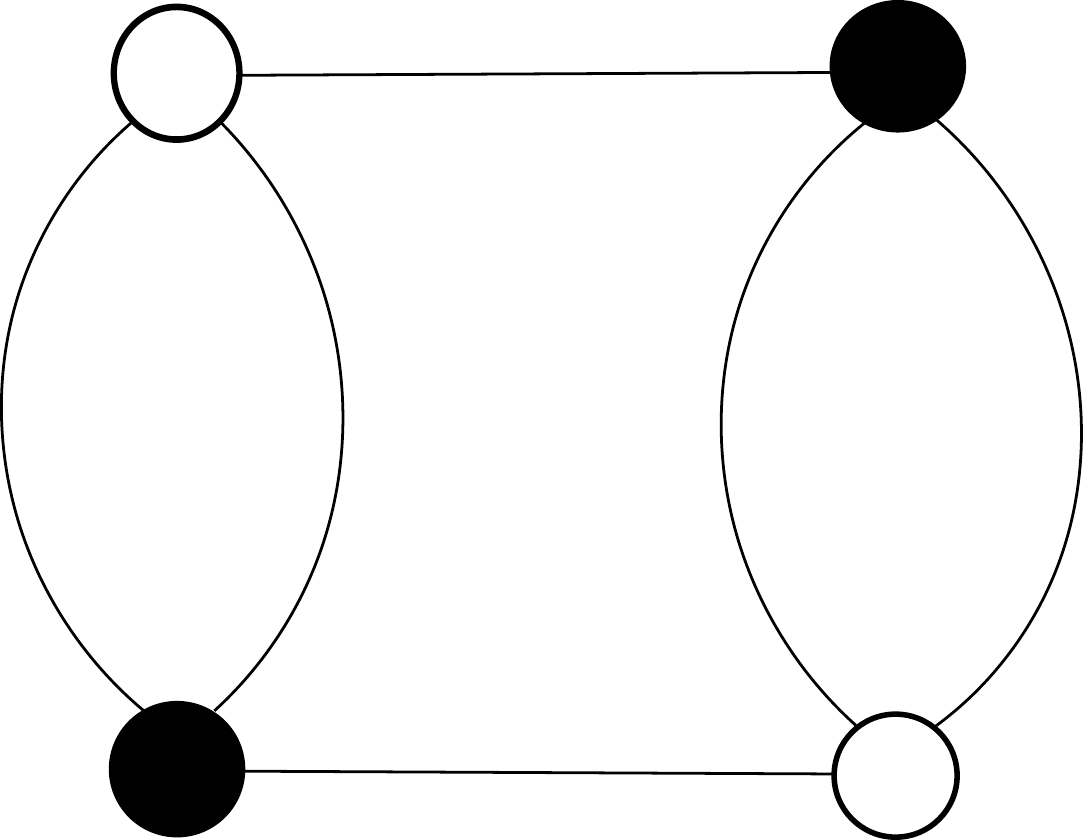}}
\big]\big \vert_{1 \; {\rm{cut}}}
%-
+
\big[
\lambda^2_{\includegraphics[width=0.7cm]{eq2point1}}
\big]\big \vert_{3 \; {\rm{cuts}}}
\nonumber \\
&&
+
{1 \over N}  
\big[
3 \;
\lambda_{\includegraphics[width=0.7cm]{eq2point2}}
\big]\big \vert_{2 \; {\rm{cuts}}}
+
{1 \over N^3}
\big[
\lambda_{{\includegraphics[width=0.7cm]{eq2point1}} \; {\includegraphics[width=0.7cm]{eq2point1}}}
\big]\big \vert_{3 \; {\rm{cuts}}}
\,.
\label{eq:D3deqStart}
\eea

\bea
\nonumber \\ \cr
{\partial \over \partial t}  \lambda_{\includegraphics[width=0.7cm]{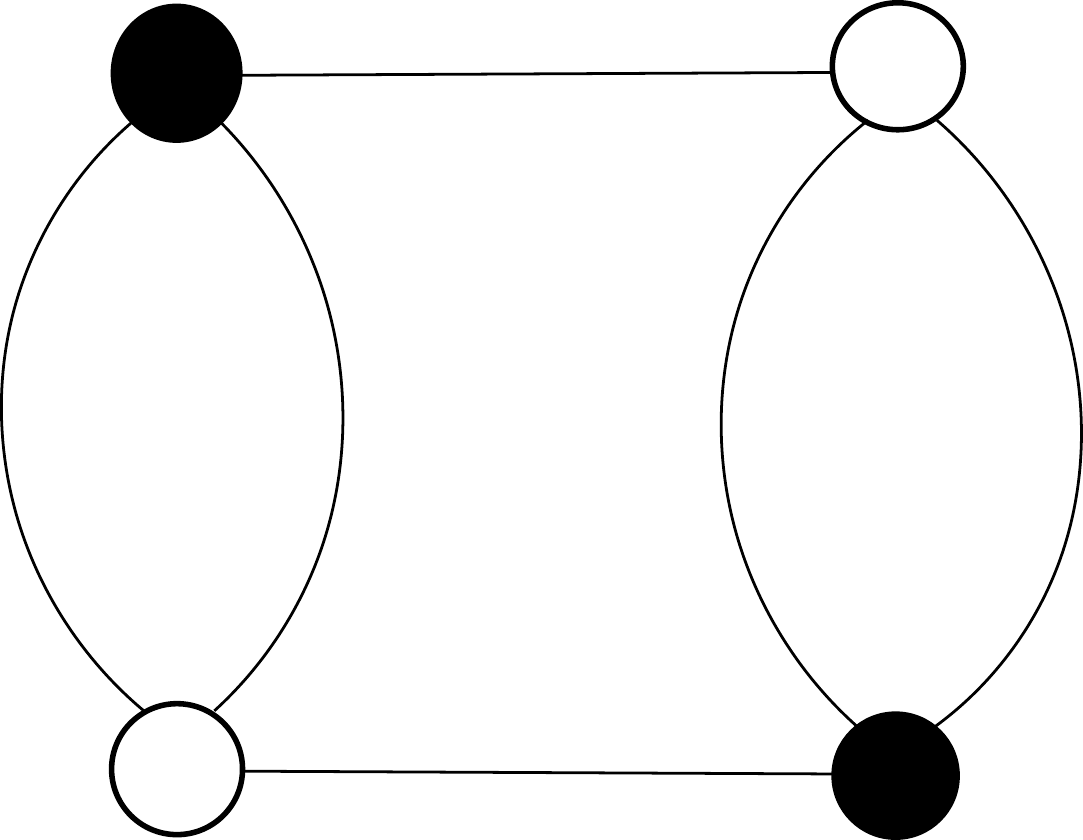}}
&=&
\big[
\lambda_{\includegraphics[width=1.4cm]{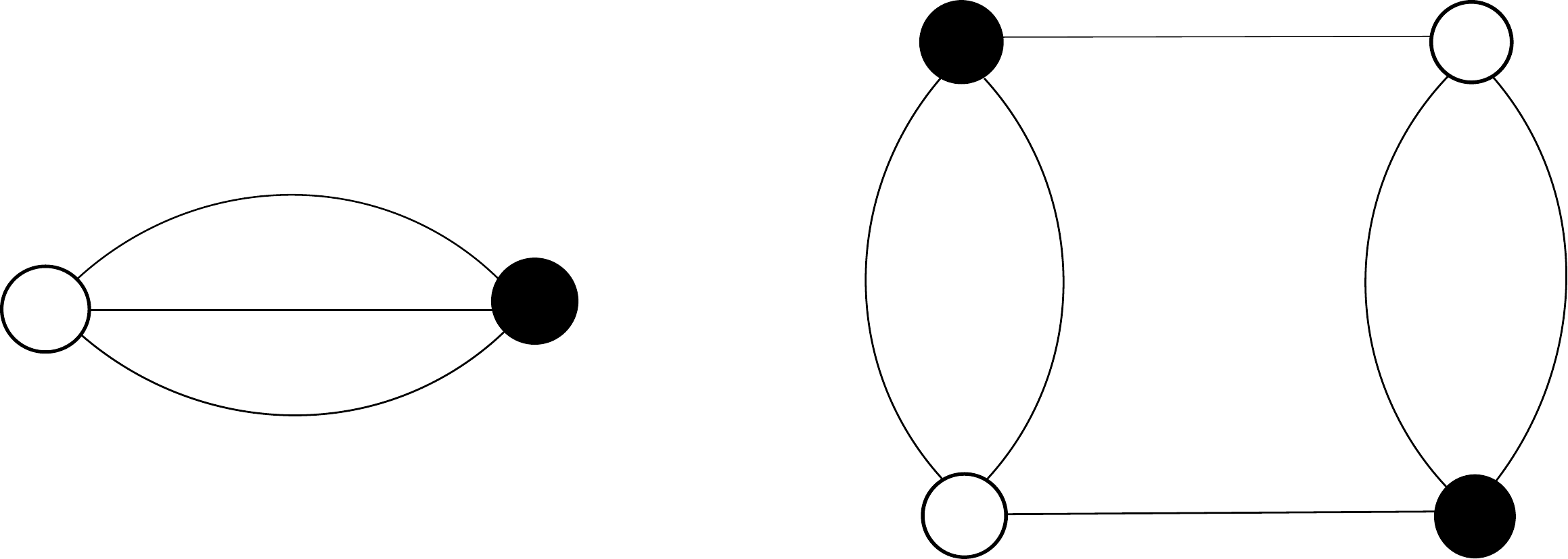}}
\big]\big \vert_{0 \; {\rm{cut}}}
+
\big[
4 \;
\lambda_{\includegraphics[width=0.7cm]{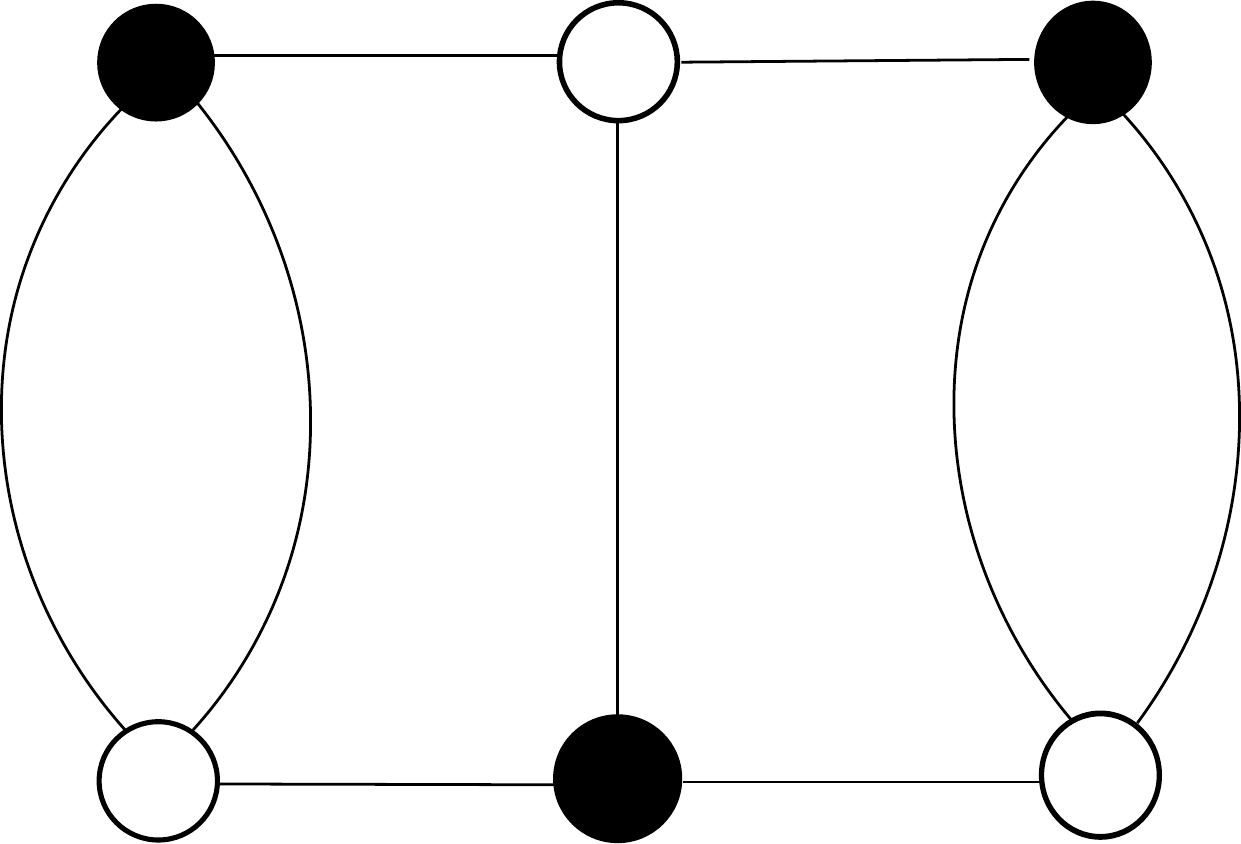}}
+
2 \;
\lambda_{\includegraphics[width=0.7cm]{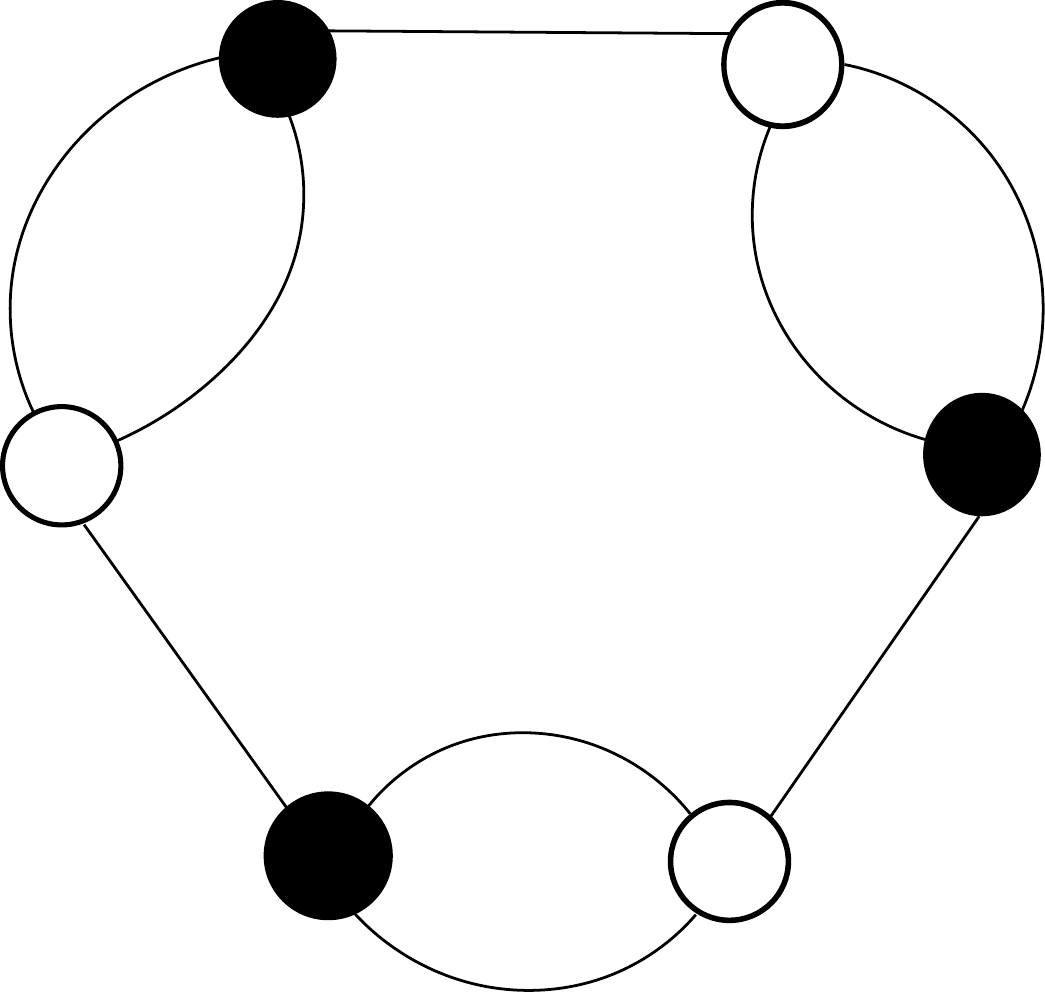}}
\big]\big \vert_{1 \; {\rm{cut}}}
+
\big[
4 \;
\lambda_{\includegraphics[width=0.7cm]{eq2point1}}
\lambda_{\includegraphics[width=0.7cm]{eq4point1}}
\big]\big \vert_{3 \; {\rm{cuts}}}
\nonumber \\
&&
+
{1 \over N} 
\big[
8 \;
\lambda_{\includegraphics[width=0.7cm]{eq62point1}}
+
2 \;
\lambda_{\includegraphics[width=0.7cm]{eq61point1}}
+
2 \;
\lambda_{\includegraphics[width=0.7cm]{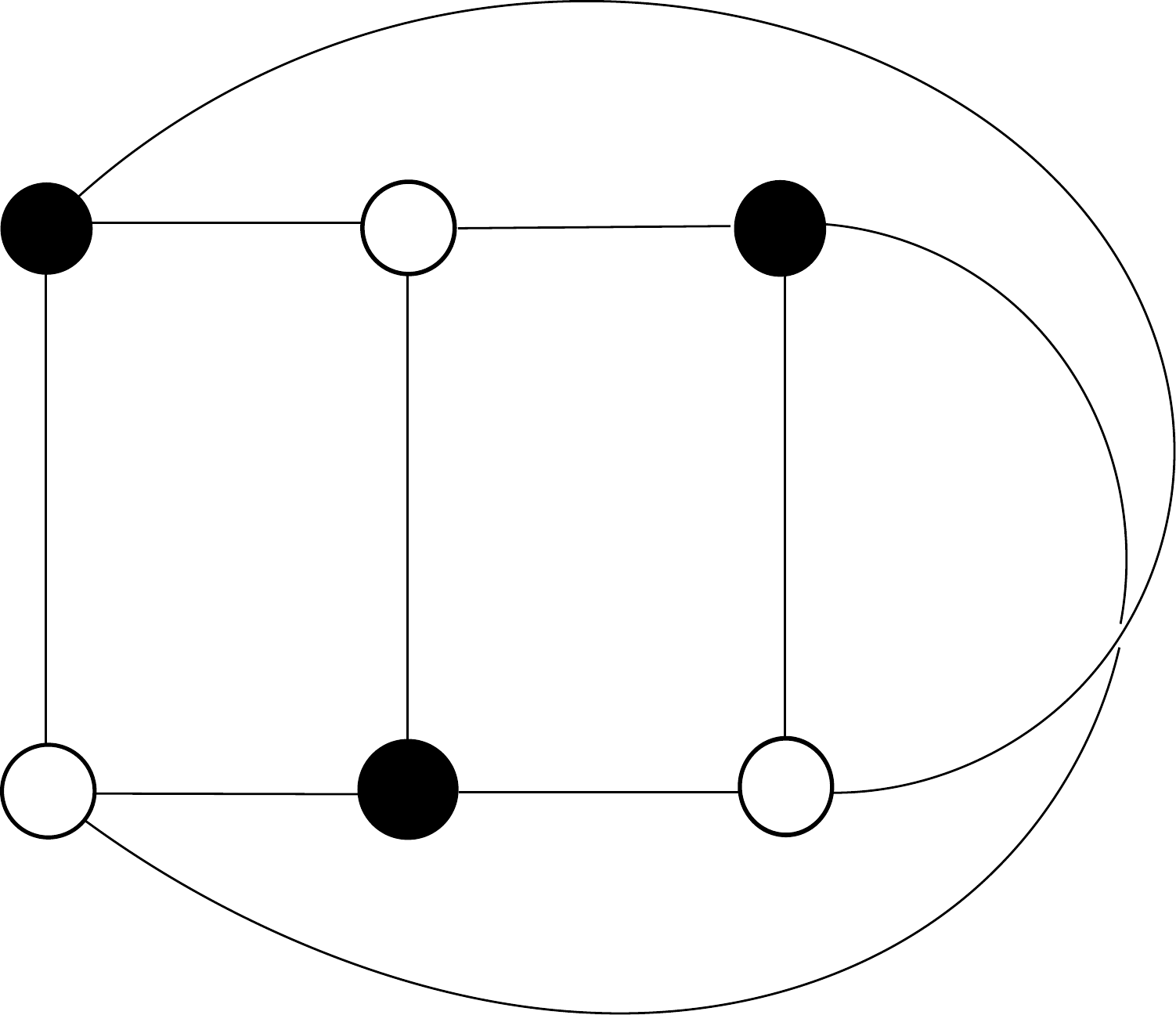}}
\big]\big \vert_{2 \; {\rm{cuts}}}
\nonumber \\
&&
+
{1 \over N^2}
\big[ 
4
\; \lambda_{\includegraphics[width=0.7cm]{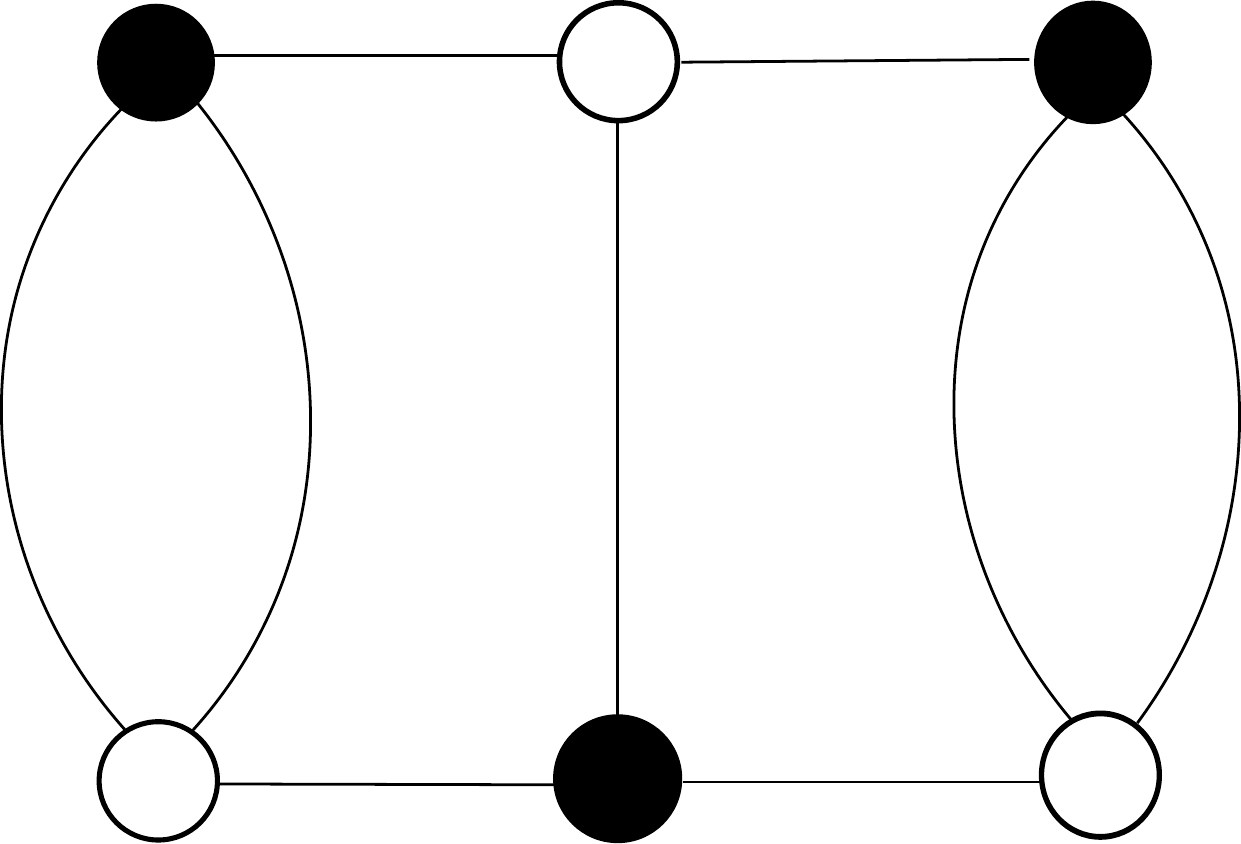}}
\big]\big \vert_{3 \; {\rm{cuts}}}
+
{1 \over N^3}
\big[
4
\;\lambda_{\includegraphics[width=1.4cm]{eq4point2}}
\big]\big \vert_{3 \; {\rm{cuts}}}
\,.
\eea

\bea
\nonumber \\ \cr
{\partial \over \partial t}  \lambda_{{\includegraphics[width=0.7cm]{eq2point1}} \; {\includegraphics[width=0.7cm]{eq2point1}}}
&=&
\big[
\lambda_{{\includegraphics[width=0.7cm]{eq2point1}} \; {\includegraphics[width=0.7cm]{eq2point1}}\; {\includegraphics[width=0.7cm]{eq2point1}}}
\big]\big \vert_{0 \; {\rm{cut}}}
+
\big[
6 \;
\lambda_{\includegraphics[width=1.4cm]{eq4point2}}
\big]\big \vert_{1 \; {\rm{cut}}}
+
\big[
6 \;
\lambda_{\includegraphics[width=0.7cm]{eq62point1}}
\big]\big \vert_{2 \; {\rm{cuts}}}
\nonumber \\
&&
+
\big[
4 \;
\lambda_{{\includegraphics[width=0.7cm]{eq2point1}} \; {\includegraphics[width=0.7cm]{eq2point1}}}
\lambda_{\includegraphics[width=0.7cm]{eq2point1}} 
\big]\big \vert_{3 \; {\rm{cuts}}}
\nonumber \\
&&
+
{1 \over N} 
\Big \{
\big[
6 \;
\lambda_{\includegraphics[width=1.4cm]{eq4point2}}
\big]\big \vert_{2 \; {\rm{cuts}}}
+
\big[
6 \;
\lambda_{\includegraphics[width=0.7cm]{eq61point1}}
\big]\big \vert_{3 \; {\rm{cuts}}}
\Big \}
\nonumber \\
&&
+
{1 \over N^3} 
\big[
2\; 
\lambda_{{\includegraphics[width=0.7cm]{eq2point1}} \; {\includegraphics[width=0.7cm]{eq2point1}}\; {\includegraphics[width=0.7cm]{eq2point1}}}
\big]\big \vert_{3 \; {\rm{cuts}}}
\,.
\eea

\bea
\nonumber \\ \cr
{\partial \over \partial t}  \lambda_{\includegraphics[width=0.7cm]{eq61point1}}
\!\!&=& \!\!
\big[
 \lambda_{\includegraphics[width=1.4cm]{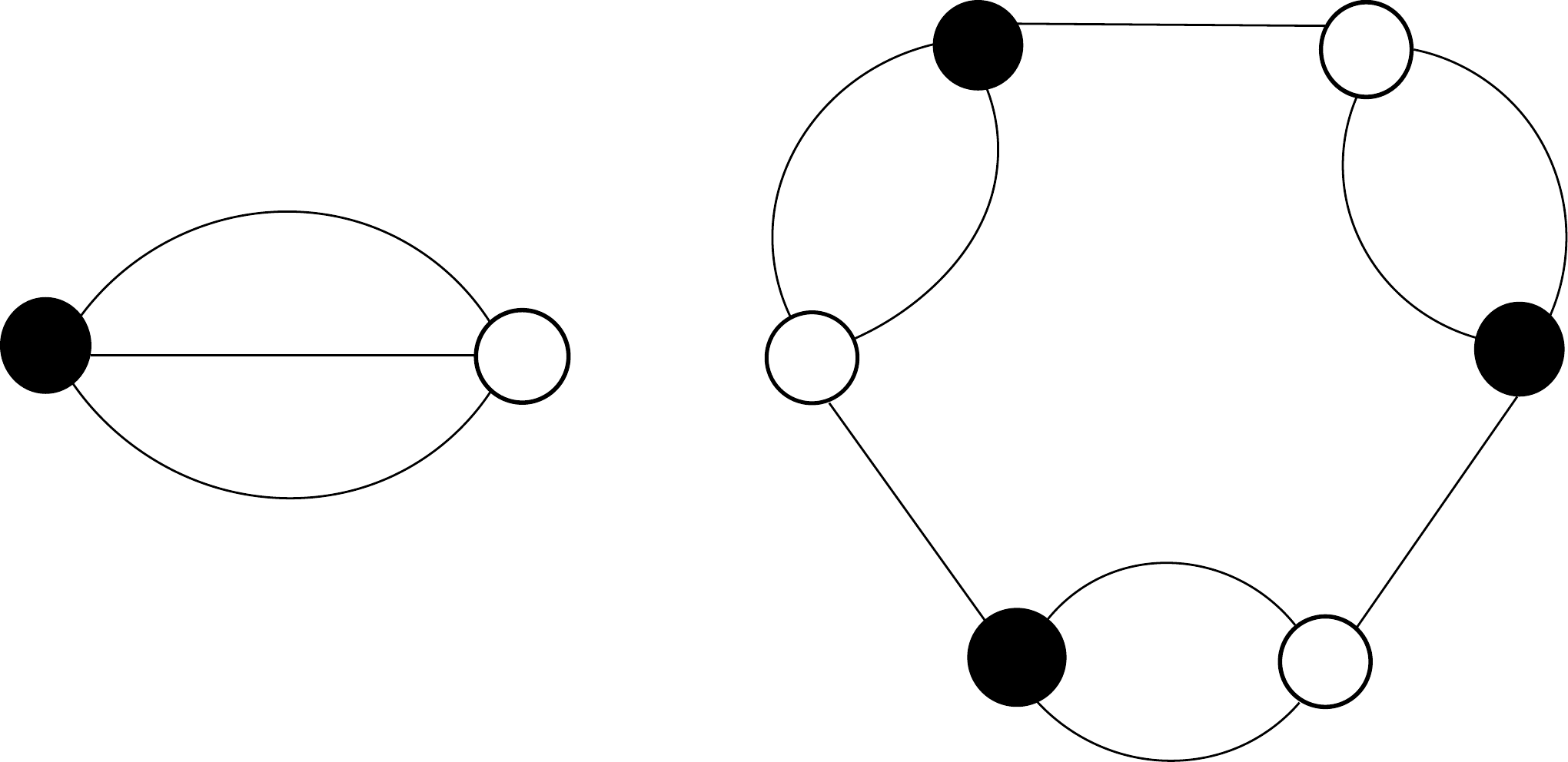}}
\big]\big \vert_{0 \; {\rm{cut}}}
\! + \!
\big[
6 \;
\lambda_{\includegraphics[width=1.1cm]{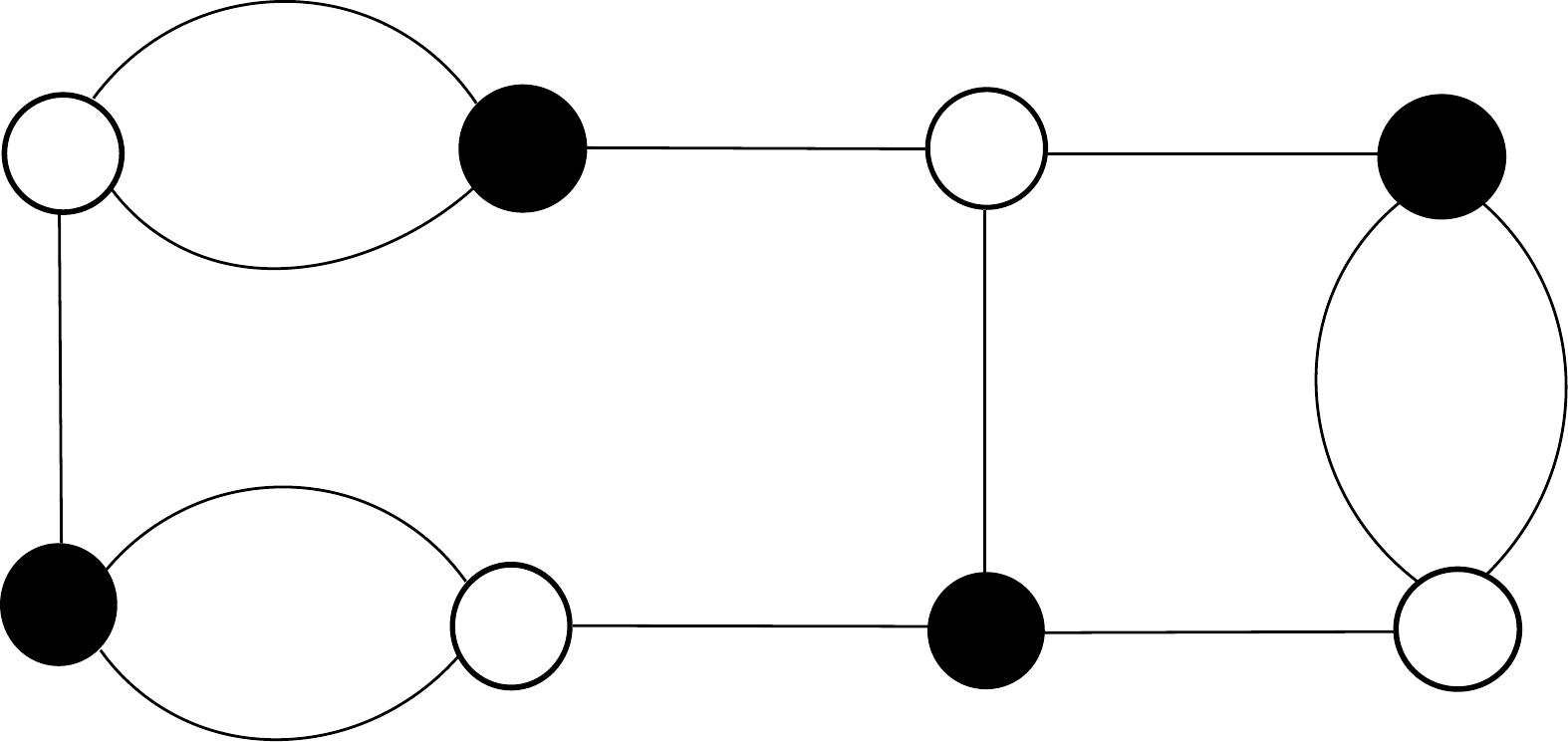}}
\! + \!
3 \;
\lambda_{\includegraphics[width=0.7cm]{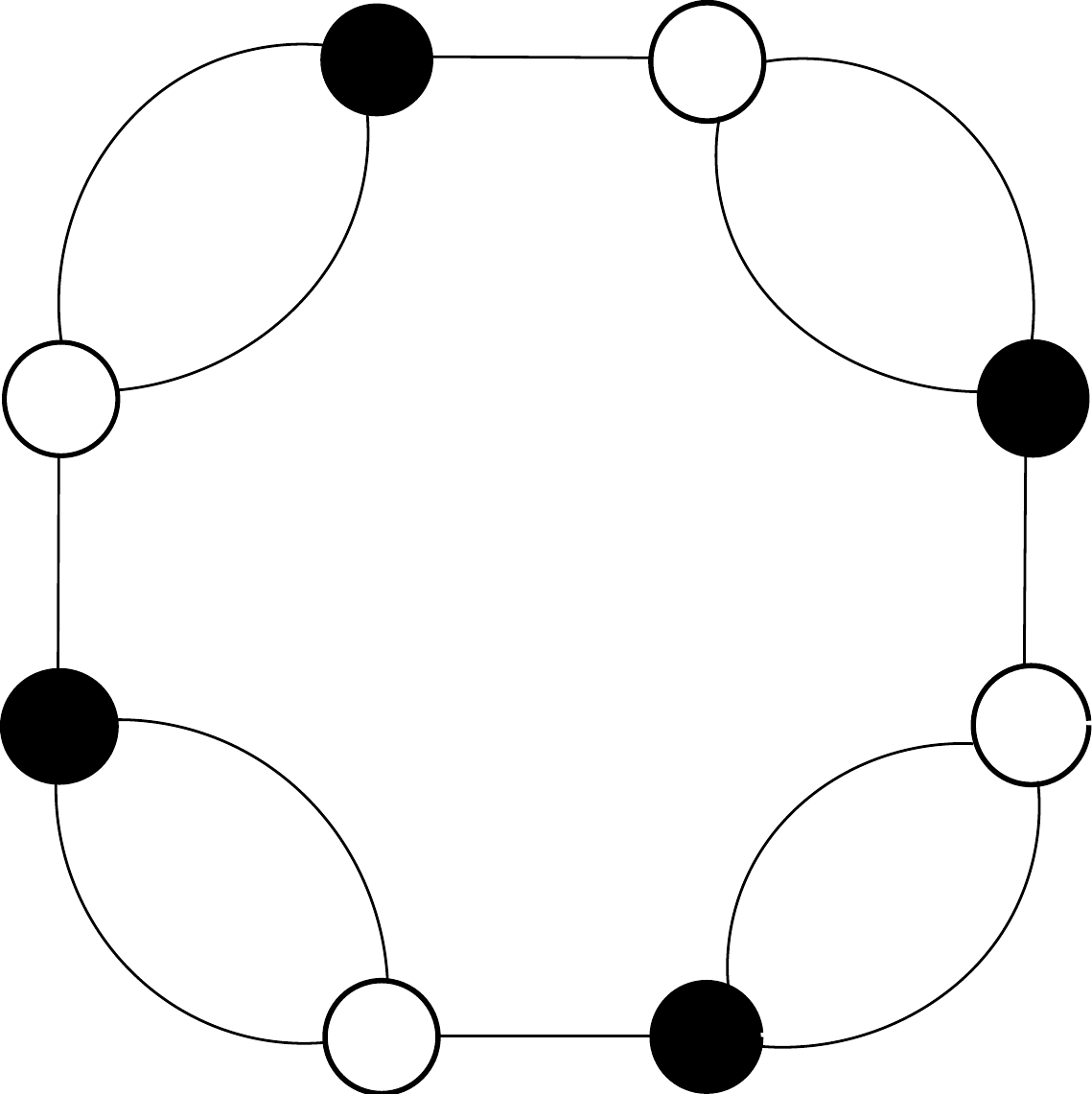}}
\big]\big \vert_{1 \; {\rm{cut}}}
\! + \!
\big[
6 \;
\lambda_{\includegraphics[width=0.7cm]{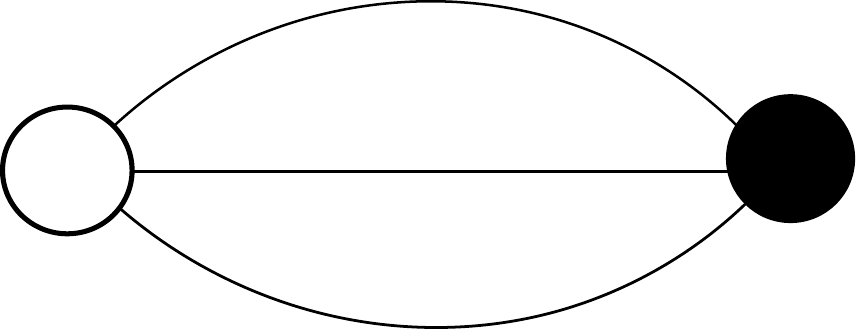}}
\lambda_{\includegraphics[width=0.7cm]{eq62point1}}
+
3 \;
\lambda^2_{\includegraphics[width=0.7cm]{eq4point1}}
\big]\big \vert_{3 \; {\rm{cuts}}}
\nonumber \\
&&
+
{1 \over N} 
\big[
3 \;
 \lambda_{\includegraphics[width=0.7cm]{eq61point6}}
+
6 \;
 \lambda_{\includegraphics[width=1.0cm]{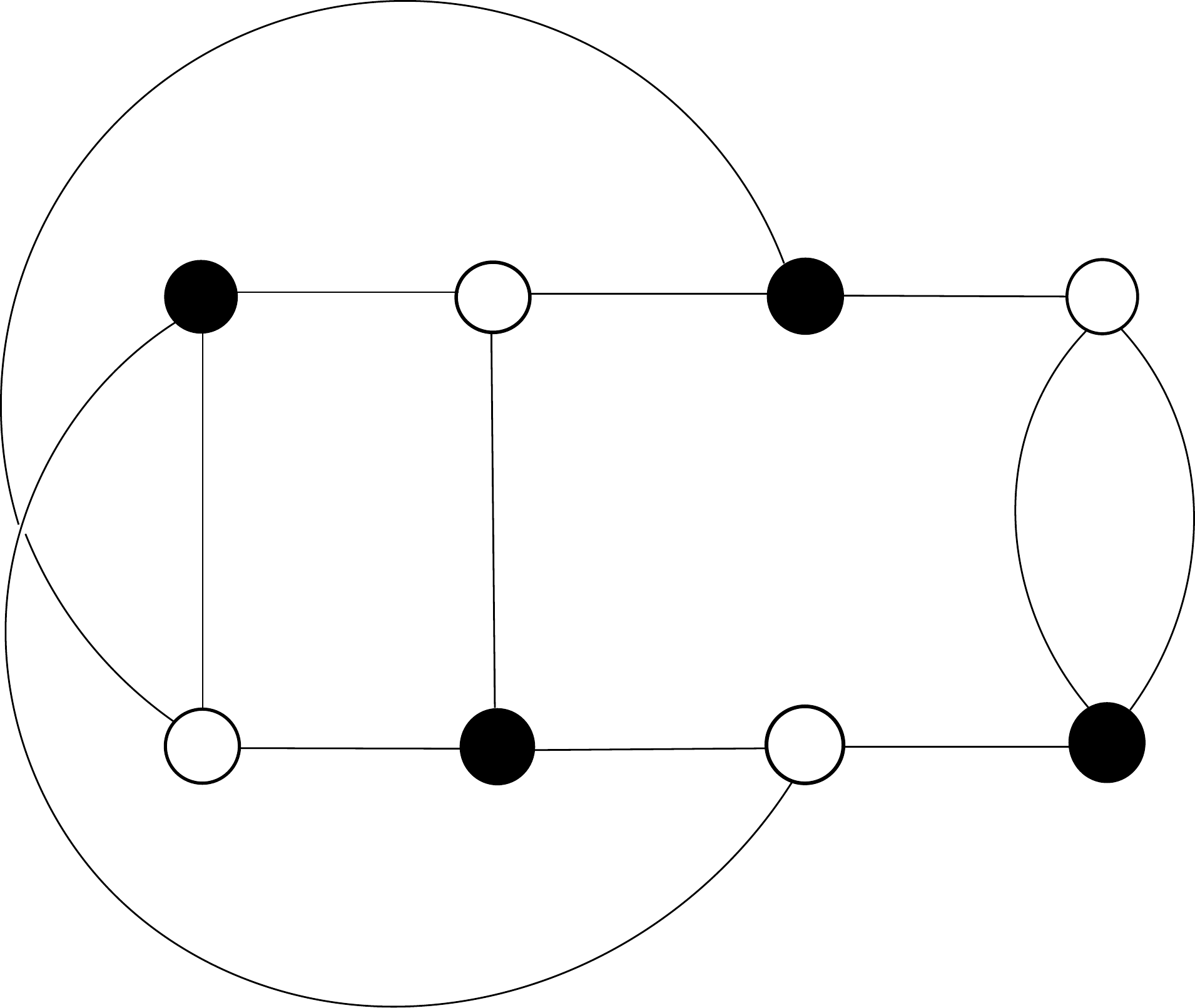}}
+
12 \;
 \lambda_{\includegraphics[width=1.1cm]{eq61point4}}
+ 
6 \;
 \lambda_{\includegraphics[width=1.0cm]{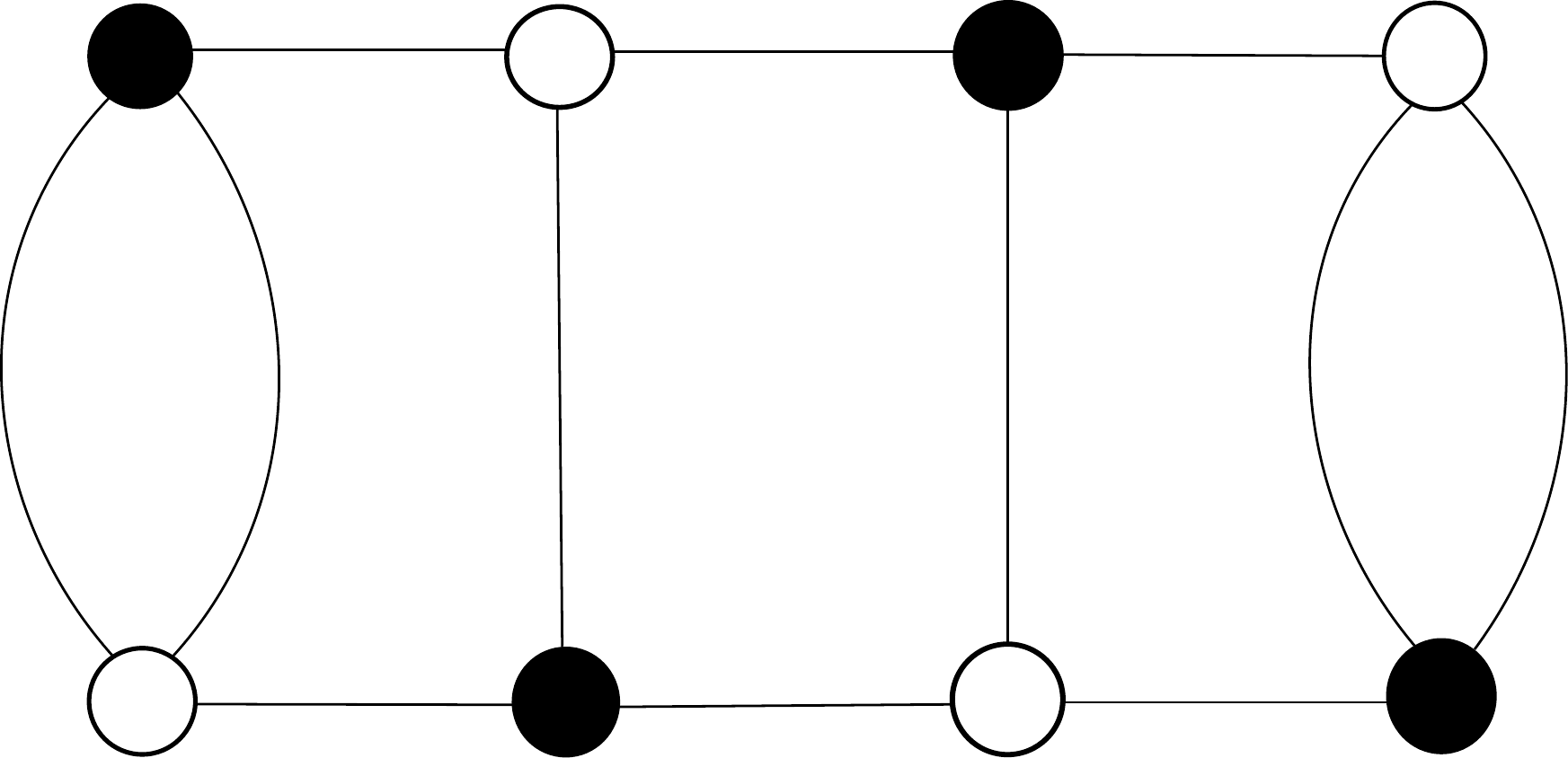}}
\big]\big \vert_{2 \; {\rm{cuts}}}
\nonumber \\
&&
+
{1 \over N^2} 
\big[ 
12 \;
\; \lambda_{\includegraphics[width=1.0cm]{eq61point8}}
+
6 \; \lambda_{\includegraphics[width=0.7cm]{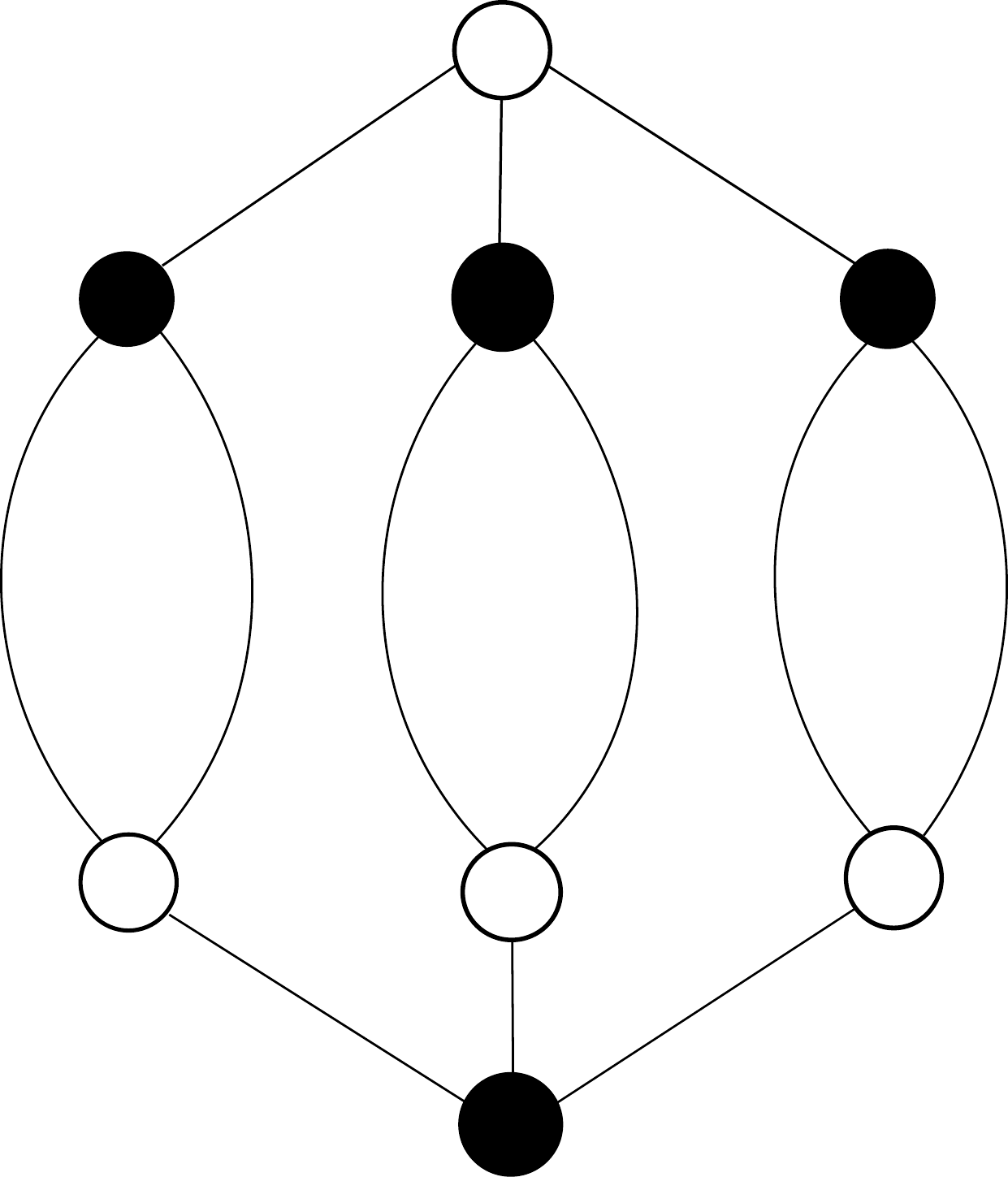}}
\big]\big \vert_{3 \; {\rm{cuts}}}
+
{1 \over N^3} 
\big[
6\;
\lambda_{\includegraphics[width=1.4cm]{eq61point2}}
+
3
\; \lambda_{\includegraphics[width=1.4cm]{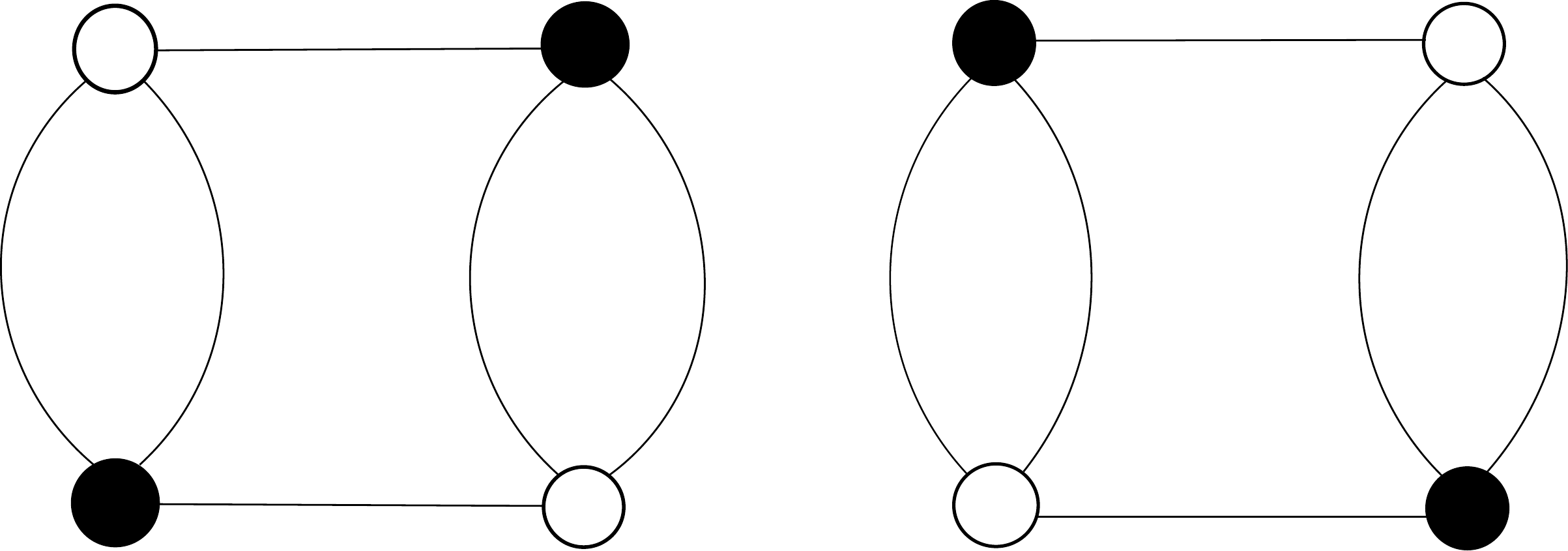}}
\big]\big \vert_{3 \; {\rm{cuts}}}
\,.
\eea

\bea
{\partial \over \partial t} 
\lambda_{\includegraphics[width=0.9cm]{eq62point1}}
&=&
\big[  
\;\lambda_{\includegraphics[width=1.6cm]{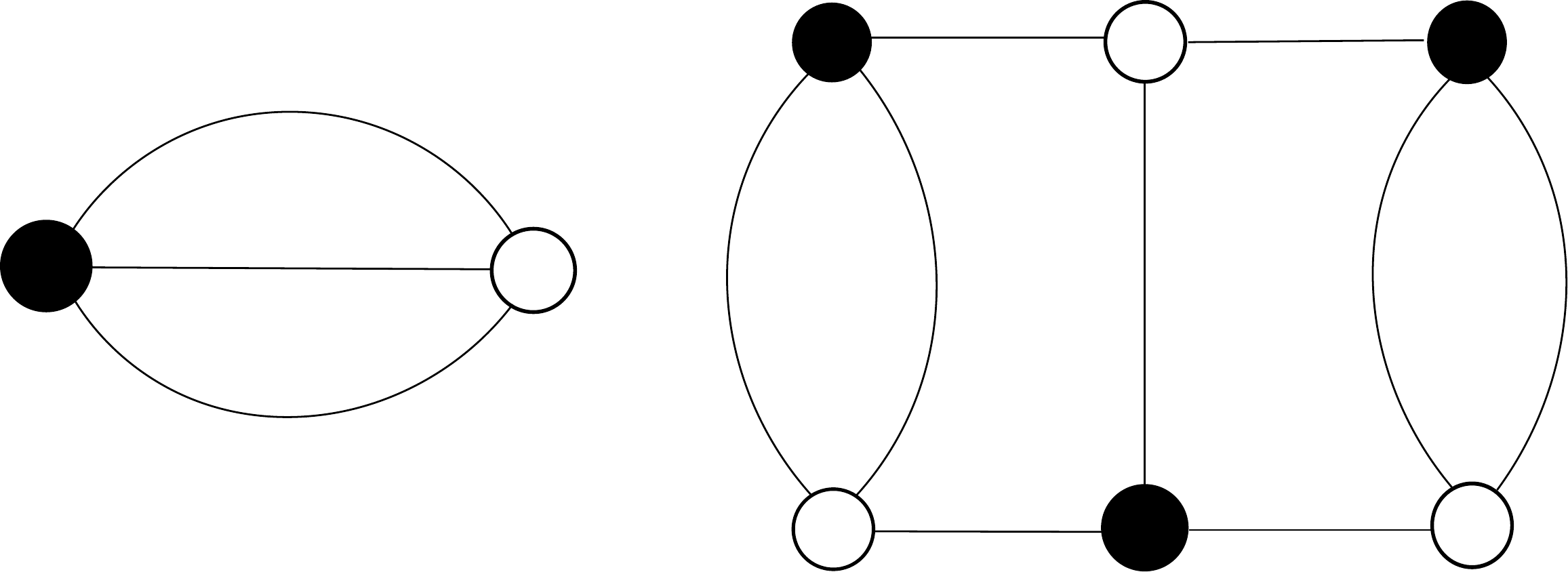}}
\big]\big \vert_{0 \; {\rm{cut}}}
+
\big[ 
4
\;\lambda_{\includegraphics[width=1.0cm]{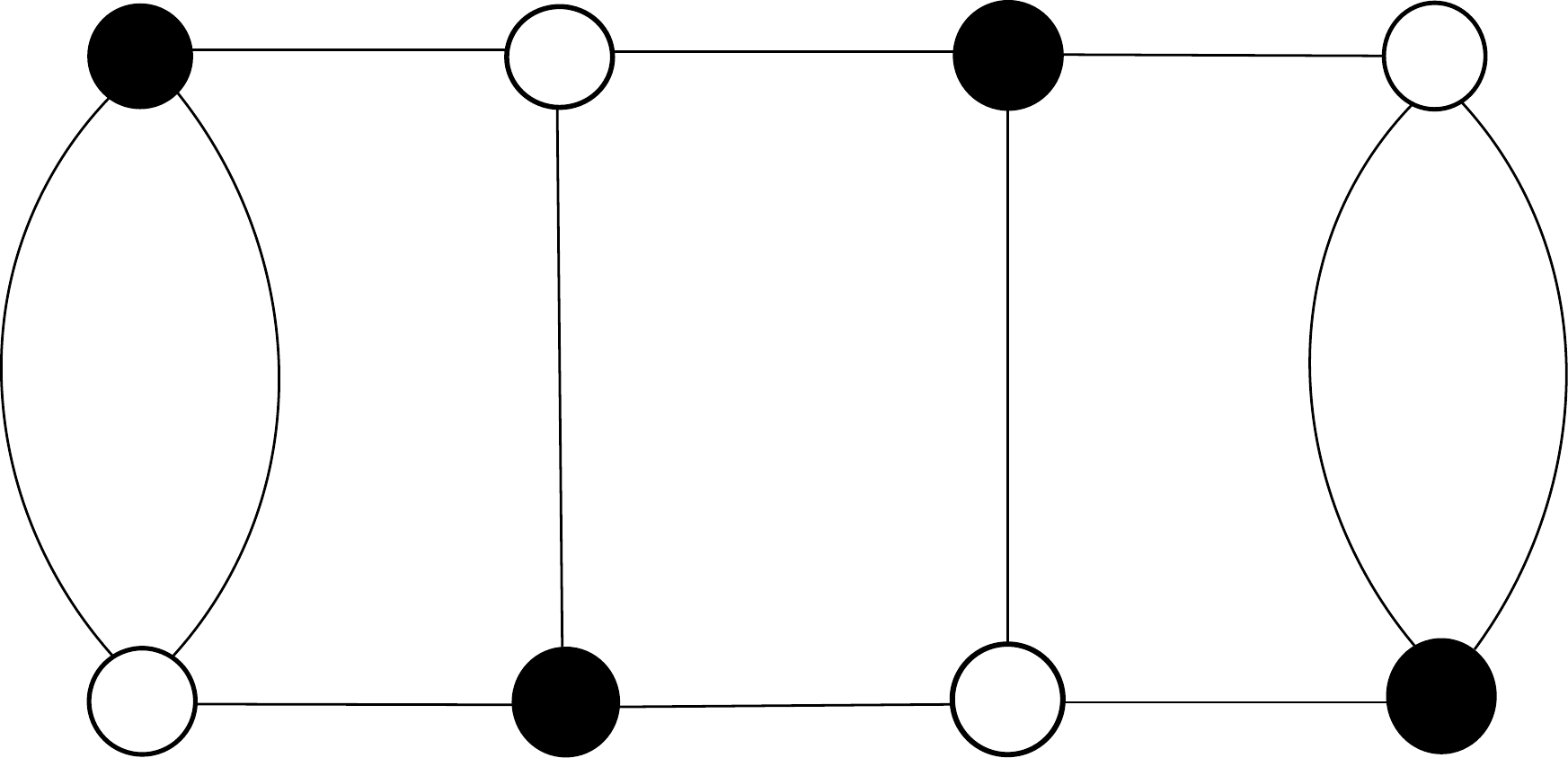}}
+
\lambda_{\includegraphics[width=0.7cm]{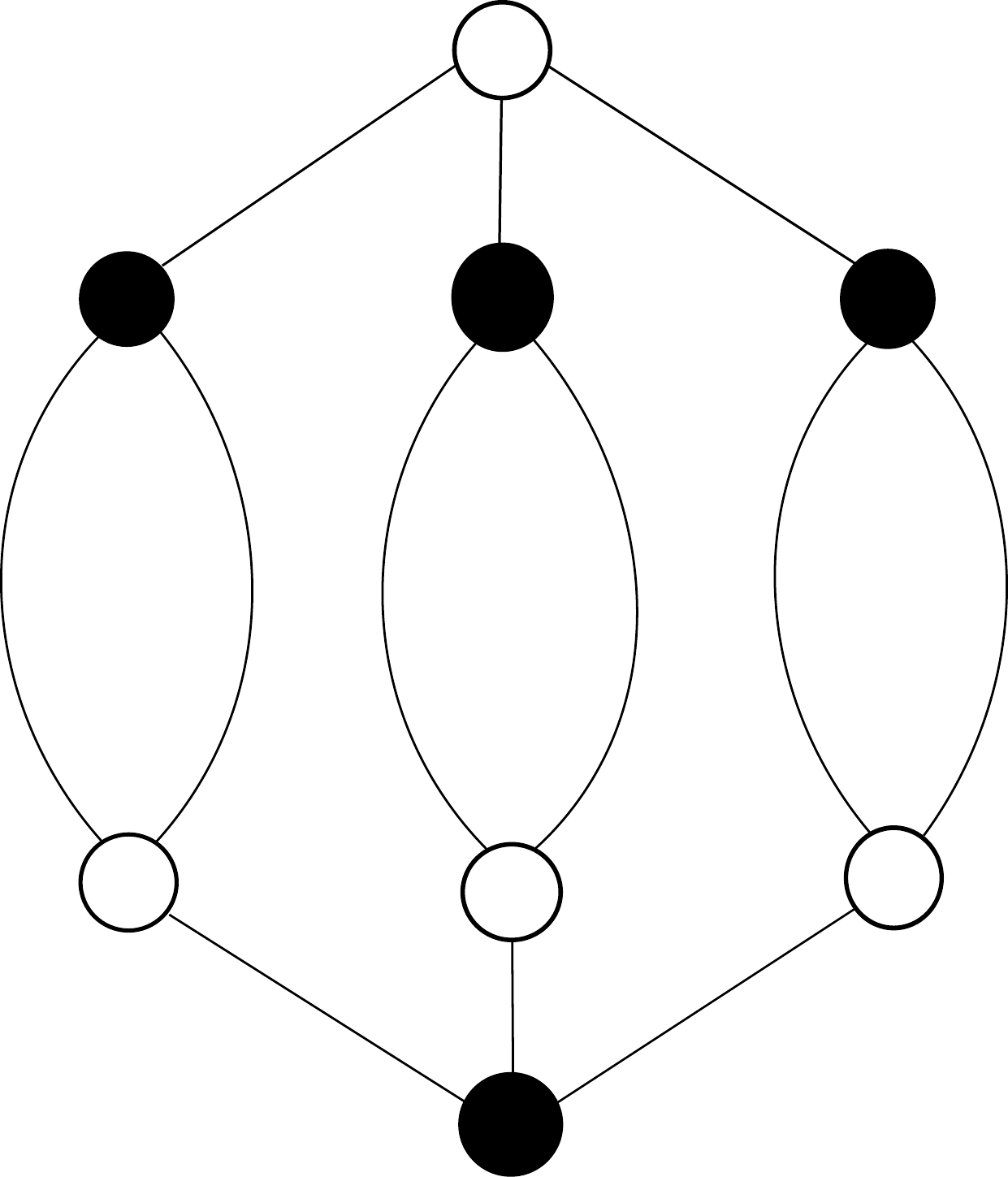}}
+ 
4
\; \lambda_{\includegraphics[width=1.0cm]{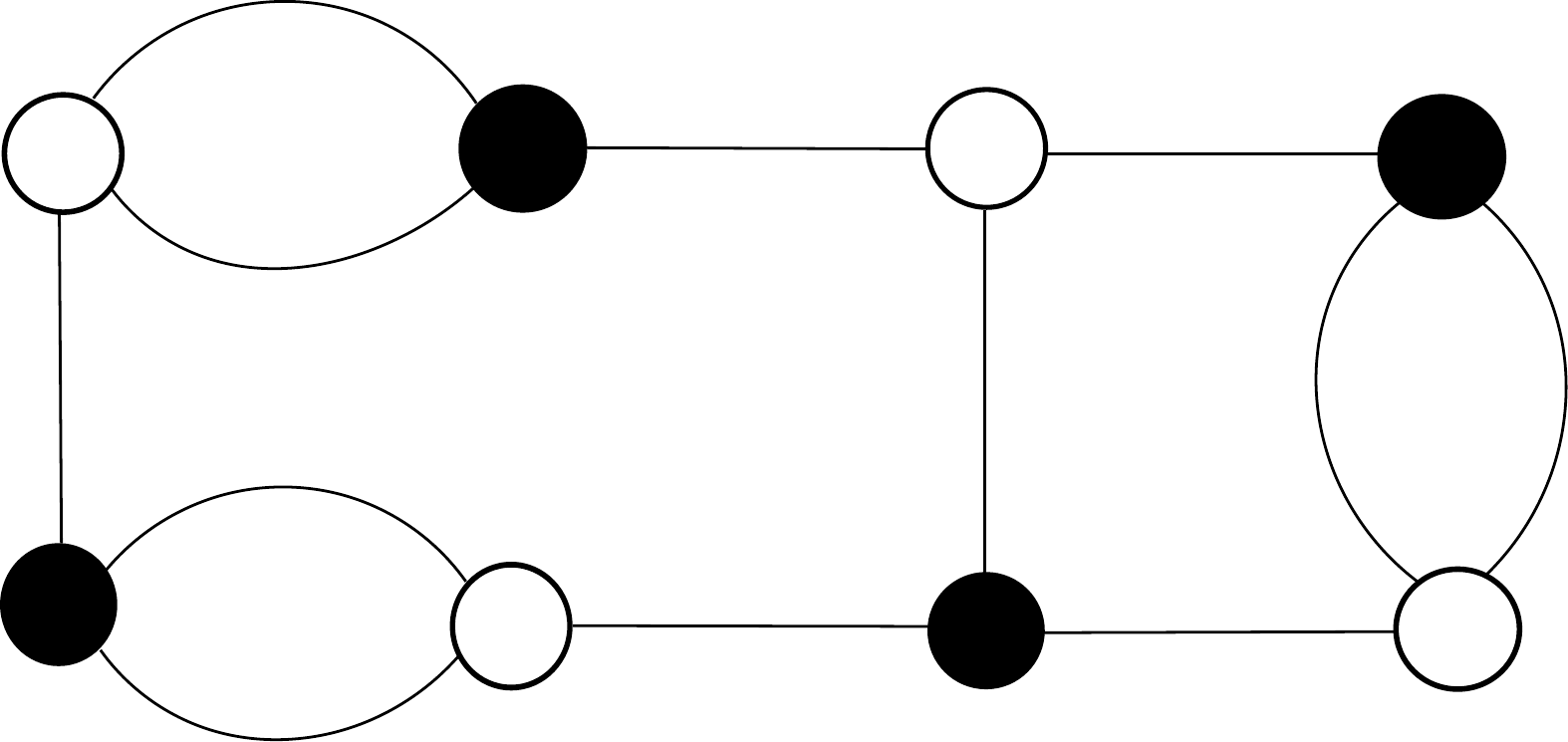}}
\big]\big \vert_{1 \; {\rm{cut}}}
\nonumber \\
&&
+ 
\big[
6\;
\lambda_{\includegraphics[width=0.7cm]{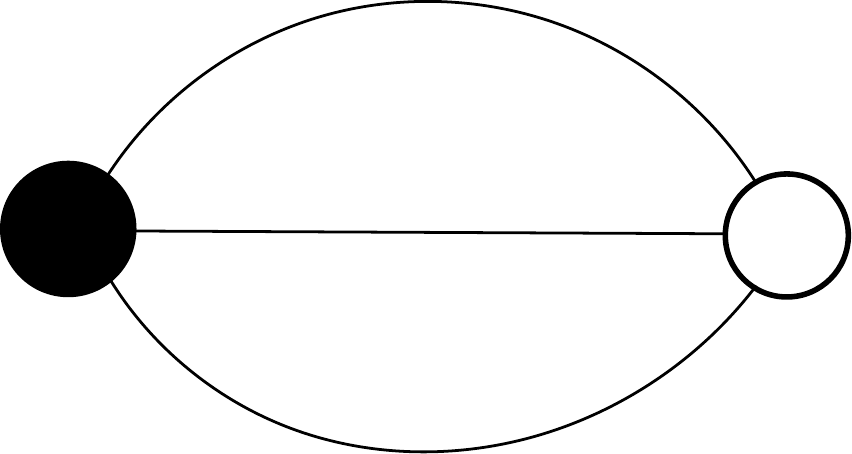}}
\;\lambda_{\includegraphics[width=0.7cm]{eq62point1}}
+
2 \;
\lambda^2_{\includegraphics[width=0.7cm]{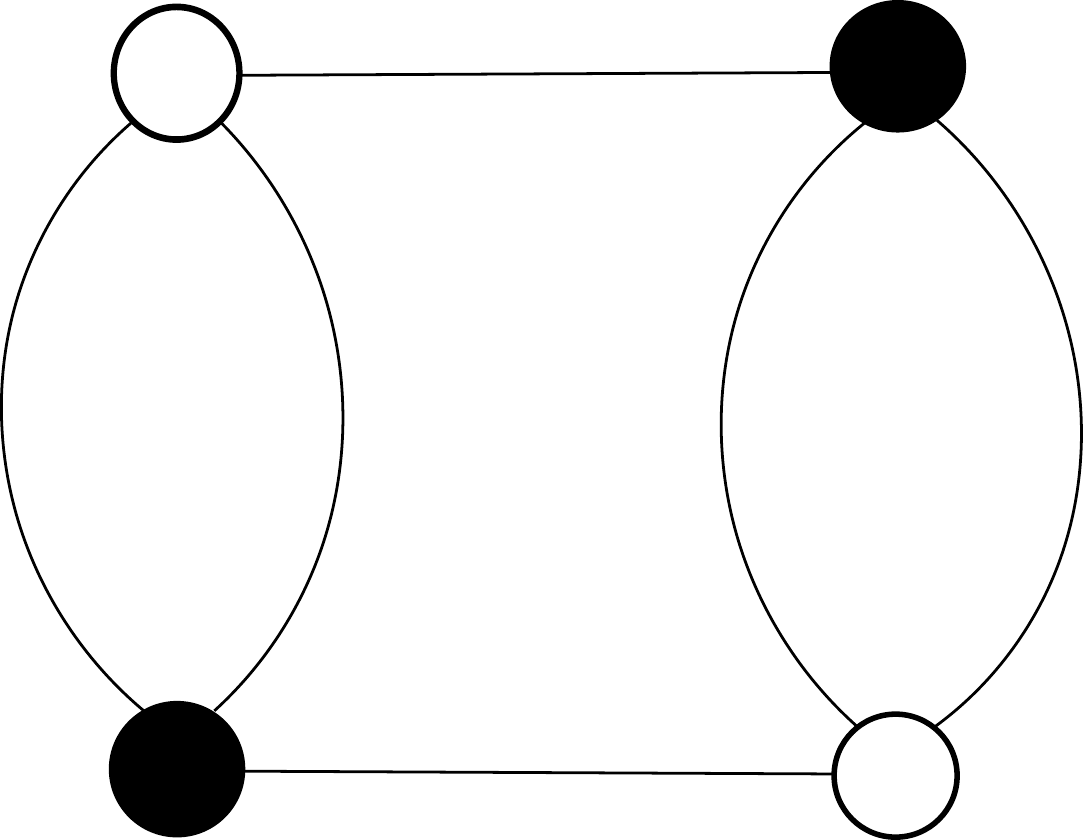}}
\big]\big \vert_{3 \; {\rm{cuts}}}
\nonumber \\
&&
+ 
{1 \over N}
\big[ 
6
\; \lambda_{\includegraphics[width=1.0cm]{eq62point5}}
+
6
\; \lambda_{\includegraphics[width=1.0cm]{eq62point6}}
+
3
\; \lambda_{\includegraphics[width=0.9cm]{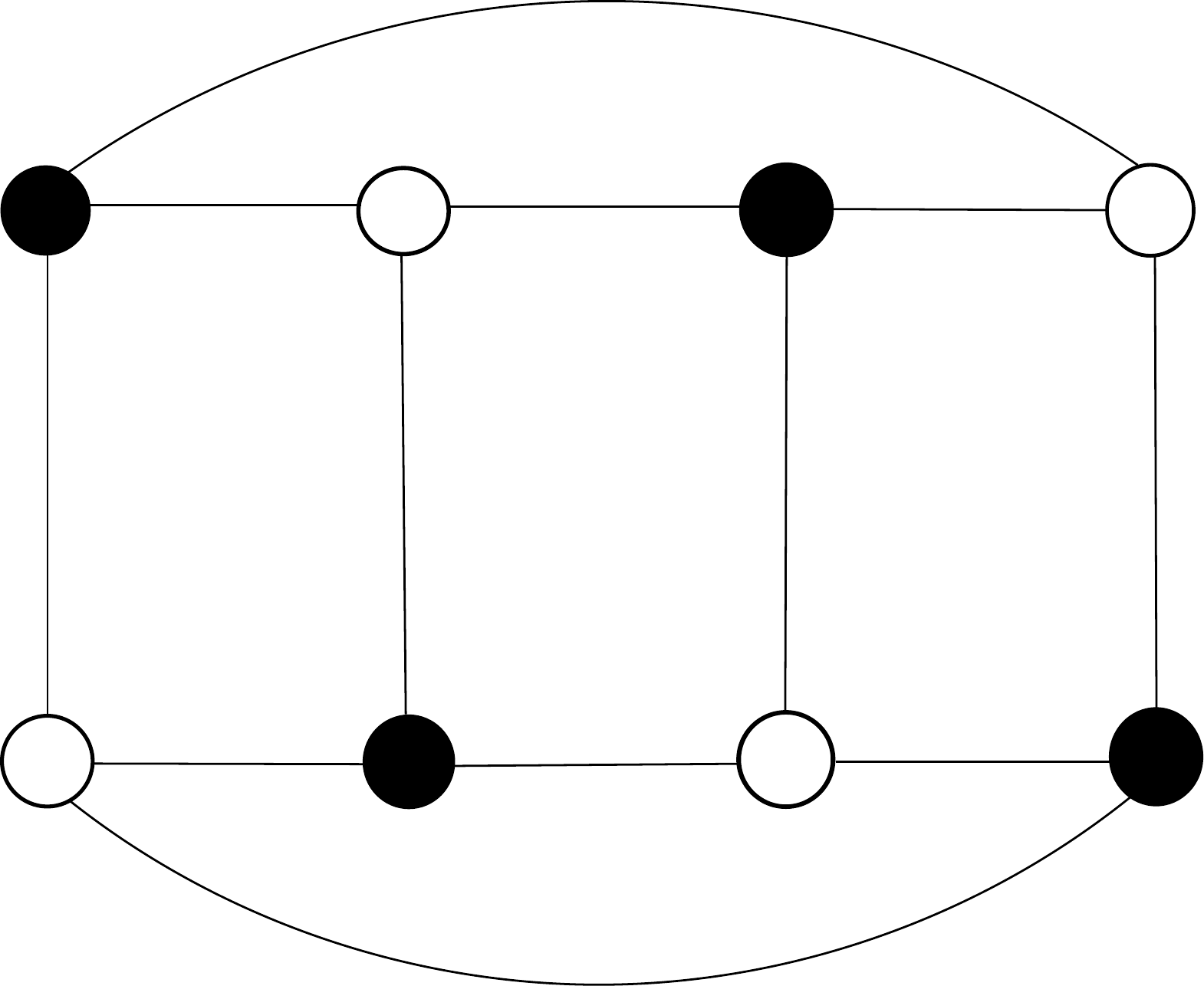}}
+
10
\; \lambda_{\includegraphics[width=1.0cm]{eq62point4}}
+
2
\; \lambda_{\includegraphics[width=0.8cm]{eq62point10}}
\big]\big \vert_{2 \; {\rm{cuts}}}
\nonumber \\
&&
+
{1 \over N^2}
\big[ 
6 \;
\; \lambda_{\includegraphics[width=1.0cm]{eq62point4}}
+
\; \lambda_{\includegraphics[width=0.9cm]{eq62point7}}
+
8
\; \lambda_{\includegraphics[width=1.0cm]{eq62point5}}
\big]\big \vert_{3 \; {\rm{cuts}}}
\nonumber \\
&&
+
{1 \over N^3}
\big[ 
6\;
\lambda_{\includegraphics[width=1.6cm]{eq62point2}}
+
2
\;\lambda_{\includegraphics[width=1.4cm]{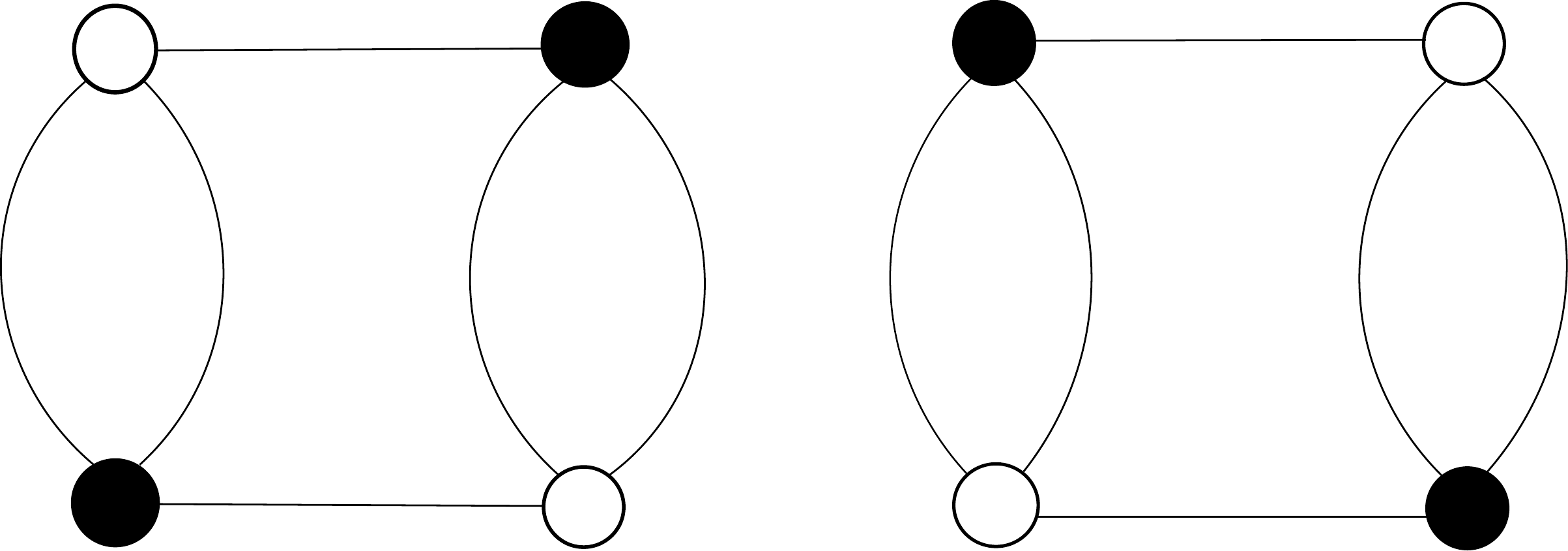}}
\big]\big \vert_{3 \; {\rm{cuts}}}
\,.
\nonumber \\
\eea

\bea
{\partial \over \partial t} 
\lambda_{{\includegraphics[width=0.7cm]{eq2point1}} \; {\includegraphics[width=0.7cm]{eq2point1}}\; {\includegraphics[width=0.7cm]{eq2point1}}}
\!\! \! \!&=& \!\!\!\!
\big[ 
\lambda_{{\includegraphics[width=0.7cm]{eq2point1}} \; {\includegraphics[width=0.7cm]{eq2point1}}\; {\includegraphics[width=0.7cm]{eq2point1}}\; {\includegraphics[width=0.7cm]{eq2point1}}}
\big]\! \big \vert_{0 \; {\rm{cut}}}
\! + \!\!
\big[
9 \;
\lambda_{\includegraphics[width=2.1cm]{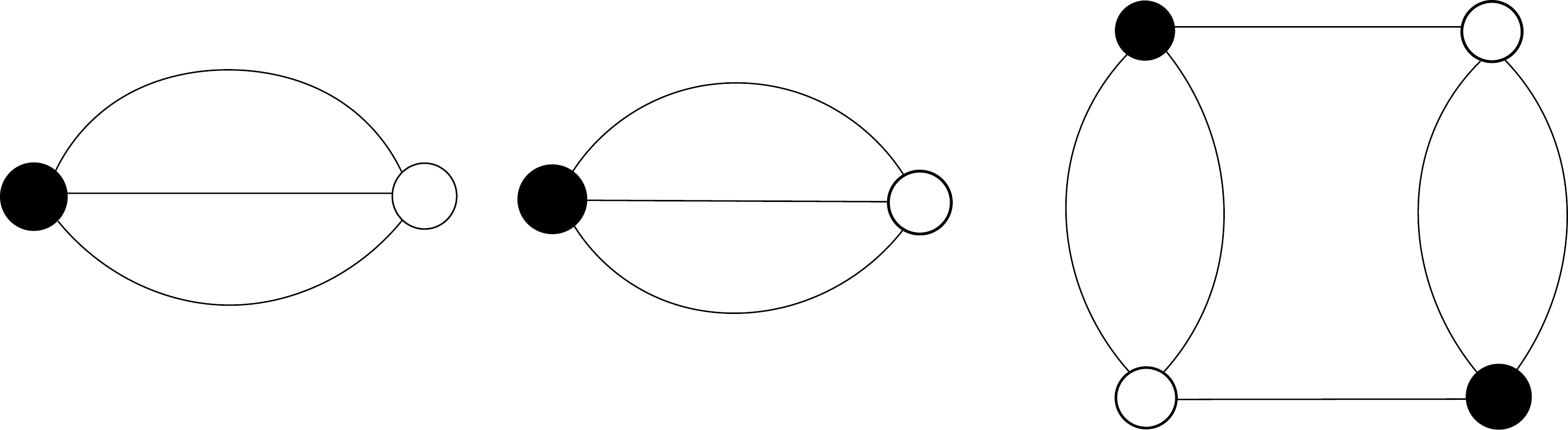}}
\big]\!\big \vert_{1 \; {\rm{cut}}}
\!+\!\!
\big[
18 \;
\lambda_{\includegraphics[width=1.6cm]{eq62point2}}
\big]\!\big \vert_{2 \; {\rm{cuts}}}
\nonumber \\
&&
+
\big[
6 \;
\lambda_{\includegraphics[width=0.7cm]{eq61point9}}
+
6 \;
\lambda_{{\includegraphics[width=0.7cm]{eq2point1}} \; {\includegraphics[width=0.7cm]{eq2point1}}\; {\includegraphics[width=0.7cm]{eq2point1}}}
\lambda_{{\includegraphics[width=0.7cm]{eq2point1}}}
+
6 \;
\lambda^2_{{\includegraphics[width=0.7cm]{eq2point1}} \; {\includegraphics[width=0.7cm]{eq2point1}}}
\big]\big \vert_{3 \; {\rm{cuts}}}
\nonumber \\
&&
+
{1 \over N}
\Big \{
\big[
9 \;
\lambda_{\includegraphics[width=2.1cm]{eq4point22}}
\big]\big \vert_{2 \; {\rm{cuts}}}
+
\big[
18 \;
\lambda_{\includegraphics[width=1.4cm]{eq61point2}}
\big]\big \vert_{3 \; {\rm{cuts}}}
\Big \}
\nonumber \\
&&
+
{1 \over N^3}
\big[
3\;
\lambda_{{\includegraphics[width=0.7cm]{eq2point1}} \; {\includegraphics[width=0.7cm]{eq2point1}}\; {\includegraphics[width=0.7cm]{eq2point1}}\; {\includegraphics[width=0.7cm]{eq2point1}}}
\big]\big \vert_{3 \; {\rm{cuts}}}
\,.
\eea

\bea
\nonumber \\ \cr
{\partial \over \partial t} \lambda_{\includegraphics[width=1.4cm]{eq4point2}}
\!\!\!&=&\!\!\!
\big[
\lambda_{\includegraphics[width=2.1cm]{eq4point22}}
\big]\big \vert_{0 \; {\rm{cut}}}
+ 
\big[
3 \;
\lambda_{\includegraphics[width=1.4cm]{eq61point10}}
+
4 \;
\lambda_{\includegraphics[width=1.4cm]{eq62point2}} 
+
2 \; 
\lambda_{\includegraphics[width=1.4cm]{eq61point2}} 
\big]\big \vert_{1 \; {\rm{cut}}}
\nonumber \\
&&
\!\!\!\!\!+
\big[
8 \;
\lambda_{\includegraphics[width=1.0cm]{eq61point8}}
+
4 \;
\lambda_{\includegraphics[width=1.0cm]{eq61point4}} 
\big] \!\big \vert_{2 \; {\rm{cuts}}}
\!+\!
\big[
6\;
\lambda_{\includegraphics[width=1.4cm]{eq4point2}} \;
\lambda_{\includegraphics[width=0.7cm]{eq2point1}}
+
4
\lambda_{{\includegraphics[width=0.7cm]{eq2point1}} \; {\includegraphics[width=0.7cm]{eq2point1}}} \;
\lambda_{\includegraphics[width=0.7cm]{eq4point1}}
\big]\!\big \vert_{3 \; {\rm{cuts}}} 
\nonumber \\
&&
\!\!\!\!\!+
{1 \over N} 
\Big \{
\big[ 
3 \; 
\lambda_{\includegraphics[width=1.4cm]{eq61point10}}
+
8 \; 
\lambda_{\includegraphics[width=1.4cm]{eq62point2}} 
+
2 \; 
\lambda_{\includegraphics[width=1.4cm]{eq61point2}} 
+
2 \; 
\lambda_{\includegraphics[width=1.4cm]{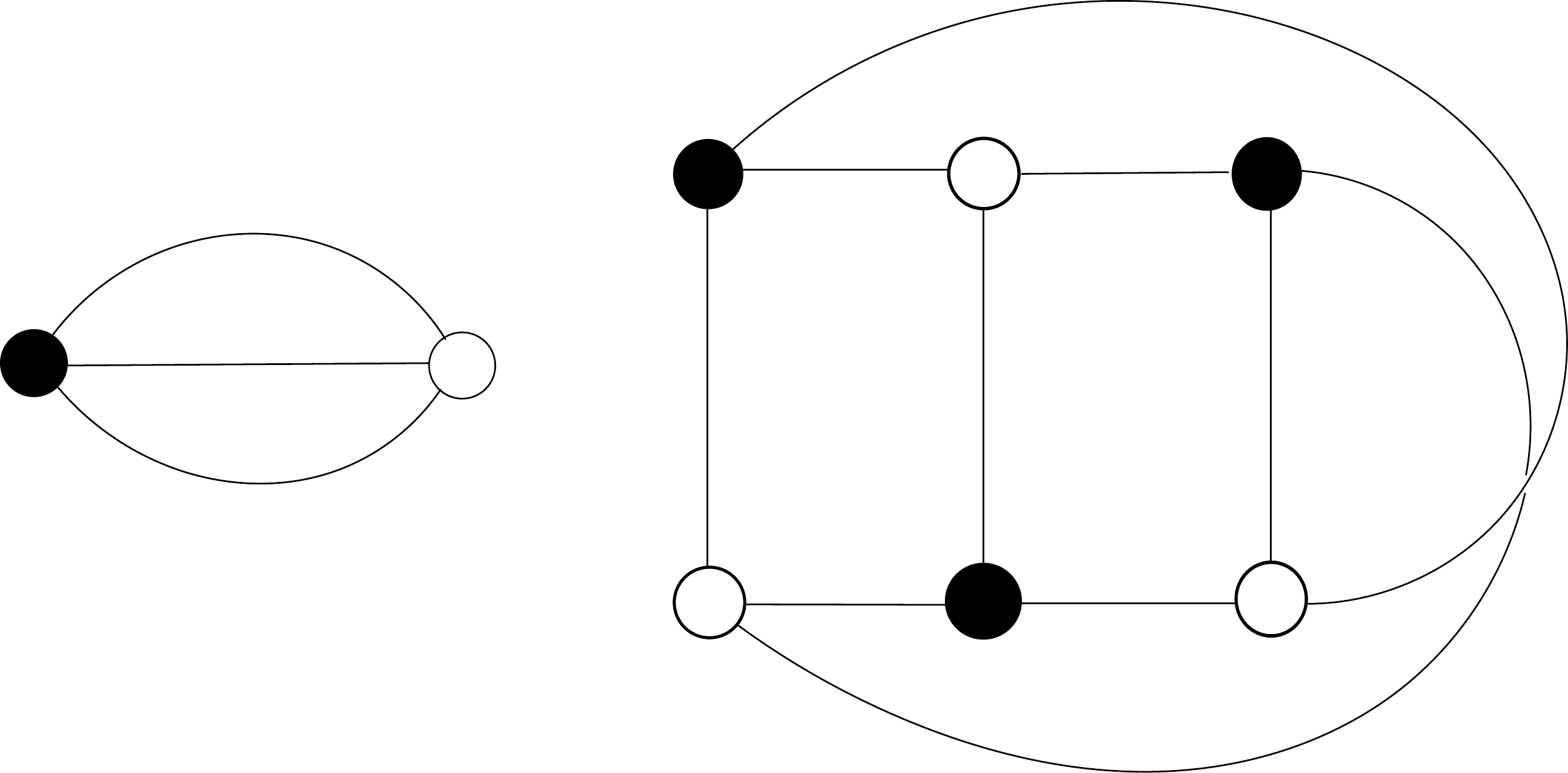}}
\big]\big \vert_{2 \; {\rm{cuts}}} 
\nonumber \\
&&
\!\!\!\!\!+
\big[ 
2 \; 
\; \lambda_{\includegraphics[width=1.0cm]{eq62point6}}
+
8 \; 
\lambda_{\includegraphics[width=0.7cm]{eq61point6}}
+
14 \; 
\lambda_{\includegraphics[width=1.0cm]{eq61point4}} 
\big]\big \vert_{3 \; {\rm{cuts}}} 
\Big \}
\nonumber \\
&&
\!\!\!\!\!+
{1 \over N^2}
\big[
4
\lambda_{\includegraphics[width=1.4cm]{eq62point2}} 
\big]\big \vert_{3 \; {\rm{cuts}}} 
+
{1 \over N^3}
\big[
5
\lambda_{\includegraphics[width=2.1cm]{eq4point22}}
\big]\big \vert_{3 \; {\rm{cuts}}} 
\,,
\label{eq:D3deqLast}
\eea

\subsection{Couplings in rank $D=4$ tensors models}

\label{example4}

\bea
{\partial \over \partial t}  \lambda_{\includegraphics[width=0.8cm]{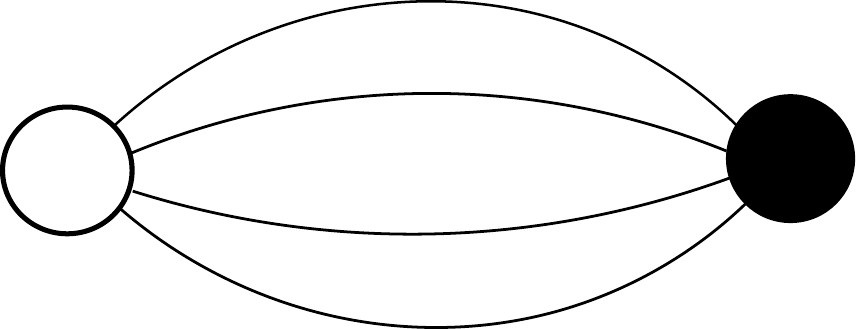}}
&=&
\big[
\lambda_{{\includegraphics[width=0.8cm]{r4eq2point1}} \; {\includegraphics[width=0.8cm]{r4eq2point1}}}
\big]\big \vert_{0 \; {\rm{cut}}}
+
\big[
4 \;
\lambda_{\includegraphics[width=0.8cm]{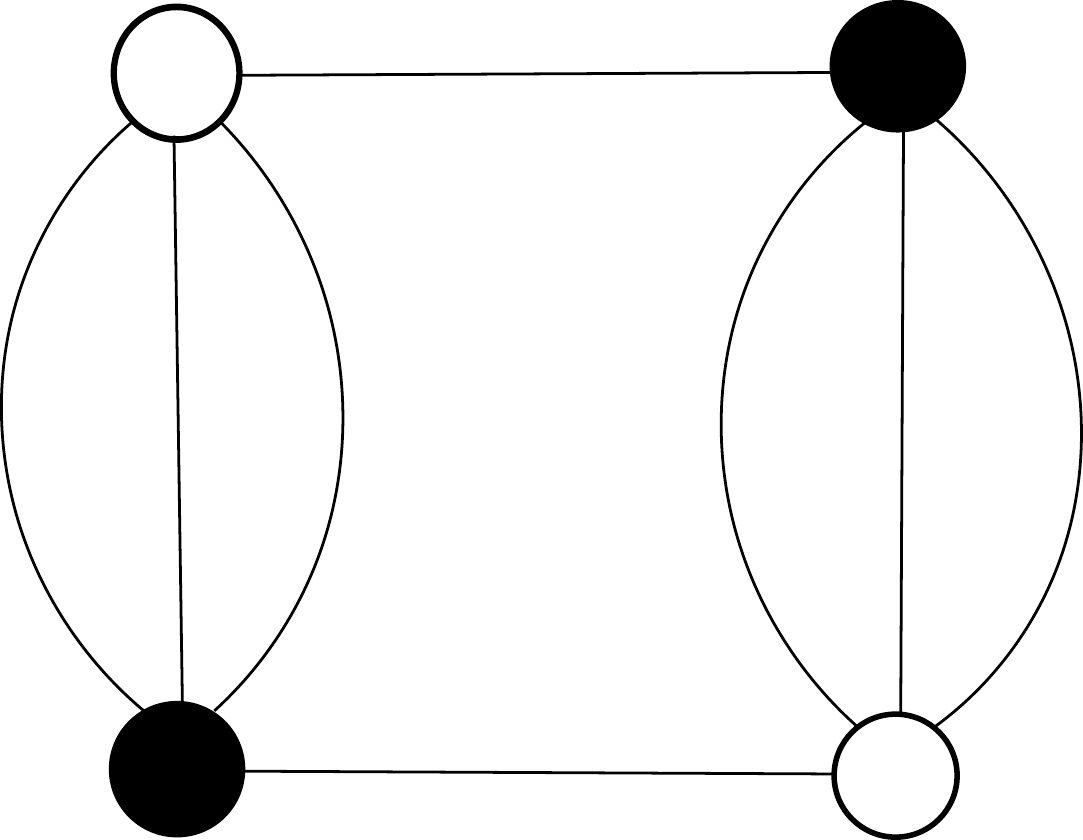}}
\big]\big \vert_{1 \; {\rm{cut}}}
+
\big[
\lambda^2_{\includegraphics[width=0.8cm]{r4eq2point1}}
\big]\big \vert_{3 \; {\rm{cuts}}}
\nonumber \\
&&
+
{1 \over N}  
\big[
6 \;
\lambda_{\includegraphics[width=0.8cm]{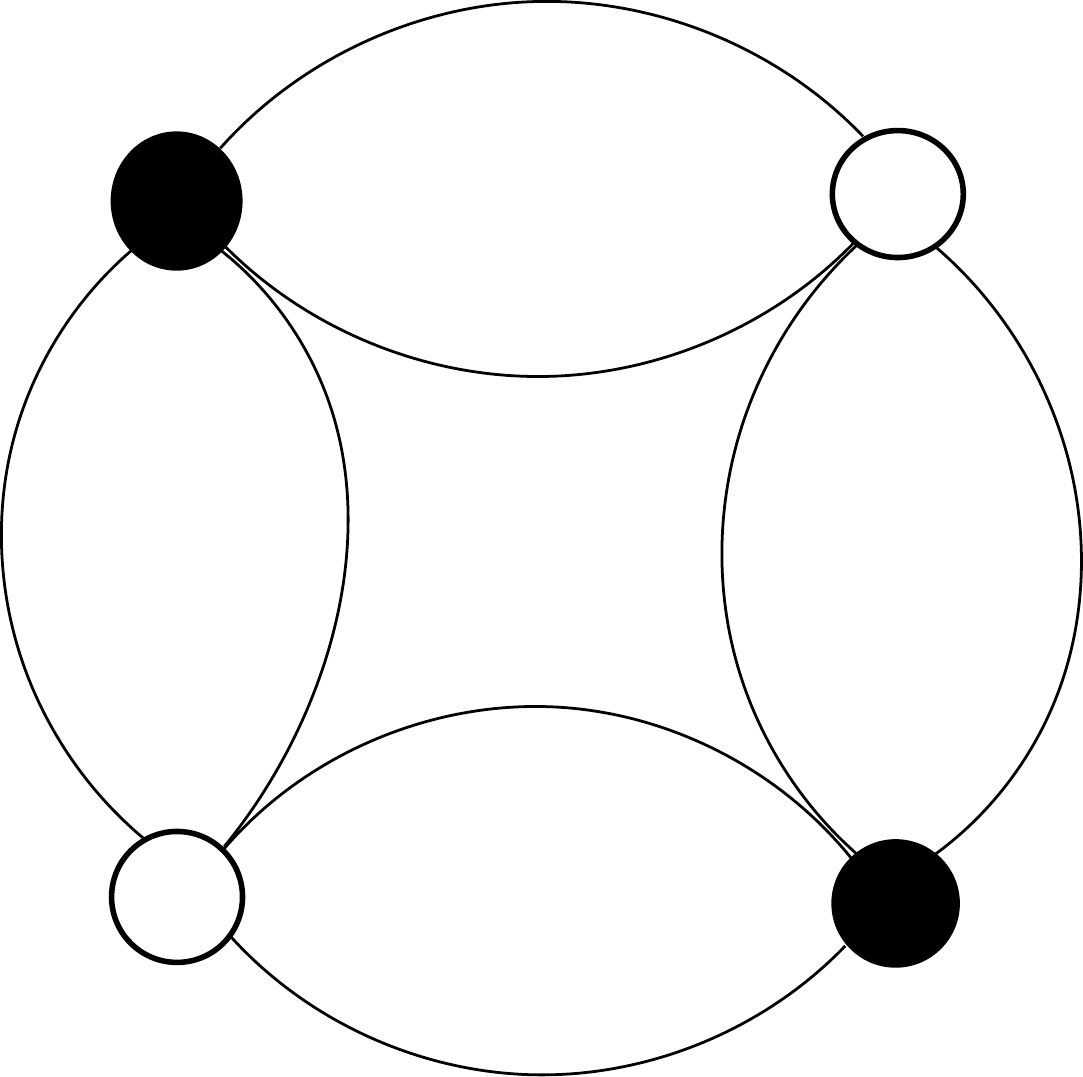}}
\big]\big \vert_{2 \; {\rm{cuts}}}
+
{1 \over N^2}
\big[
4 \;
\lambda_{\includegraphics[width=0.8cm]{r4eq2point2}}
\big]\big \vert_{3 \; {\rm{cuts}}}
+
{1 \over N^4}
\big[
\lambda_{{\includegraphics[width=0.8cm]{r4eq2point1}} \; {\includegraphics[width=0.8cm]{r4eq2point1}}}
\big]\big \vert_{4 \; {\rm{cuts}}}
\,.
\eea

\bea
\nonumber \\ \cr
{\partial \over \partial t}  \lambda_{\includegraphics[width=0.8cm]{r4eq2point2}}
&=&
\big[
\lambda_{\includegraphics[width=1.6cm]{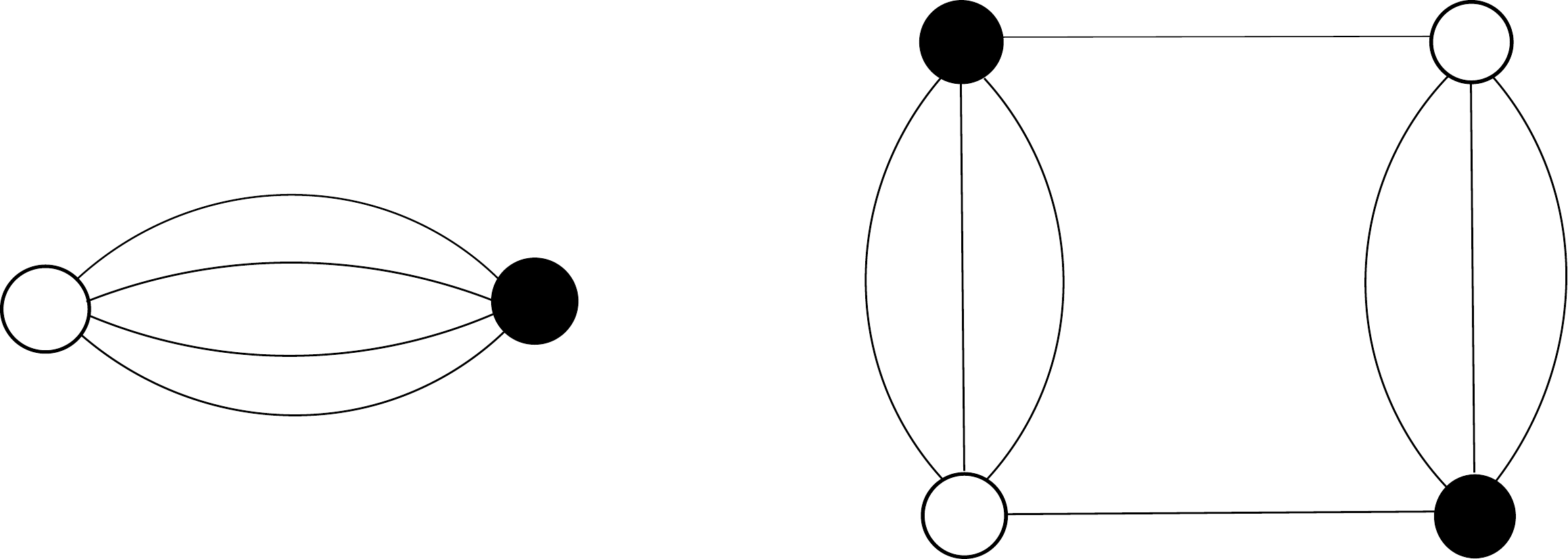}}
\big]\big \vert_{0 \; {\rm{cut}}}
+
\big[
2 \;
\lambda_{\includegraphics[width=0.8cm]{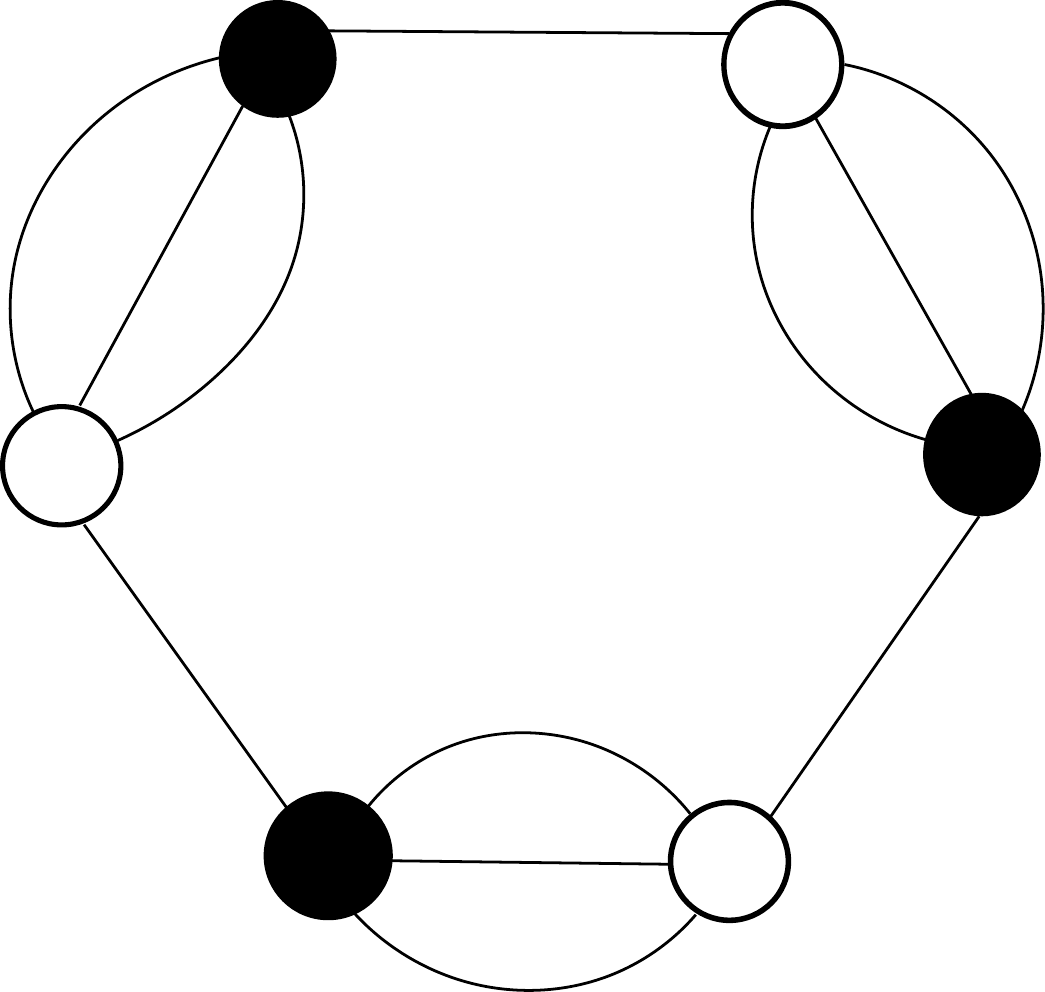}}
\big]\big \vert_{1 \; {\rm{cut}}}
+
\big[
4 \;
\lambda_{\includegraphics[width=0.8cm]{r4eq2point1}}
\lambda_{\includegraphics[width=0.8cm]{r4eq2point2}}
\big]\big \vert_{4 \; {\rm{cuts}}}
\nonumber \\
&&
+
{1 \over N}
\Big \{
\big[
6 \;
\lambda_{\includegraphics[width=0.8cm]{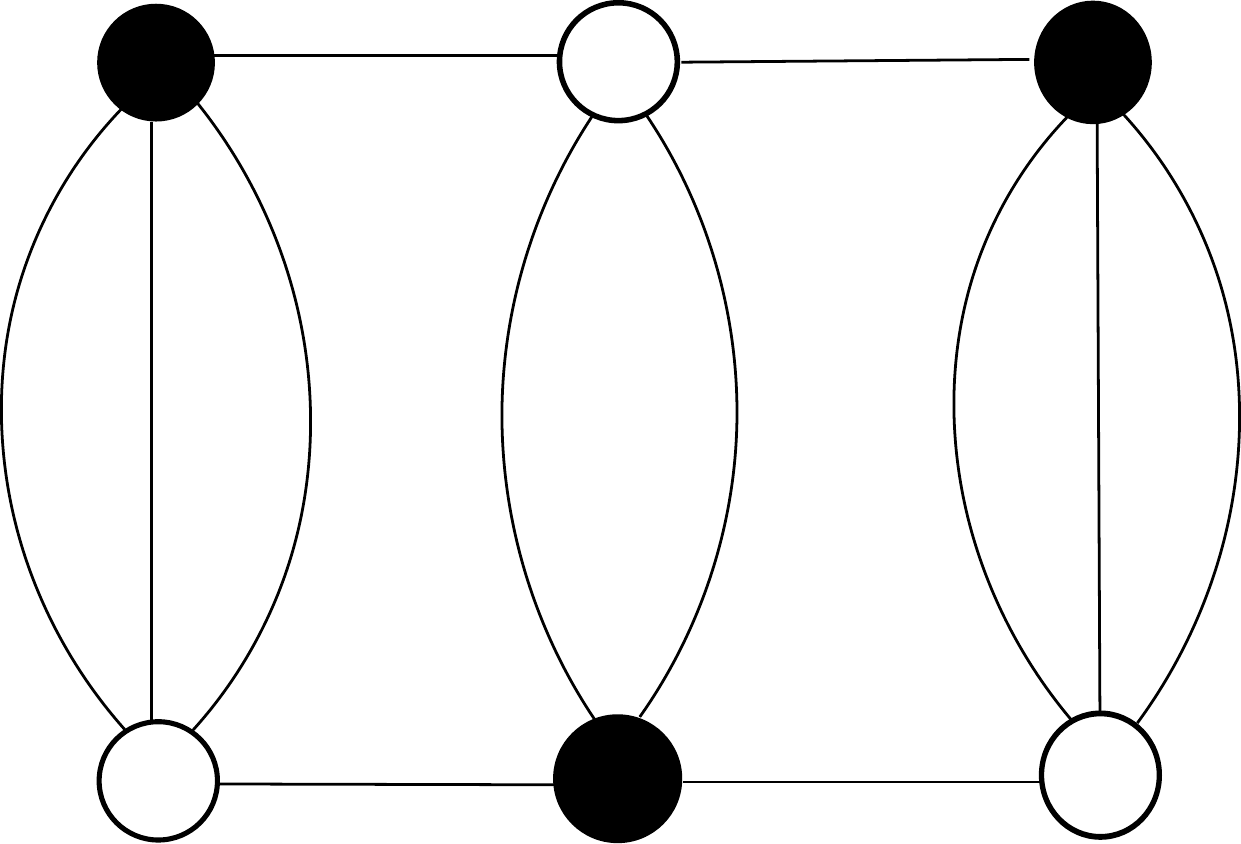}}
\big]\big \vert_{1 \; {\rm{cut}}}
+
\big[
18 \;
\lambda_{\includegraphics[width=0.8cm]{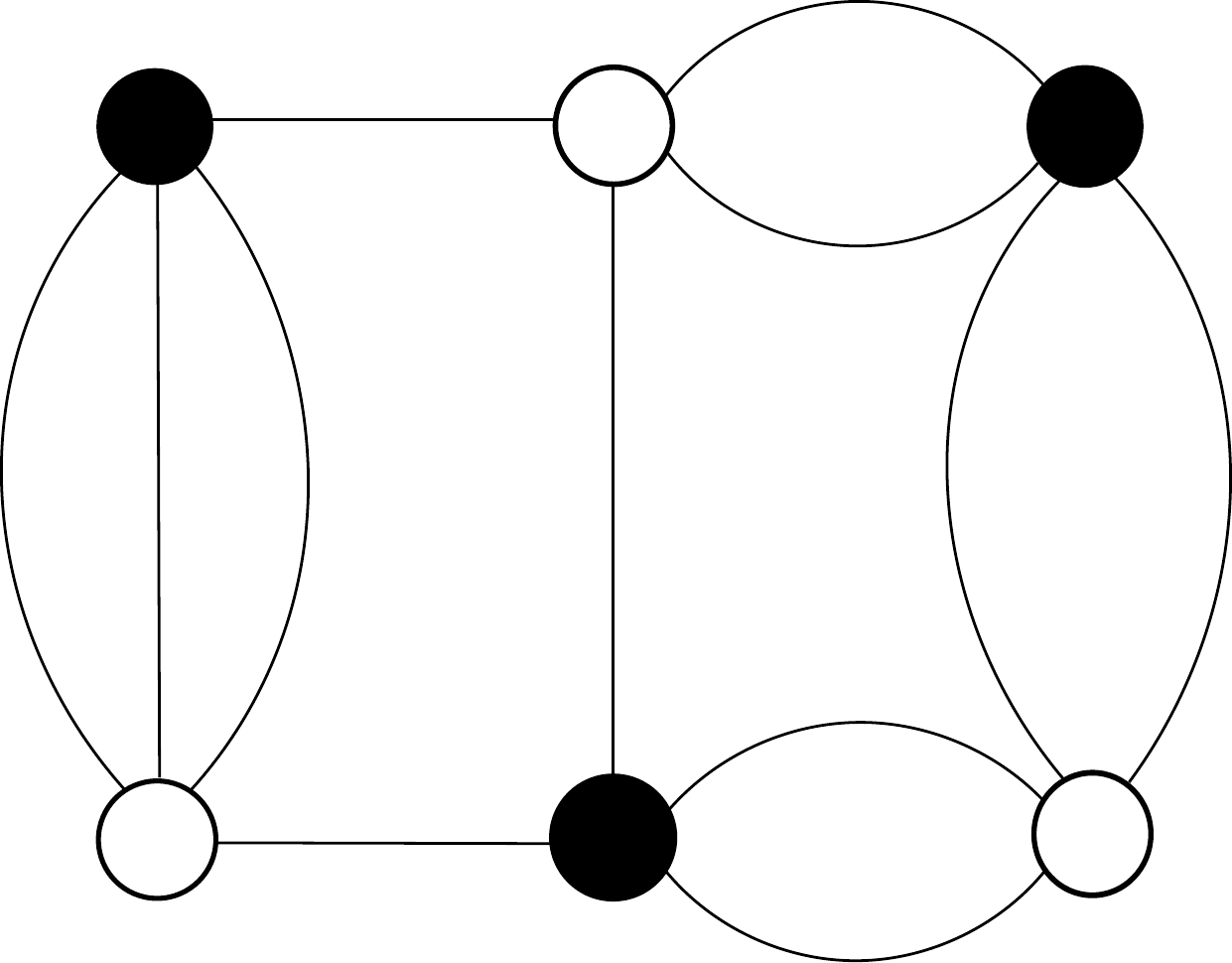}}
+
6 \;
\lambda_{\includegraphics[width=0.8cm]{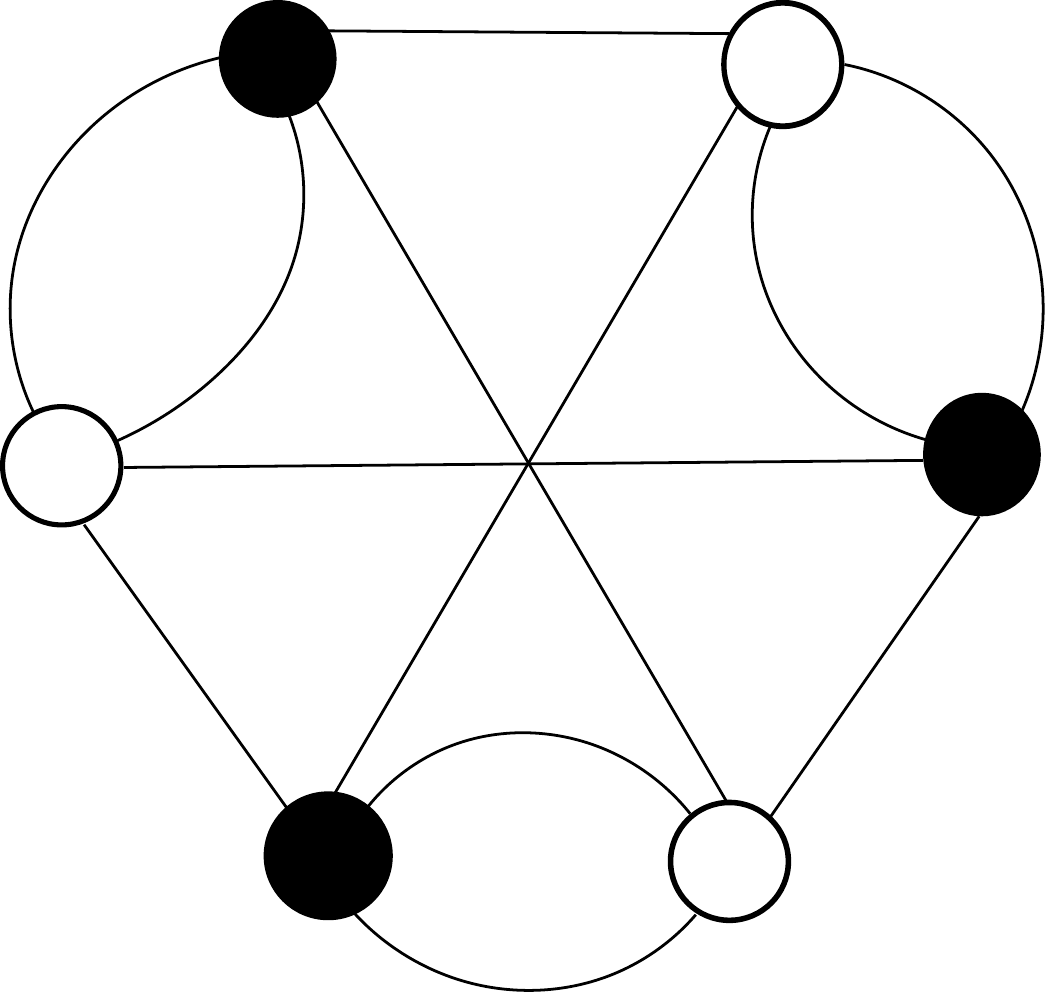}}
\big]\big \vert_{2 \; {\rm{cuts}}}
\Big \}
\nonumber \\
&&
+
{1 \over N^2}
\big[
2 \;
\lambda_{\includegraphics[width=0.8cm]{r4eq61point1}}
+
12 \;
\lambda_{\includegraphics[width=0.8cm]{r4eq62point1}}
+
12 \;
\lambda_{\includegraphics[width=0.8cm]{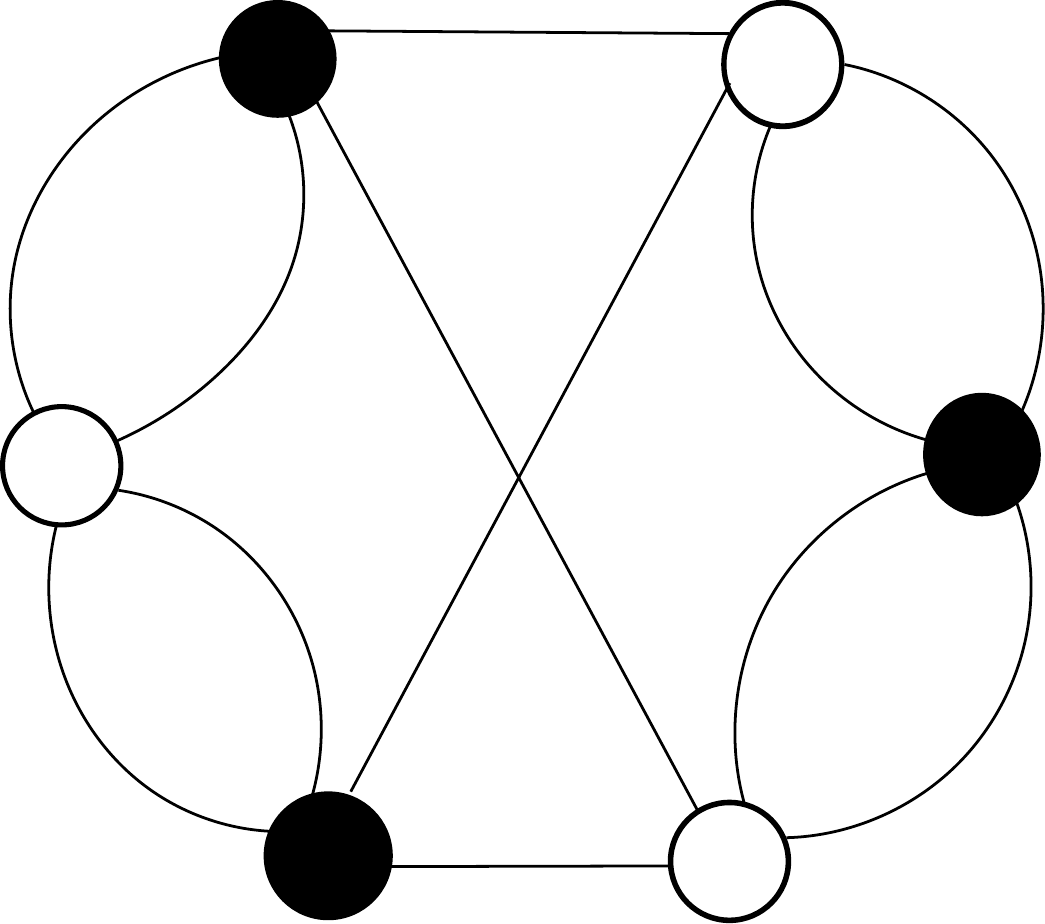}}
+
6 \;
\lambda_{\includegraphics[width=0.8cm]{r4eq63point1}}
\big]\big \vert_{3 \; {\rm{cuts}}}
\nonumber \\
&&
+
{1 \over N^3}
\big[
12 \;
\lambda_{\includegraphics[width=0.8cm]{r4eq66point1}}
\big]\big \vert_{4 \; {\rm{cuts}}}
+
{1 \over N^4}
\big[
4
\;\lambda_{\includegraphics[width=1.6cm]{r4eq4point2}}
\big]\big \vert_{4 \; {\rm{cuts}}}
\,.
\eea

\bea
\nonumber \\ \cr
{\partial \over \partial t}  \lambda_{\includegraphics[width=0.8cm]{r4eq4point1}}
&=&
\big[
\lambda_{\includegraphics[width=1.6cm]{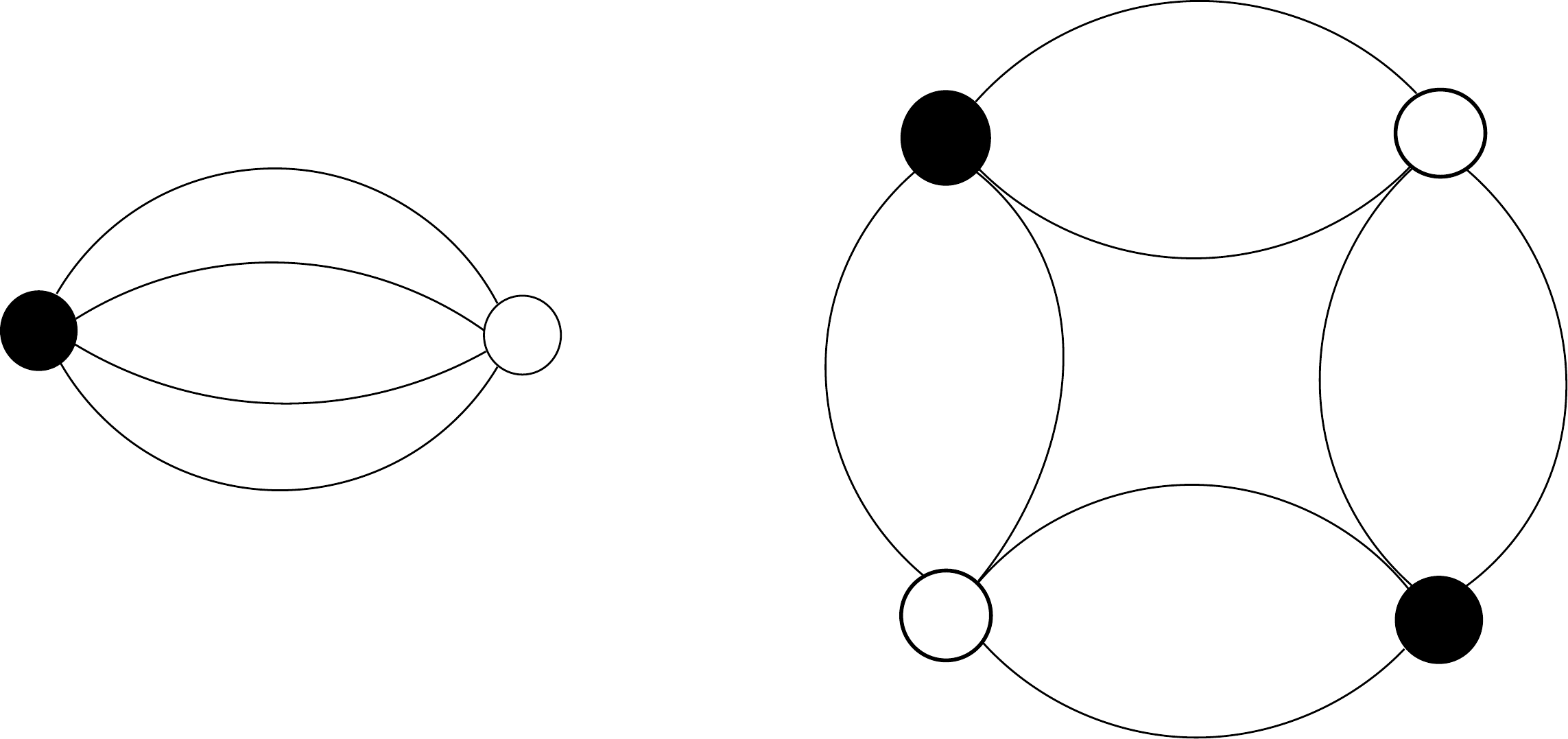}}
\big]\big \vert_{0 \; {\rm{cut}}}
+
\big[
8 \;
\lambda_{\includegraphics[width=0.8cm]{r4eq66point1}}
\big]\big \vert_{1 \; {\rm{cut}}}
+
\big[
4 \;
\lambda_{\includegraphics[width=0.8cm]{r4eq2point1}}
\lambda_{\includegraphics[width=0.8cm]{r4eq4point1}}
\big]\big \vert_{4 \; {\rm{cuts}}}
\nonumber \\
&&
+
{1 \over N}
\big[
4 \;
\lambda_{\includegraphics[width=0.8cm]{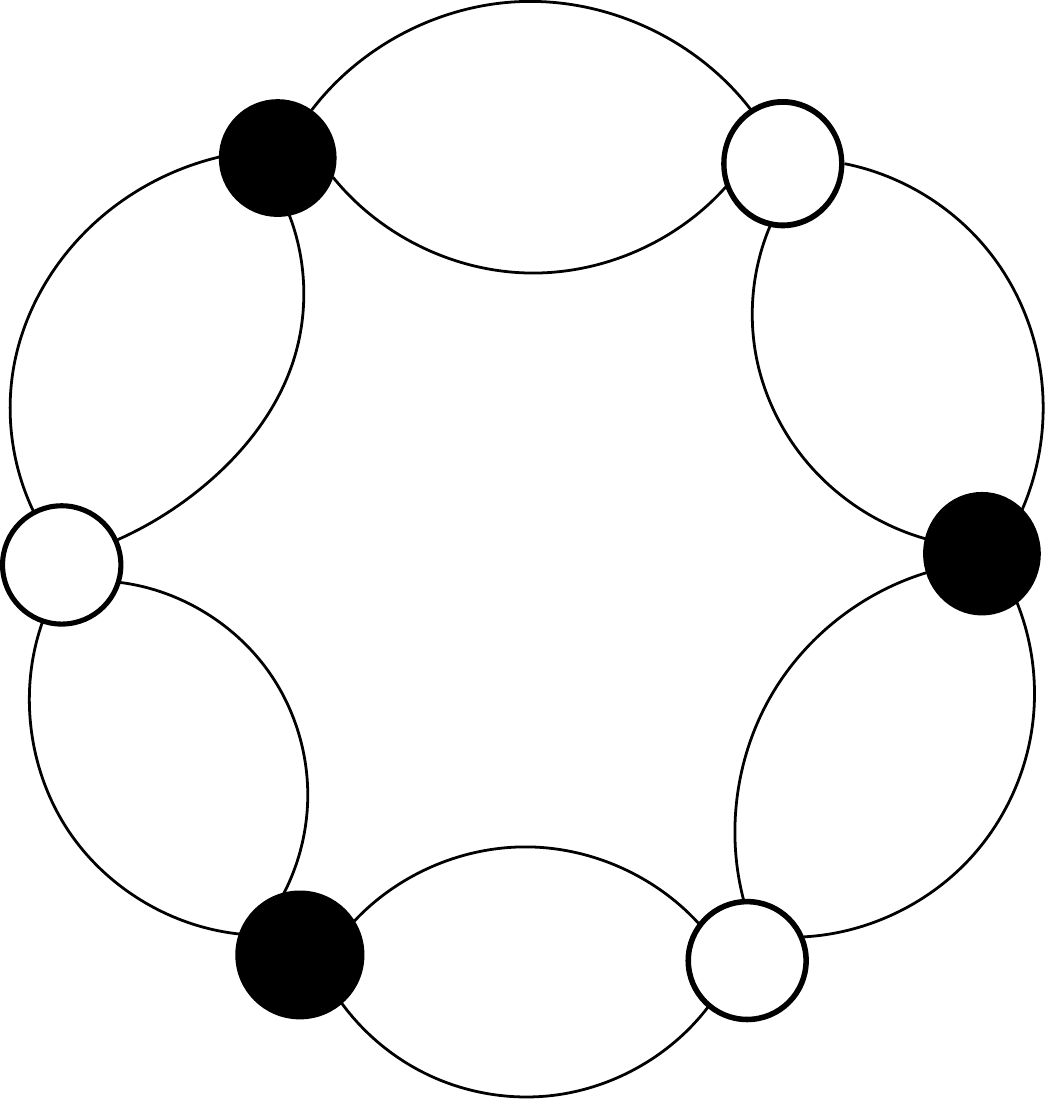}}
+
16 \;
\lambda_{\includegraphics[width=0.8cm]{r4eq64point1}}
+
4 \;
\lambda_{\includegraphics[width=0.8cm]{r4eq63point1}}
\big]\big \vert_{2 \; {\rm{cuts}}}
+
{1 \over N^2}
\big[
16 \;
\lambda_{\includegraphics[width=0.8cm]{r4eq66point1}}
+
16 \;
\lambda_{\includegraphics[width=0.8cm]{r4eq63point1}}
\big]\big \vert_{3 \; {\rm{cuts}}}
\nonumber \\
&&
+
{1 \over N^3}
\big[
8 \;
\lambda_{\includegraphics[width=0.8cm]{r4eq62point1}}
+
4\;
\lambda_{\includegraphics[width=0.8cm]{r4eq64point1}}
\big]\big \vert_{4 \; {\rm{cuts}}}
+
{1 \over N^4}
\big[
4
\;\lambda_{\includegraphics[width=1.6cm]{r4eq4point11}}
\big]\big \vert_{4 \; {\rm{cuts}}}
\,.
\eea

\bea
\nonumber \\ \cr
{\partial \over \partial t}  \lambda_{{\includegraphics[width=0.8cm]{r4eq2point1}} \; {\includegraphics[width=0.8cm]{r4eq2point1}}}
&=&
\big[
\lambda_{{\includegraphics[width=0.8cm]{r4eq2point1}} \; {\includegraphics[width=0.8cm]{r4eq2point1}}\; {\includegraphics[width=0.8cm]{r4eq2point1}}}
\big]\big \vert_{0 \; {\rm{cut}}}
+
\big[
8 \;
\lambda_{\includegraphics[width=1.6cm]{r4eq4point2}}
\big]\big \vert_{1 \; {\rm{cut}}}
+
\big[
12 \;
\lambda_{\includegraphics[width=0.8cm]{r4eq62point1}}
\big]\big \vert_{2 \; {\rm{cuts}}}
\nonumber \\
&&
+
\big[
4 \;
\lambda_{{\includegraphics[width=0.8cm]{r4eq2point1}} \; {\includegraphics[width=0.8cm]{r4eq2point1}}}
\lambda_{\includegraphics[width=0.8cm]{r4eq2point1}} 
\big]\big \vert_{4 \; {\rm{cuts}}}
\nonumber \\
&&
+
{1 \over N} 
\Big \{
\big[
12 \;
\lambda_{\includegraphics[width=1.6cm]{r4eq4point11}}
\big]\big \vert_{2 \; {\rm{cuts}}}
+
\big[
24 \;
\lambda_{\includegraphics[width=0.8cm]{r4eq66point1}}
\big]\big \vert_{3 \; {\rm{cuts}}}
\Big \}
\nonumber \\
&&
+
{1 \over N^2} 
\Big \{
\big[
8\;
\lambda_{\includegraphics[width=1.6cm]{r4eq4point2}}
\big]\big \vert_{3 \; {\rm{cuts}}}
+
\big[
8\;
\lambda_{\includegraphics[width=0.8cm]{r4eq61point1}}
+
6\;
\lambda_{\includegraphics[width=0.8cm]{r4eq65point1}}
\big]\big \vert_{4 \; {\rm{cuts}}}
\Big \}
\nonumber \\
&&
+
{1 \over N^4} 
\big[
2\;
\lambda_{{\includegraphics[width=0.8cm]{r4eq2point1}} \; {\includegraphics[width=0.8cm]{r4eq2point1}}\; {\includegraphics[width=0.8cm]{r4eq2point1}}}
\big]\big \vert_{4 \; {\rm{cuts}}}
\,.
\eea

\subsection{Solutions of the Polchinski equations in large $N$ limit}
\label{}

For a concrete illustration, we consider a specific model  given earlier in the  Section \ref{sec:largeN} with a diagonal covariance $C=\frac{t}{N^{D-1}}\delta_{I,\overline{I}}$, where $t$ is a parameter to control the flow. When $t=0$. $C^{-1}=0$ so that there is no integration at all and the  effective action equals the bare action.

 The partition function is the effective action  (given by \eqref{eq:actiont}) evaluated at 0 field.
In addition to the set of differential equations for couplings given in \eqref{eq:D3deqStart}-\eqref{eq:D3deqLast}, we have
\beq
{\partial \over \partial t} \lambda_{\emptyset} 
=
\lambda_{\includegraphics[width=0.5cm]{eq2point1}}\,,
\label{eq:D3deq0}
\eeq
where $\lambda_{\emptyset}$ is the coupling for an empty graph denoted by $\emptyset$, on which one can only perform $0$-cut, therefore obtaining the dipole graph on the {\it{r.h.s.}}.
Using these differential equations,
one can readily solve this set of differential equation iteratively with an ansatz for the partition function as it can be written as the sum of all possible vacuum diagrams,
\beq
\lambda_{\emptyset}(t) =  \lambda_{\emptyset}(0) + \sum_m a_{{\emptyset}_m}\, t^m\,,
\eeq
where $t$ is the momentum associated with the propagator (in other words, $0$ color line), and $a_m$ is the coupling associated with the $2 m$-point function.
Here, for simplicity, we will not distinguish the colors.

We remark here that the leading contribution in the large $N$ limit in the Polchinski equation selects only melonic graphs with the normalization given in \eqref{eq:actiont}.
However, once we start including the terms of ${\mathcal {O}}({1 \over N})$, nonplanar terms also appear in the evolution equation.

We rewrite the ansatze for each couplings, 
\beq
\lambda_{\emptyset}(t) =\lambda_{\emptyset}(0) +  \sum_m a_{{\emptyset}_m} \, t^m  + {\mathcal O}(t^{m+1}) \,,
\eeq
\beq
\lambda_{\includegraphics[width=0.7cm]{eq2point1}}(t) 
=\lambda_{\includegraphics[width=0.7cm]{eq2point1}}(0) 
+  \sum_{m-1} a_{{\includegraphics[width=0.7cm]{eq2point1}}_{m-1}} \, t^{m-1}  + {\mathcal O}(t^{m}) \,,
\eeq

\beq
\lambda_{{\includegraphics[width=0.7cm]{eq2point1}} \; {\includegraphics[width=0.7cm]{eq2point1}}}
(t) 
=\lambda_{{\includegraphics[width=0.7cm]{eq2point1}} \; {\includegraphics[width=0.7cm]{eq2point1}}}(0) 
+  \sum_{m-2} a_{{\includegraphics[width=0.7cm]{eq2point1}} \; {\includegraphics[width=0.7cm]{eq2point1}}_{m-2}} \, t^{m-2}  + {\mathcal O}(t^{m-1}) \,,
\eeq

\beq
\lambda_{\includegraphics[width=0.7cm]{eq4point1}}
(t) 
=\lambda_{\includegraphics[width=0.7cm]{eq4point1}}(0) 
+  \sum_{m-2} a_{{\includegraphics[width=0.7cm]{eq4point1}}_{m-2}} \, t^{m-2}  + {\mathcal O}(t^{m-1}) \,,
\eeq
{\it{etc}}, where constants $a_m$'s are to be determined in terms of the couplings at $t =0$.
The above ansatze are consistent with the orders in $t$ for each couplings observing the set of differential equations given.

\subsubsection{The leading order in large $N$ limit}
We will first only solve the set of differential equations  \eqref{eq:D3deqStart}-\eqref{eq:D3deqLast} and \eqref{eq:D3deq0} in the leading order in large $N$ limit.
Starting with $m = 1$, we obtain:
\bea
\lambda_{\emptyset} (t) &=& \lambda_{\emptyset}(0)+ \lambda_{\includegraphics[width=0.5cm]{eq2point1}}(0) t + {\mathcal O} (t^2)\,,
\nonumber \\
\lambda_{\includegraphics[width=0.5cm]{eq2point1}}(t)  &=& \lambda_{\includegraphics[width=0.5cm]{eq2point1}}(0) + {\mathcal O} (t) \,.
\eea

We proceed to $m=2$ using the above result to obtain the higher order corrections in $t$;
\bea
\lambda_{\emptyset} (t) 
&=& 
\lambda_{\emptyset}(0)
+ 
\lambda_{\includegraphics[width=0.7cm]{eq2point1}}(0) t 
+ 
{1 \over 2} 
\Big( \lambda_{{\includegraphics[width=0.7cm]{eq2point1}} \; {\includegraphics[width=0.7cm]{eq2point1}}}(0)
+ 
3
\lambda_{\includegraphics[width=0.7cm]{eq4point1}}(0)
+
\lambda^2_{\includegraphics[width=0.7cm]{eq2point1}}(0)\Big)
t^2
+
{\mathcal O} (t^3)
\nonumber \\
\lambda_{\includegraphics[width=0.7cm]{eq2point1}}(t)  
&=& 
\lambda_{\includegraphics[width=0.7cm]{eq2point1}}(0) 
+
\Big(
\lambda_{{\includegraphics[width=0.7cm]{eq2point1}} \; {\includegraphics[width=0.7cm]{eq2point1}}}(0)
+
3
\lambda_{\includegraphics[width=0.7cm]{eq4point1}}(0)
+
\lambda^2_{\includegraphics[width=0.7cm]{eq2point1}}(0)\Big)
t
+ 
{\mathcal O} (t^2) \,,
\nonumber \\
\lambda_{{\includegraphics[width=0.7cm]{eq2point1}} \; {\includegraphics[width=0.7cm]{eq2point1}}}
(t) 
&=&
\lambda_{{\includegraphics[width=0.7cm]{eq2point1}} \; {\includegraphics[width=0.7cm]{eq2point1}}}(0)
+ 
{\mathcal O} (t) \,, 
\nonumber \\
\lambda_{\includegraphics[width=0.7cm]{eq4point1}}
(t) 
&=&
\lambda_{\includegraphics[width=0.7cm]{eq4point1}}(0)
+ 
{\mathcal O} (t) \,.
\eea

Including up to $m =3$ in the coupling for vacuum, we obtain:
\bea
\lambda_{\emptyset} (t) 
&=& 
\lambda_{\emptyset}(0)
+ 
\lambda_{\includegraphics[width=0.7cm]{eq2point1}}(0) t 
+ 
{1 \over 2} 
\Big( \lambda_{{\includegraphics[width=0.7cm]{eq2point1}} \; {\includegraphics[width=0.7cm]{eq2point1}}}(0) 
+ 
3
\lambda_{\includegraphics[width=0.7cm]{eq4point1}}(0)
+
\lambda^2_{\includegraphics[width=0.7cm]{eq2point1}}(0)\Big)
t^2
\cr\cr
&&+
{1 \over 3} \Big(
{1 \over 2} \lambda_{{\includegraphics[width=0.7cm]{eq2point1}} \; {\includegraphics[width=0.7cm]{eq2point1}}\; {\includegraphics[width=0.7cm]{eq2point1}}}(0)
+ {9 \over 2} \lambda_{\includegraphics[width=1.4cm]{eq4point2}} (0)
+ 9 \lambda_{\includegraphics[width=0.7cm]{eq62point1}}(0)
+ 3 \lambda_{\includegraphics[width=0.7cm]{eq61point1}}(0)
\Big.
\cr\cr
\Big.
&&
+
3 \lambda_{{\includegraphics[width=0.7cm]{eq2point1}} \; {\includegraphics[width=0.7cm]{eq2point1}}}\lambda_{\includegraphics[width=0.7cm]{eq2point1}}(0)
+
9 \lambda_{\includegraphics[width=0.7cm]{eq4point1}}(0)
\lambda_{\includegraphics[width=0.7cm]{eq2point1}}(0)
+
\lambda^3_{\includegraphics[width=0.7cm]{eq2point1}}(0)
\Big)
t^3
+
{\mathcal O} (t^4)
\cr\cr
\lambda_{\includegraphics[width=0.7cm]{eq2point1}}(t)  
&=& 
\lambda_{\includegraphics[width=0.7cm]{eq2point1}}(0) 
+
\Big(
\lambda_{{\includegraphics[width=0.7cm]{eq2point1}} \; {\includegraphics[width=0.7cm]{eq2point1}}}(0)
+
3
\lambda_{\includegraphics[width=0.7cm]{eq4point1}}(0)
+
\lambda^2_{\includegraphics[width=0.7cm]{eq2point1}}(0)\Big)
t
\cr\cr
&&+
\Big(
{1 \over 2} \lambda_{{\includegraphics[width=0.7cm]{eq2point1}} \; {\includegraphics[width=0.7cm]{eq2point1}}\; {\includegraphics[width=0.7cm]{eq2point1}}}(0)
+ {9 \over 2} \lambda_{\includegraphics[width=1.4cm]{eq4point2}} (0)
+ 9 \lambda_{\includegraphics[width=0.7cm]{eq62point1}}(0)
+ 3 \lambda_{\includegraphics[width=0.7cm]{eq61point1}}(0)
\Big.
\cr\cr
\Big.
&&
+
3 \lambda_{{\includegraphics[width=0.7cm]{eq2point1}} \; {\includegraphics[width=0.7cm]{eq2point1}}}\lambda_{\includegraphics[width=0.7cm]{eq2point1}}(0)
+
9 \lambda_{\includegraphics[width=0.7cm]{eq4point1}}(0)
\lambda_{\includegraphics[width=0.7cm]{eq2point1}}(0)
+ \lambda^3_{\includegraphics[width=0.7cm]{eq2point1}}(0)
\Big)
t^2
+ 
{\mathcal O} (t^3) \,,
\cr\cr
\lambda_{{\includegraphics[width=0.7cm]{eq2point1}} \; {\includegraphics[width=0.7cm]{eq2point1}}}
(t) 
&=&
\lambda_{{\includegraphics[width=0.7cm]{eq2point1}} \; {\includegraphics[width=0.7cm]{eq2point1}}}(0)
\cr\cr
&&+
\Big(
\lambda_{{\includegraphics[width=0.7cm]{eq2point1}} \; {\includegraphics[width=0.7cm]{eq2point1}}\; {\includegraphics[width=0.7cm]{eq2point1}}}(0)
+6 \lambda_{\includegraphics[width=1.4cm]{eq4point2}} (0)
+6 \lambda_{\includegraphics[width=0.7cm]{eq62point1}}(0)
\Big.
\cr\cr
\Big.
&&+
4 \lambda_{{\includegraphics[width=0.7cm]{eq2point1}} \; {\includegraphics[width=0.7cm]{eq2point1}}}\lambda_{\includegraphics[width=0.7cm]{eq2point1}}(0)
\Big)
t 
+
{\mathcal O} (t^2) \,, 
\cr\cr
\lambda_{\includegraphics[width=0.7cm]{eq4point1}}
(t) 
&=&
\lambda_{\includegraphics[width=0.7cm]{eq4point1}}(0)
\cr\cr
&&+
\Big(
\lambda_{\includegraphics[width=1.4cm]{eq4point2}} (0)
+ 4 \lambda_{\includegraphics[width=0.7cm]{eq62point1}}(0)
+ 2 \lambda_{\includegraphics[width=0.7cm]{eq61point1}}(0)
+
4 \lambda_{\includegraphics[width=0.7cm]{eq4point1}}\lambda_{\includegraphics[width=0.7cm]{eq2point1}}(0)
\Big)
t
+ 
{\mathcal O} (t^2) \,, 
\cr\cr
\lambda_{{\includegraphics[width=0.7cm]{eq2point1}} \; {\includegraphics[width=0.7cm]{eq2point1}}\; {\includegraphics[width=0.7cm]{eq2point1}}}(t)
&=&
\lambda_{{\includegraphics[width=0.7cm]{eq2point1}} \; {\includegraphics[width=0.7cm]{eq2point1}}\; {\includegraphics[width=0.7cm]{eq2point1}}}(0)
+ 
{\mathcal O} (t) \,, 
\cr\cr
\lambda_{\includegraphics[width=0.7cm]{eq61point1}}(t)
&=&
\lambda_{\includegraphics[width=0.7cm]{eq61point1}}(0)
+ 
{\mathcal O} (t) \,, 
\nonumber \\
\lambda_{\includegraphics[width=0.7cm]{eq62point1}}(t)
&=&
\lambda_{\includegraphics[width=0.7cm]{eq62point1}}(0)
+ 
{\mathcal O} (t) \,, 
\nonumber \\
\lambda_{\includegraphics[width=1.4cm]{eq4point2}}(t)
&=&
\lambda_{\includegraphics[width=1.4cm]{eq4point2}}(0)
+ 
{\mathcal O} (t) \,.
\eea

\subsubsection{To ${\mathcal {O}}({1 \over N})$}

Performing a similar calculation as for the leading order in $N$, up to $m =3$ in the coupling for vacuum, we obtain:
\bea
\lambda_{\emptyset} (t) 
&=& 
\lambda_{\emptyset}(0)
+ 
\lambda_{\includegraphics[width=0.7cm]{eq2point1}}(0) t 
\cr\cr
&&
+ 
{1 \over 2} 
\Big [
 \lambda_{{\includegraphics[width=0.7cm]{eq2point1}} \; {\includegraphics[width=0.7cm]{eq2point1}}}(0) 
+ 
\left(3 + {1 \over N} 3\right)
\lambda_{\includegraphics[width=0.7cm]{eq4point1}}(0)
+
\lambda^2_{\includegraphics[width=0.7cm]{eq2point1}}(0)
\Big ]
t^2
\cr\cr
&&+
{1 \over 3} 
\Bigg [
{1 \over 2} \lambda_{{\includegraphics[width=0.7cm]{eq2point1}} \; {\includegraphics[width=0.7cm]{eq2point1}}\; {\includegraphics[width=0.7cm]{eq2point1}}}(0)
+ \left({9 \over 2} + {1 \over N} {9 \over 2}\right) \lambda_{\includegraphics[width=1.4cm]{eq4point2}} (0)
\Big.
\cr\cr
\Big.
&&
+ \left( 9 + {1 \over N}18\right) 
\lambda_{\includegraphics[width=0.7cm]{eq62point1}}(0)
+ \left( 3 + {1 \over N}9\right) 
 \lambda_{\includegraphics[width=0.7cm]{eq61point1}}(0)
+
3 \lambda_{{\includegraphics[width=0.7cm]{eq2point1}} \; {\includegraphics[width=0.7cm]{eq2point1}}}\lambda_{\includegraphics[width=0.7cm]{eq2point1}}(0)
\Big.
\cr\cr
\Big.
&&
+
\left( 9 + {1 \over N}9\right) 
 \lambda_{\includegraphics[width=0.7cm]{eq4point1}}(0)
\lambda_{\includegraphics[width=0.7cm]{eq2point1}}(0)
+ \lambda^3_{\includegraphics[width=0.7cm]{eq2point1}}(0)
+ {3 \over N}
 \lambda_{\includegraphics[width=0.7cm]{eq4point6}}(0)
\Bigg ]
t^3
+
{\mathcal O} (t^4)
\nonumber \\
\lambda_{\includegraphics[width=0.7cm]{eq2point1}}(t)  
&=& 
\lambda_{\includegraphics[width=0.7cm]{eq2point1}}(0) 
+
\Big [
\lambda_{{\includegraphics[width=0.7cm]{eq2point1}} \; {\includegraphics[width=0.7cm]{eq2point1}}}(0)
+
\left(3 + {1 \over N} 3\right)
\lambda_{\includegraphics[width=0.7cm]{eq4point1}}(0)
-
\lambda^2_{\includegraphics[width=0.7cm]{eq2point1}}(0)
\Big ]
t
\cr\cr
&&+
\Bigg [
{1 \over 2} \lambda_{{\includegraphics[width=0.7cm]{eq2point1}} \; {\includegraphics[width=0.7cm]{eq2point1}}\; {\includegraphics[width=0.7cm]{eq2point1}}}(0)
+ \left({9 \over 2} + {1 \over N} {9 \over 2}\right) \lambda_{\includegraphics[width=1.4cm]{eq4point2}} (0)
\Big.
\cr\cr
\Big.
&&
+ \left( 9 + {1 \over N}18\right) 
\lambda_{\includegraphics[width=0.7cm]{eq62point1}}(0)
+ \left( 3 + {1 \over N}9\right) 
 \lambda_{\includegraphics[width=0.7cm]{eq61point1}}(0)
+
3 \lambda_{{\includegraphics[width=0.7cm]{eq2point1}} \; {\includegraphics[width=0.7cm]{eq2point1}}}\lambda_{\includegraphics[width=0.7cm]{eq2point1}}(0)
\Big.
\cr\cr
\Big.
&&
+
\left( 9 + {1 \over N}9\right) 
\lambda_{\includegraphics[width=0.7cm]{eq4point1}}(0)
\lambda_{\includegraphics[width=0.7cm]{eq2point1}}(0)
+ \lambda^3_{\includegraphics[width=0.7cm]{eq2point1}}(0)
+ {3 \over N}
 \lambda_{\includegraphics[width=0.7cm]{eq4point6}}(0)
\Bigg ]
t^2
+ 
{\mathcal O} (t^3) \,,
\nonumber \\
\lambda_{{\includegraphics[width=0.7cm]{eq2point1}} \; {\includegraphics[width=0.7cm]{eq2point1}}}
(t) 
&=&
\lambda_{{\includegraphics[width=0.7cm]{eq2point1}} \; {\includegraphics[width=0.7cm]{eq2point1}}}(0)
+
\Bigg [
\lambda_{{\includegraphics[width=0.7cm]{eq2point1}} \; {\includegraphics[width=0.7cm]{eq2point1}}\; {\includegraphics[width=0.7cm]{eq2point1}}}(0)
+\left(6 + {1 \over N} 6\right)
 \lambda_{\includegraphics[width=1.4cm]{eq4point2}} (0)
\Bigg.
\cr\cr
\Bigg.
&&
+6 \lambda_{\includegraphics[width=0.7cm]{eq62point1}}(0)
+
4 \lambda_{{\includegraphics[width=0.7cm]{eq2point1}} \; {\includegraphics[width=0.7cm]{eq2point1}}}\lambda_{\includegraphics[width=0.7cm]{eq2point1}}(0)
+ {1 \over N} 6 
\lambda_{\includegraphics[width=0.7cm]{eq61point1}} 
\Bigg ]
t 
+
{\mathcal O} (t^2) \,, 
\cr\cr
\lambda_{\includegraphics[width=0.7cm]{eq4point1}}
(t) 
&=&
\lambda_{\includegraphics[width=0.7cm]{eq4point1}}(0)
+
\Bigg [
\lambda_{\includegraphics[width=1.4cm]{eq4point2}} (0)
+ \left(4 + {1 \over N} 8 \right)
 \lambda_{\includegraphics[width=0.7cm]{eq62point1}}(0)
\Bigg.
\cr\cr
\Bigg.
&&
+ \left(2  + {1 \over N} 2\right)
\lambda_{\includegraphics[width=0.7cm]{eq61point1}}(0)
+
4 \lambda_{\includegraphics[width=0.7cm]{eq4point1}}\lambda_{\includegraphics[width=0.7cm]{eq2point1}}(0)
+
{1 \over N} 2 
\lambda_{\includegraphics[width=0.7cm]{eq4point6}}(0)
\Bigg ]
t
+ 
{\mathcal O} (t^2) \,, 
\cr\cr
\lambda_{{\includegraphics[width=0.7cm]{eq2point1}} \; {\includegraphics[width=0.7cm]{eq2point1}}\; {\includegraphics[width=0.7cm]{eq2point1}}}(t)
&=&
\lambda_{{\includegraphics[width=0.7cm]{eq2point1}} \; {\includegraphics[width=0.7cm]{eq2point1}}\; {\includegraphics[width=0.7cm]{eq2point1}}}(0)
+ 
{\mathcal O} (t) \,, 
\cr\cr
\lambda_{\includegraphics[width=0.7cm]{eq61point1}}(t)
&=&
\lambda_{\includegraphics[width=0.7cm]{eq61point1}}(0)
+ 
{\mathcal O} (t) \,, 
\cr\cr
\lambda_{\includegraphics[width=0.7cm]{eq62point1}}(t)
&=&
\lambda_{\includegraphics[width=0.7cm]{eq62point1}}(0)
+ 
{\mathcal O} (t) \,, 
\cr\cr
\lambda_{\includegraphics[width=1.4cm]{eq4point2}}(t)
&=&
\lambda_{\includegraphics[width=1.4cm]{eq4point2}}(0)
+ 
{\mathcal O} (t)
\cr\cr
\lambda_{\includegraphics[width=0.7cm]{eq4point6}}(t)
&=&
\lambda_{\includegraphics[width=0.7cm]{eq4point6}}(0)
+ 
{\mathcal O} (t) \,.
\eea

\end{document}